\documentclass[a4paper]{article}
\usepackage{epsfig}
\usepackage{graphicx}
\newcommand{\gapprox}{\stackrel{>}{\sim}}
\newcommand{\lapprox}{\stackrel{<}{\sim}}
\begin{document}
\begin{titlepage}

\title{{\bf Solution of the Bogoliubov-de Gennes Equations at Zero Temperature Throughout the BCS-BEC Crossover: Josephson and Related Effects }}
\vspace{2cm}
\par
\author{A. Spuntarelli, P. Pieri, and G.C. Strinati
\\  Dipartimento di Fisica, 
\\ Universit\`{a} di Camerino, I-62032 Camerino, Italy}
\vspace{1cm}

\date{}
\maketitle
\vspace{1cm}

\begin{abstract}
The BCS-BEC crossover has received much attention lately, owing especially to its experimental realization with trapped ultracold Fermi atoms.
Theoretically, the two limiting situations, of paired fermions described by BCS theory in weak coupling and of composite bosons undergoing Bose-Einstein condensation (BEC) in strong coupling, can be connected with continuity throughout the crossover.
This evolution encompasses the unitary limit at intermediate values of the coupling, where the scattering length for two-fermion scattering diverges.  
Several quantities have been measured experimentally and calculated theoretically in this context over the last several years, with the notable exception of the Josephson and related effects. 
This is in spite of the fact that the Josephson effect is intimately associated with the spontaneous breaking of  the phase of the complex order parameter which unifies superconductivity and superfluidity. 

In the present paper we aim at filling (at least partially) this gap and investigate the evolution of the Josephson and related effects throughout the BCS-BEC crossover, by performing a systematic numerical solution of the (time-independent) Bogoliubov-de Gennes equations at zero temperature in a \emph{fully self-consistent} fashion.
We consider a stationary and uniform current flowing in the presence of a three-dimensional barrier with a slab geometry. This extended geometry is specifically required to reach the BEC limit of the crossover, where the formation of composite bosons in terms of their fermionic constituents requires consideration of wave vectors with components along all three dimensions. 
In addition, we regard the fermionic attraction to extend unmodified over the barrier region, a situation that typically applies to ultracold Fermi atoms.
The fully self-consistent solution of the Bogoliubov-de Gennes equations in such an extended geometry and coupling range represents a non-trivial numerical calculation.
The numerical strategies and algorithms we have adopted will therefore be described in detail, with the aim of easing further independent studies.

Several results are obtained by the present calculation. 
The profiles of the magnitude and phase of the gap parameter across the barrier are determined under a variety of conditions.
We find that the Josephson current is considerably enhanced at about unitarity for all barriers we have considered.
A related enhancement is also found in the contribution to the total current from the Andreev bound states, which stem from the depression of the gap profile about the barrier.
The Josephson current-phase characteristics (relating the total current $J$ to the phase difference $\delta \phi$ across the barrier) turn out to evolve from the standard $J \propto \sin \delta \phi$ relation to $J \propto \cos \delta \phi/2$, when the height of the barrier is decreased at fixed coupling or the coupling is decreased for given barrier.
For vanishing barrier height, we find that the critical Josephson current approaches the limiting value predicted by the Landau criterion, which is determined by either pair-breaking or sound-mode excitations depending on the coupling value.  
In the BCS limit, we reveal the presence of Friedel oscillations in the oscillatory modulations of the gap and density profiles.
In this limit, we also emphasize the special role played by the Andreev bound state in determining the critical Josephson current in the presence of a barrier.
Finally, the stability of the two branches, out of which the Josephson characteristics are composed, is analyzed by calculating the energy required to produce a given spatial profile of the gap parameter.
\end{abstract}

\end{titlepage}

\newpage
\tableofcontents

\newpage

\begin{center}
{\bf List of symbols:}
\end{center}
\vspace{-0.3cm}

\noindent
$a_{B},a_{F}$ = bosonic (B) and fermionic (F) scattering lengths
\vspace{0.01cm}

\noindent
$E_{F} = k_{F}^{2}/(2m)$ = Fermi energy
\vspace{0.01cm}

\noindent
$\mathcal{E}$ = total energy of the system
\vspace{0.01cm}

\noindent
$-g$ = bare coupling strength of the fermionic attraction
\vspace{0.01cm}

\noindent
$J$ = Josephson current
\vspace{0.01cm}

\noindent
$L$ = width of a rectangular barrier
\vspace{0.01cm}

\noindent
$k_{F}$ = Fermi wave vector
\vspace{0.01cm}

\noindent
$(k_{\perp},\mathbf{k}_{\parallel})$ = wave-vector components orthogonal and parallel to the barrier
\vspace{0.01cm}

\noindent
$m, m_{B}$ = fermion and boson (B) mass
\vspace{0.01cm}

\noindent
$n(x)$ = profile of the fermionic number density 
\vspace{0.01cm}

\noindent
$n_{0}$ = bulk value of the fermionic number density away from the barrier
\vspace{0.01cm}

\noindent
$q$ = wave vector associated with the supercurrent
\vspace{0.01cm}

\noindent
$s$ = sound velocity
\vspace{0.01cm}

\noindent
$V(x)$ = one-body potential representing the barrier 
\vspace{0.01cm}

\noindent
$V_{0}$ = height of the barrier 
\vspace{0.01cm}

\noindent
$(u_{\nu},v_{\nu})$ = eigen-solution components of the BdG equations
\vspace{0.01cm}

\noindent
$x$ = spatial coordinate orthogonal to the barrier
\vspace{0.01cm}

\noindent
$\mathcal{Z}$ = strength of a Dirac-delta barrier
\vspace{0.01cm}

\noindent
$\delta \phi$ = phase difference accumulated by the gap parameter across the barrier
\vspace{0.01cm}

\noindent
$\Delta(x)$ = spatial profile of the (complex) gap parameter
\vspace{0.01cm}

\noindent
$\Delta_{0}$ = bulk magnitude of the gap parameter away from the barrier
\vspace{0.01cm}

\noindent
$\varepsilon_{0} = (m a_{F}^{2})^{-1}$ = two-fermion binding energy
\vspace{0.01cm}

\noindent
$\mu$ = fermionic chemical potential in the presence of a supercurrent
\vspace{0.01cm}

\noindent
$\mu_{0}$ = fermionic chemical potential in the absence of a supercurrent
\vspace{0.01cm}

\noindent
$\mu_{B}$ = chemical potential of composite bosons
\vspace{0.01cm}

\noindent
$\sigma$ = width of a Gaussian barrier
\vspace{0.01cm}

\noindent
$2 \phi(x)$ = spatial profile of the phase of the gap parameter
\vspace{0.01cm}

\noindent
$\Phi(x)$ = condensate wave function solution of the GP equation

\newpage
\section{Introduction}

This paper deals with the solution of the (time-independent) Bogoliubov-de Gennes (BdG) equations for a system of fermions with a mutual attraction of arbitrary strength.
The BdG equations [see Eqs.(\ref{BdG-equations}) below] are the analog of the Schr\"{o}dinger equation for two-component fermionic wave functions in the presence of a superconducting gap parameter \cite{BdG}, and generalize to spatially dependent situations the BCS mean-field approch 
for superconducting fermions \cite{BCS}.
Although these equations can, in principle, be utilized at any temperature in the superconducting phase, our interest here in extending their applicability to large values of the mutual attraction limits, in practice, their use to zero temperature for reasons that will be clarified below.

In particular, we shall be interested in phenomena involving macroscopic quantum coherence which are made manifest by the presence of inhomogeneities in the sample. 
Typically, we will consider a potential barrier that partitions the superconducting sample in two halves, such that a Josephson current can be established between the two sides of the barrier. 
The description of the Josephson and related effects will then constitute the principal concern of this paper.
By our approach, we will be able to explore these effects for parameter ranges that could not be 
explored before, specifically, when the mutual fermionic attraction giving rise to superconductivity is made arbitrarily large. 
From a technical point of view, this program requires us to refine to a considerable extent the self-consistent solution of the BdG equations, in such a way that one is able to describe the evolution from the weak-coupling limit (which has been considered so far in the literature) to the strong-coupling limit, when the original fermionic identity of the constituent particles is apparently lost in favor of bosonic entities.

Quite generally, we shall refer to the (dc) Josephson \emph{effect} as the occurrence of a stationary superfluid current $J$ flowing across a barrier of \emph{arbitrary} shape in an otherwise homogeneous system, such that a well-defined relation is established between the value of $J$ and the difference $\delta \phi$ accumulated by the phase of the superfluid order parameter across the barrier. This current-phase relation is expected to assume different forms, depending on the type and shape of the barrier \cite{GKI-RMP-2004}. These different forms can be regarded as defining different \emph{regimes} of the same underlying physical effect. In this respect, the analysis presented in this paper will enlarge the range of regimes which can be attained by the Josephson effect, by considering that the fermionic attraction can be varied in such a way that the description of  the supercurrent flow evolves with continuity from a BCS description in terms of fermions and Cooper pairs to a BEC description in terms of condensate composite bosons.

Our emphasis in such an evolution of the original fermionic system when the strength of the mutual attraction is increased, relates to what is nowadays known as the BCS-BEC crossover, from a standard (BCS) description when the mutual attraction is sufficiently \emph{weak\/}, to a description in terms of composite bosons which undergo Bose-Eistein condensation (BEC) when the mutual attraction is sufficiently \emph{strong\/}.
The interest in this evolution has been considerably stimulated by the observation that high-temperature cuprate superconductors have coherence lengths much smaller than conventional superconductors, and more recently by the full experimental realization of the BCS-BEC crossover with ultracold Fermi atoms.

We will thus begin by giving a short account of this crossover especially in the context of the physics of ultracold Fermi atoms, a field which has been very much active over the last several years and offers many opportunities to come for exploring previously not accessible physical regimes.
We refer the reader to Refs.\cite{BDZ-RMP-2008,GPS-RMP-2008} for recent reviews of this field
[see also Ref.\cite{Levin-PR-2005} for a more specific review of a theoretical approach to the problem].
In the rest of the paper we shall then proceed, more specifically, to a detailed description of the Josephson and related effects throughout the BCS-BEC crossover, thus filling a notable gap in the literature.
By this calculation, several novel results will be obtained and their physical implications discussed at some length.

\newpage
\vspace{0.3cm}
\begin{center}
\begin{large}
{\bf 1a. The BCS-BEC crossover and the physics of ultracold Fermi atoms}
\end{large}
\end{center}
\addcontentsline{toc}{subsection}{1a. The BCS-BEC crossover and the physics of ultracold Fermi atoms}
\vspace{0.1cm}

In conventional superconductors (like Al, Hg, Sn, $\cdots$), electrons with opposite spin pair to form \emph{Cooper pairs\/} at temperature below the superconducting critical temperature $T_{c}$.
In these superconductors, the average size $\xi_{\mathrm{pair}}$ of the pairs is much larger than the average interparticle distance $k_{F}^{-1}$ (where $k_{F}$ is the Fermi wave vector),
such that the product $k_{F} \, \xi_{\mathrm{pair}}$ turns out to be about $10^{3}-10^{5}$.
The Cooper pairs are thus largely overlapping and it is not appropriate to regard them as spin-zero bosons (in this case, in fact, it is more appropriate to refer to the correlation between fermions with opposite spins at a certain distance).
This consideration has been central to the BCS theory of superconductivity \cite{BCS} and has made it possible its successful application to conventional superconductors, because the large spatial overlap among Cooper pairs justifies a description based on a ``mean-field'' analysis of the system.
In this respect, the superconducting transition in conventional superconductors should be regarded as rather exceptional among second-order phase transitions, to the extent that a mean-field treatment can accurately account for the experimental observations.

This framework has changed considerably with the discovery of high-temperature cuprate superconductors, for which $k_{F} \, \xi_{\mathrm{pair}}$ is of the order $5-10$ only.
The BCS assumption of largely overlapping fermion pairs is thus not fully realistic for these materials, requiring the BCS theory to be suitably modified.
Specifically, for these novel superconductors the coupling between fermions seems to be somewhat \emph{intermediate\/} between  situations with highly overlapping Cooper pairs and with composite bosons spatially separated from each other.
[Here and in the following, the term ``composite'' bosons relates to the fact that the dissociation temperature of these entitites is comparable with the critical temperature for condensation, while for ``point-like'' bosons (like
$^{4}\mathrm{He}$) the two temperature scales differ considerably.]
 
These considerations have promoted the development of a theoretical approach which is able to connect the BCS limit as described above in terms of Cooper pairs with the Bose-Einstein condensation of composite bosons.
Since the evolution between these two (BCS and BEC) limits occurs with continuity without the occurrence of an intervening phase transition, the phenomenon is referred to as the BCS-BEC \emph{crossover\/}. 

There actually exist some theoretical papers \cite{Eagles,Leggett,NSR} which have dealt with this crossover beforehand, motivated by the need of producing a rational description of the evolution between the BCS and BEC limits.
Subsequently with the discovery of high-temperature cuprate superconductors, the interest in the BCS-BEC crossover has surged \cite{Randeria-90,Haussmann,PS-1994,Zwerger,Levin}.
It is, however, only with the advent of the experiments on ultracold trapped atoms that the physics of 
the BCS-BEC crossover has been promoted to the attention of the scientific community at large, being of interest in such apparently different fields like atomic and molecular physics, condensed matter physics, nuclear physics, elementary particle physics, and astrophysics.

On the experimental side, the main novelty introduced by the use of ultracold trapped Fermi atoms to explore the BCS-BEC crossover is the possibility of varying essentially at will the attractive interaction between fermions of different species \cite{FF-resonances-e}, since it is just the occurrence of this attraction that makes it possible the formation of Cooper pairs in the medium and of composite bosons in vacuum, starting from the isolated fermionic species. 
[We anticipate that the electronic spin quantum number of superconductors is now replaced by a different quantum number associated with the atomic hyperfine levels of the Fermi atoms.]
It is just this possibility that allows ultracold Fermi atoms to be considered as \emph{prototype\/} systems in Nature with respect to others for which this possibility is strongly limited.

In practice, in ultracold Fermi atoms the attractive interaction is effectively varied by the use of \emph{Fano-Feshbach resonances\/} \cite{FF-resonances-t}, which are characterized by a resonant coupling between a two-atom scattering state with vanishing energy and a bound state in a closed channel.
By suitably scanning the position of the bound state with respect of this zero-energy level through the variation of a magnetic field, the corresponding \emph{scattering length\/} $a_{F}$ can be varied arbitrarily, passing from negative values before the bound-state is formed in the two-body problem, to positive values once the bound state has appeared through the resonance \cite{Fano-Rau}.

The BCS-BEC crossover has been realized in this way with $^{6}\mathrm{Li}$ and $^{40}\mathrm{K}$ ultracold Fermi atoms.
The Fano-Feshbach resonances utilized for the purpose are sufficiently ``broad'' that the fermionic many-body problem can be effectively described in a simplified manner by a single-channel Hamiltonian with an instantaneous interaction \cite{SPS-2005}.
For ``narrow'' resonances, on the other hand, the single-channel description turns out not to be accurate, and it is not even possible in this case to follow adequately the evolution between the BCS and BEC regimes with the fermionic densities currently available in the traps \cite{SPS-2005}.

Several are the experiments which have been realized so far with ultracold trapped Fermi atoms.
Among those, we mention:
(i) The initial production of composite bosons \cite{Jin-2003-I,Hulet-2003};
(ii) The Bose-Einstein condensation of composite bosons \cite{Jin-2003-II,Grimm-2003,Ketterle-2003};
(iii) The evidence of a condensate on the BCS side of the crossover \cite{Jin-2004,Ketterle-2004,Ketterle-2005-I};
(iv) The evidence of a pairing gap \cite{Grimm-2004-I,Jin-2005-I,Ketterle-2007-I};
(v) The measurement of collective modes \cite{Thomas-2004,Grimm-2004-II};
(vi) The evidence for a component in the closed channel \cite{Hulet-2005};
(vii) The ``shot noise'' atomic correlations \cite{Jin-2005-II};
(viii) The evidence for vortices throughout the whole crossover \cite{Ketterle-2005-II};
(ix) The introduction of imbalanced spin populations \cite{Ketterle-2006,Hulet-2006};
(x) The study of transport properties \cite{Inguscio} and band structures \cite{Esslinger-2005} in optical lattices;
(xi) The measurement of thermodynamic properties \cite{Thomas-2007};
(xii) The measurement of the critical velocity for superfluid flow \cite{Ketterle-2007};
(xiii) The measurement of single-particle spectral features \cite{Jin-2008}.

On the theoretical side, an adequate description of the whole crossover bridging the BCS and BEC regimes represents a difficult problem which has not been fully solved yet.
By its very definition, a \emph{single\/} crossover theory should be able to capture the two limiting behaviors
which aims at connecting (in the present context, they are the BCS theory on the one side and the description of a dilute Bose system on the other side.)
Here, the main difficulty is represented by the need of describing a many-body system in the absence of a small parameter, whose presence would allow one to adequately control the theoretical approximations.
Such a small parameter (identified by the product $k_{F} |a_{F}|$) actually exists separately in the two BCS and BEC regimes, where it can be used to control the approximations to which the given crossover theory reduces for \emph{dilute\/} Fermi and Bose systems, separately.
The lack of a small parameter thus concerns only the ``intermediate'' region between the two regimes (which is centered about the so-called \emph{unitary limit\/} where $|a_{F}|$ diverges) where the product $k_{F} |a_{F}|$ can by far exceed unity.
A possible strategy in this context consists in restricting as much as possible the extension of the intermediate region where a stringent control of the approximations is not possible, through a successive refinement of the quality of the approximations to which the single crossover theory reduces in the two BCS and BEC regimes that are contiguous to the intermediate region.
This strategy has been adopted, for instance, by the diagrammatic approach of Refs.\cite{PS-2000,PPS-2004} which includes ``pairing fluctuations'' beyond mean field, an approach 
that has produced good numerical results also in the intermediate (crossover) region and even at finite temperatures.
Note that the inclusion of pairing fluctuations is especially required to describe the BEC regime at finite temperatures, for which the density of non-condensed bosons acquires a major role.
By this approach, fluctuations effects beyond mean field are initially considered for a homogeneous system and then extended to the presence of a trapping potential that confines the interacting Fermi gas via a local-density approximation, whereby the system is considered to be locally homogeneous.

When dealing with the Josephson and related effects throughout the BCS-BEC crossover of interest in the present paper, however, the need arises for dealing with inhomogeneities in the system (such as those due to the presence of a barrier separating  two superconducting regions) beyond the local-density approximation.
In this case, the inclusion of fluctuations beyond mean field proves much harder than for the corresponding homogeneous case, so that one may be willing to limit to a mean-field approach when performing practical calculations.
That this may not be too severe a limitation, however, stems from the consideration that, in the presence of a spatially dependent external potential, the outcomes of a mean-field calculation should already be highly nontrivial, to the extent that the imprint of the excitation spectrum of a quantum-mechanical inhomogeneous system can be already found in its ground state \cite{Gross}.
In the following, we shall thus limit ourselves to consider the BdG equations \emph{at zero temperature\/} and explicitly verify this subtle physical expectation from the results of our numerical calculations.

When calculating physical quantities across the BCS-BEC crossover for the homogeneous case, one may often rely on analytic results in the two BCS and BEC limits, against which the results of the fully numerical calculations can be confronted.
A similar need for confronting with analytic results shows up when solving numerically the BdG equations in the presence of inhomogeneities.
Special care will thus be devoted in the following to identify alternative and independent procedures, which provide independent numerical tests for the results of the BdG equations both in the BCS and BEC limits. 

The simplest description of the BCS-BEC crossover occurs at the mean-field level for a homogeneous system in the zero-temperature limit.
In this case, the usual BCS equation for the gap parameter $\Delta_{0}$ is supplemented by the equation for the density $n$ which specifies the fermionic chemical potential $\mu_{0}$, and the two coupled equations are solved simultaneously for given density and coupling.
These two equations read:

\begin{equation}
\int \! \frac{d\mathbf{k}}{(2 \pi)^{3}} \, \left( \frac{1}{2 E_{\mathbf{k}}} \, - \, \frac{m}{\mathbf{k}^{2}} \right) \, = \, - \, \frac{m}{4 \, \pi \, a_{F}}                                 \label{gap-homogeneous}
\end{equation}

\begin{equation}
n \, = \, \int \! \frac{d\mathbf{k}}{(2 \pi)^{3}} \, \left(1 \, - \, 
\frac{\xi_{\mathbf{k}}}{E_{\mathbf{k}}} \right)                         \label{density-homogeneous}
\end{equation}

\noindent
where $\mathbf{k}$ is a wave vector, $m$ the fermion mass, $\xi_{\mathbf{k}}=\mathbf{k}^{2}/(2m) - \mu_{0}$, and
$E_{\mathbf{k}} =\sqrt{\xi_{\mathbf{k}}^{2} + \Delta_{0}^{2}}$.
Note that the gap equation (\ref{gap-homogeneous}) has been suitably regularized to remedy for the presence of an ultraviolet divergence that originates from the use of a contact potential to represent the fermionic attractive interaction.
We shall further comment on this point in sub-section 1b below.

Adopting the Fermi wave vector $k_{F}=(3 \pi^{2} n)^{1/3}$ as the unit of
$|\mathbf{k}|$, one sees from the right-hand side of Eq.(\ref{gap-homogeneous}) that the 
quantity $(k_{F} a_{F})^{-1}$ plays the role of the \emph{coupling parameter\/} of the theory.
Depending on the sign of $a_{F}$, this parameter ranges from 
$(k_{F}\, a_{F})^{-1} \lapprox -1$ characteristic of the weak-coupling BCS regime when $a_{F} < 0$, to $(k_{F}\, a_{F})^{-1} \gapprox +1$ characteristic of the strong-coupling BEC regime when $a_{F} > 0$, across the value $(k_{F}\, a_{F})^{-1}=0$ at unitarity when
$|a_{F}|$ diverges.
In practice, the ``crossover region'' of most interest is limited to the interval 
$-1 \lapprox (k_{F}\, a_{F})^{-1} \lapprox +1$, with the approximate values of the boundaries depending on the specific physical quantity at hand.

\vspace{-0.2cm}
\begin{center}
\begin{figure}[htc]
\includegraphics[angle=0,width=6.5in]{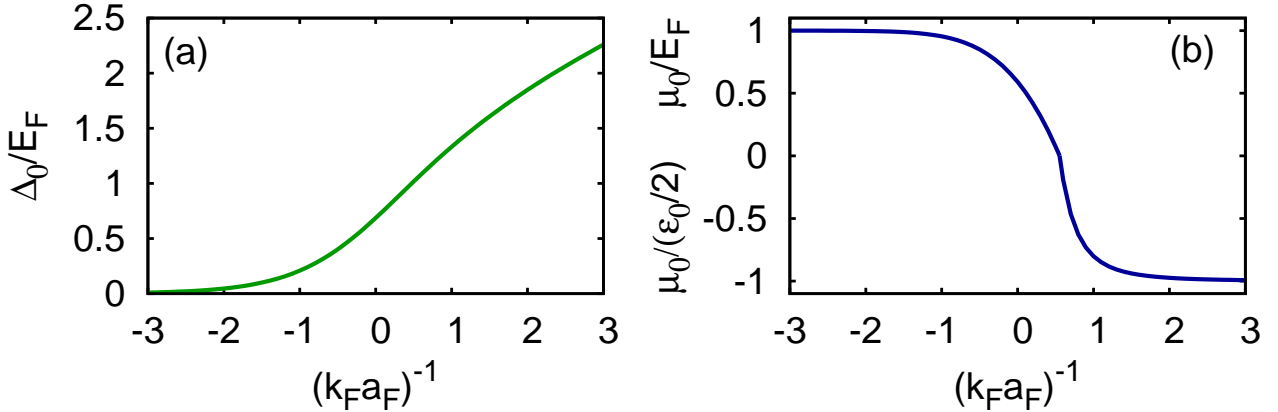}
\caption{(a) Gap parameter $\Delta_{0}$ and (b) chemical potential $\mu_{0}$ for a homogeneous system vs the coupling
              parameter $(k_{F} a_{F})^{-1}$, evaluated within the mean-field approximation at zero temperature across the BCS-BEC crossover.}
\label{fig1}
\end{figure}
\end{center} 

The equations (\ref{gap-homogeneous}) and (\ref{density-homogeneous}) can be solved analytically for $\Delta_{0}$ and $\mu_{0}$ vs $(k_{F}\, a_{F})^{-1}$ in terms of elliptic integrals, as shown in Ref.\cite{MPS}. These solutions are plotted in Fig.\ref{fig1}, where we note that a different normalization has been adopted for $\mu_{0}$ depending on its sign (namely, the Fermi energy $E_{F}=k_{F}^{2}/(2m)$ when $\mu_{0}>0$ and the two-fermion binding energy $\varepsilon_{0} = (m a_{F}^{2})^{-1}$ when $\mu_{0}<0$).
While $\Delta_{0}$ is a monotonically increasing function of $(k_{F}\, a_{F})^{-1}$, $\mu_{0}$ is seen to drop rather sharply from the BCS to the BEC values across the crossover region of limited extension.
In reverse, one may say that it is just this characteristic behavior of the chemical potential to drive the BCS-BEC crossover from the presence of an underlying Fermi surface
(represented by $E_{F}$) to the occurrence of pre-formed fermion pairs (represented by $\varepsilon_{0}$).

The values of $\Delta_{0}$ and $\mu_{0}$ reported in Fig.\ref{fig1} for a homogeneous system will also be of use for the solution that we shall undertake of BdG equations in the presence of a barrier of \emph{finite\/} width.
This is because the value $\Delta_{0}$ will be asymptotically reached far away from the barrier by the magnitude of the spatially dependent gap parameter, while the value $\mu_{0}$ will remain unaffected by the presence of the finite barrier which perturbs the density profile only locally. 
 
 
\vspace{0.3cm}
\begin{center}
\begin{large}
{\bf 1b. Single-channel Hamiltonian and regularization of the fermionic attractive interaction}
\end{large}
\end{center}
\addcontentsline{toc}{subsection}{1b. Single-channel Hamiltonian and regularization of the fermionic attractive interaction}
\vspace{0.1cm}

It is worth commenting more extensively at the outset on the use of a contact potential to represent 
the attractive fermionic interaction, as well on the procedure we have adopted to regularize the divergences that may arise in physical quantities when using this kind of potential.

We have already mentioned that the attractive interaction between two Fermi atoms 
($^{6}\mathrm{Li}$ and $^{40}\mathrm{K}$ are commonly used in experiments) is provided by 
a molecular mechanism referred to as Fano-Feshbach resonance \cite{FF-resonances-t}.
This mechanism is characterized by a resonant coupling between a two-atom scattering state with vanishing energy in an open channel and a bound (molecular) state in a closed channel.
Upon varying the position of the bound state with respect to the threshold of the open channel (which is obtained, in practice, by varying an applied magnetic field that acts differently on the open and closed channels), the value of the scattering length $a_{F}$ at the threshold of the open channel can be changed essentially at will, passing from negative values in the absence of the bound-state 
to positive values once the bound state has formed through the resonance  \cite{Fano-Rau}.
In all cases, an \emph{attraction\/} has effectively developed between the two Fermi atoms with different internal states (which are conventionally assimilated to ``spin up'' and ``spin down'' states).
\begin{center}
\begin{figure}[htc]
\hspace{3truecm}\includegraphics[angle=0,width=3.5in]{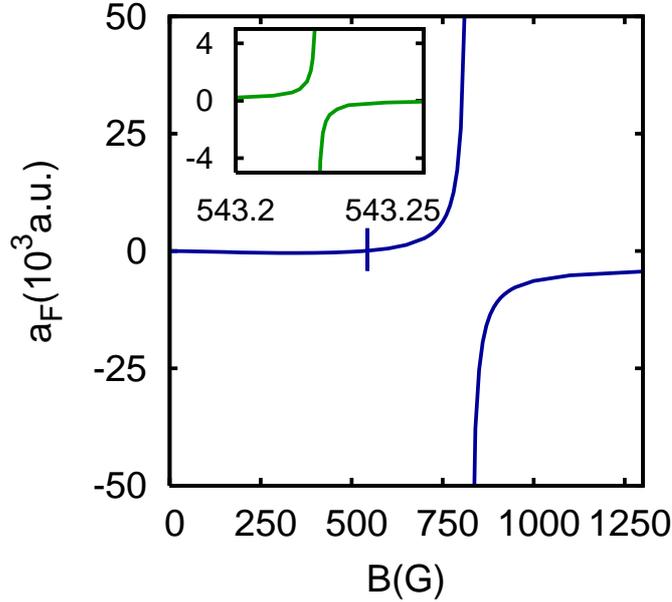}
\caption{Scattering length $a_{F}$ (in atomic units) for the collision of two fermionic $^{6}\mathrm{Li}$ atoms 
              vs the magnetic field $\mathrm{B}$ (in Gauss) that detunes the scattering from the close channel 
              (the figure is adapted from Ref.\cite{SPS-2005}). 
             The inset show the behavior of $a_{F}$ for the narrow resonance.} 
\label{fig2}
\end{figure}
\end{center} 
As an example, the scattering length for collisions of two $^{6}\mathrm{Li}$ atoms is reported in Fig.\ref{fig2} vs the magnetic field 
$\mathrm{B}$. 
Note that in this case $a_{F}$ diverges at the position of the ``broad'' resonance at 834 G and of the ``narrow'' resonance at 543 G.

Provided $a_{F}$ is the \emph{only\/} relevant parameter which characterizes the low-energy resonant scattering, the effective attraction between two fermionic atoms of different spin components at spatial positions $\mathbf{r}$ and $\mathbf{r'}$ can be conveniently represented by a two-body contact potential of the form $ - g \, \delta(\mathbf{r}-\mathbf{r'})$ with negative strength 
($g > 0$).
In this case, some care should be exerted when relating the value of $g$ to the scattering length $a_{F}$ which is meant to represent.
This relation is obtained by solving the integral equation in the low-energy limit for the two-body t-matrix associated with the contact potential \cite{RT}, which reads:
\begin{equation}
\frac{m}{4 \, \pi \, a_{F}} \, = \, - \, \frac{1}{g} \, + \, \int^{k_{0}} \! \frac{d\mathbf{k}}{(2 \pi)^{3}} \,\,
 \frac{m}{\mathbf{k}^{2}} \,\, .                                                   \label{regularization}
\end{equation}

\noindent
In this expression, the delta-function potential has been suitably regularized by introducing an ultraviolet cutoff $k_{0}$ in the (otherwise divergent) integral on the right-hand side.
Equation (\ref{regularization}) provides a prescription to relate the cutoff $k_{0}$ and the strength $g$ to a given value of $a_{F}$, by letting $k_{0} \rightarrow \infty$ and $g \rightarrow 0$ in such a way that \cite{PS-2000}:
\begin{equation}
g \, = \, \frac{2 \, \pi^{2}}{m \, k_{0}} \, + \, \frac{\pi^{2}}{m \, a_{F} \, k_{0}^{2}}  \,\, .
                                                                                                                          \label{g-k_o-vs-a_F}
\end{equation}

\noindent
In some circumstances, Eq.(\ref{regularization}) as it stands can be directly used to regularize expressions with the same kind of divergence, as it was already done for the gap equation (\ref{gap-homogeneous}). 

With these provisions, one says that the scattering problem for two Fermi atoms, which had originally involved (at least) the two channels connected via the Fano-Feshbach resonance, has been effectively modeled by a \emph{single-channel Hamiltonian\/} with a two-body point-contact potential.
The ensuing treatment of the many-body Hamiltonian is in a sense ``universal'', insofar as it will not depend on microscopic details that would distinguish different physical systems \cite{Braaten-2008}.
We recall that a similar attitude was adopted when the microscopic BCS theory of superconductivity was formulated in terms of the Gorkov point-contact potential, which retains, however, a finite upper cutoff given by the Debye frequency and is therefore limited 
to dealing with weak-coupling situations \cite{BCS}.

It was shown in Ref.\cite{SPS-2005} that $a_{F}$ can indeed be considered as the only relevant parameter characterizing the low-energy scattering, whenever the Fano-Feshbach resonance is sufficiently ``broad'' that other characteristic lengths (like the so-called ``effective range'') can be neglected for all practical purposes.
In these cases, it turns out that the probability of finding the two-atom system in the closed channel 
is sufficiently small, such that only the open channel can be retained for an effective description of the system.
The presence of the closed channel remains, of course, essential for an \emph{ab initio\/} determination of the scattering length $a_{F}$ itself as a function of the magnetic field $\mathrm{B}$, like in the calculation shown 
in Fig.\ref{fig2}.
In the single-channel model, on the other hand, the correspondence between the values of $a_{F}$ and $\mathrm{B}$ can only be fed in as an external information.
This is the model that we are going to adopt in the following for a numerical solution of the BdG equations across the BCS-BEC crossover in the presence of a barrier.


\vspace{0.3cm}
\begin{center}
\begin{large}
{\bf 1c. Summary of the main results about the Josephson effect throughout the BCS-BEC crossover}
\end{large}
\end{center}
\addcontentsline{toc}{subsection}{1c. Summary of the main results about the Josephson effect throughout the BCS-BEC crossover}
\vspace{0.1cm} 

When dealing with ultracold Fermi atoms, typical values for the width of the barriers that can be introduced 
in the system are about 5-8 $\mu$m and the largest spatial extension of the atomic cloud is about a few hundreds 
$\mu$m.
Under these conditions, the external magnetic field driving the Fano-Feshbach resonance cannot be 
varied in the interior of the barrier region, in such a way to make the inter-particle attraction to vanish there.
For these reasons, we shall limit in the following to consider an SsS barrier, with the same kind of superfluid extending on the two sides (S) of the barrier and with the fermionic attraction stretching unmodified over the barrier (s) region.
This differs from the SnS situation encountered while dealing with the Josephson effect in more conventional condensed-matter superconductors, whereby a normal (n) region lacking the fermionic attraction is embedded in an otherwise infinite superconductor (S).

An additional property which could be readily considered for ultracold Fermi atoms is the density imbalance between the two spin populations \cite{Ketterle-2006,Hulet-2006}.
This imbalance acts as an effective magnetic field which affects the spin but not the orbital degrees 
of freedom, and may thus let one to reveal effects which are obscured in condensed-matter systems, even as far as the Josephson and related effects are concerned.
In this paper, we shall limit to consider balanced pupulations only (as it was already done in Eqs.(\ref{gap-homogeneous}) and (\ref{density-homogeneous})), and defer consideration of imbalanced populations to a subsequent study.

This paper will thus be devoted to a \emph{systematic\/} self-consistent solution of the time-independent BdG equations with an SsS barrier and equal spin populations throughout the BCS-BEC crossover
\cite{Spuntarelli-07}.
Focus on this crossover will inevitably require us to avoid approximations which are specific of the weak-coupling (BCS) regime as usually considered in the BdG literature.
In addition, to acquire definite confidence in our numerical calculations, we will test them by comparison with calculations based on alternative approaches which are specific to the weak-coupling (BCS) and strong-coupling (BEC) regimes.
In particular, the connection which emerges in the strong-coupling limit between the fermionic BdG equations
and the Gross-Pitaevskii (GP) equation for composite bosons, that was established at a formal level in Ref.\cite{PS-2003} and will be verified here by numerical calculations, will enable us to connect phenomena of macroscopic quantum coherence occurring in fermionic and bosonic systems, by following their evolution from the one to the other end.
Especially in this limit, the need for a self-consistent solution of the equations (both BdG and GP) will be evident.

Besides the spatial profiles of the (magnitude and phase of the) order parameter, our calculations will 
produce the Josephson ``current vs phase characteristics'' for different couplings and barrier (height and width) 
values.
In particular, the value of the current will be analyzed in terms of a complete set of wave functions which are solutions of the BdG equations, in such a way that its partition between the contributions of the bound and continuum 
states will naturally emerge from the calculation.
Physically, the presence of these bound states (called the Andreev-Saint-James states \cite{Andreev-states}) is related to the depression of the magnitude of the order parameter about the repulsive barrier, and has been strictly associated with the very occurrence of the Josephson effect in the literature dealing with weak coupling only \cite{Kummel-1994}.

Quite generally, we find that the maximum Josephson current which is attainable for given barrier as well as the contribution of the Andreev bound states are maximal at (about) unitarity, which thus appears the coupling range to be preferred for studying the occurrence of macroscopic quantum phenomena.
We will attribute this feature to the reduced effect at unitarity of the elementary excitations in the system (namely, the pair-breaking excitations on the weak-coupling side and the sound-mode excitations on the strong-coupling side), which would act to destroy the superfluid flow.

The occurrence of these two types of excitations is explicitly revealed in the limiting situation of vanishing barrier height, whereby the infinitesimal barrier acts as a seed for triggering the collapse of the superfluid flow according to the Landau criterion \cite{AGD}.
We will show how our self-consistent calculation is able to detect the presence of this upper boundary for the maximum value of the Josephson current across the BCS-BEC crossover, by signalling the lost of the self-consistent solution when approaching from below the boundaries related to the two (pair-breaking and sound mode) types of excitations.
These two boundaries, which cross each other at (about) unitarity, delimit the region of stability of the stationary Josephson flow
in a current vs coupling ``phase diagram''.
The occurrence of these boundaries has been confirmed experimentally by a recent study performed at MIT
\cite{Ketterle-2007}.
In addition to leading to an identification of these boundaries, our calculation determines also how the maximum current approaches its Landau limiting value for vanishing barrier, for each coupling across the BCS-BEC crossover.

Several interesting features are further revealed by our calculation away from unitarity.
On the weak-coupling side, the profiles of the gap parameter and density appear modulated by the occurrence of Friedel oscillations of period $2 k_{F}$, and the presence of Andreev bound states below the threshold of pair-breaking excitations relates to the maximum Josephson current in the presence of a finite barrier.
On the strong-coupling side, on the other hand, we compare the numerical solutions of the BdG equations with the corresponding independent solution of the GP equation for the composite bosons, and reveal how characteristic features related to the composite nature of the bosons emerge, when the barrier width is sufficiently small with respect to the bosonic healing length that the composite bosons cannot be accomodated within the barrier region.

Finally, the \emph{stability} of the self-consistent solutions obtained by our numerical calculations will be studied by calculating the corresponding total energy (both in the presence and in the absence of the barrier) associated with the required boundary conditions.
This analysis will lead us eventually to relate the onset of the Landau criterion (when the supercurrent exceeds a certain threshold) with the energy instability of the self-consistent solutions of the BdG equations against the inclusion of fluctuations beyond mean field.

Additional material is confined to the Appendices, and includes both technical details as well as some clarifications about subtle aspects of the Josephson effect.

\vspace{0.5cm}  
\section{Implementing the self-consistent solution of the BdG equations}

The Bogoliubov-de Gennes equations embody the BCS mean-field theory of superconductivity for a system of fermions with a mutual attraction and subject to an external potential.
At zero temperature, solving for these equations corresponds to finding a (variational) broken-symmetry ground state of the many-body system in an inhomogeneous situation.
Although mean-field like, this solution turns, however, out to be highly non-trivial, since quite generally in the presence of a non-trivial geometry the imprint of the excitation spectrum of a quantum-mechanical system can be already found in its ground state \cite{Gross}. 

In this Section, we set up the solution of the BdG equations for a specific geometry appropriate to realize a Josephson link with a uniform current.
As we allow the mutual fermionic attraction to take arbitrary values, we shall avoid adopting a number of approximations often used in the literature, which are peculiar to the weak-coupling (BCS) limit of this attraction and fail accordingly to account for the formation of composite bosons in the opposite strong-coupling (BEC) limit.
Since the ensuing numerical solution of the BdG equations will unavoidably be quite involved, we shall rely on definite benchmarks with which to compare our numerical results in the two limiting (BCS and BEC) situations.
These will be identified with the delta-like barrier and the Gross-Pitaevskii equation, in the order, to be discussed separately below.
To avoid overwhelming here the reader with details, some of them will be deferred to the Appendices A and B.


\vspace{0.3cm}
\begin{center}
\begin{large}
{\bf 2a. The Bogoliubov-de Gennes equations}
\end{large}
\end{center}
\addcontentsline{toc}{subsection}{2a. The Bogoliubov-de Gennes equations}
\vspace{0.1cm}

The fermionic BdG equations read \cite{BdG}:
\begin{equation}
\left( 
\begin{array}{cc}
\mathcal{H}(\mathbf{r}) & \Delta(\mathbf{r})            \\
\Delta(\mathbf{r})^{*}  & - \mathcal{H}(\mathbf{r})  
\end{array} 
\right)
\left( \begin{array}{c}
u_{\nu}(\mathbf{r}) \\
v_{\nu}(\mathbf{r}) 
\end{array} 
\right) 
= \epsilon_{\nu}
\left( \begin{array}{c}
u_{\nu}(\mathbf{r}) \\
v_{\nu}(\mathbf{r}) 
\end{array} 
\right)  \,\, .                                          \label{BdG-equations} 
\end{equation}

\noindent
Here, $\mathcal{H}(\mathbf{r}) = - \nabla^{2}/(2m) + V(\mathbf{r}) - \mu$ where $V(\mathbf{r})$ is the external potential and $\mu$ the chemical potential (we set $\hbar = 1$ throughout).
The local gap parameter $\Delta(\mathbf{r})$ is determined via the \emph{self-consistent condition}:
\begin{equation}
\Delta(\mathbf{r}) = g \sum_{\nu} u_{\nu}(\mathbf{r}) 
v_{\nu}(\mathbf{r})^{*}
\left[ 1 - 2 f_{F}(\epsilon_{\nu}) \right]                                   \label{self-consistency}
\end{equation}

\noindent
where $f_{F}(\epsilon)=(e^{\epsilon/(k_{B}T)} +1)^{-1}$ is the Fermi function at temperature 
$T$ ($k_{B}$ being Boltzmann constant) and $g$ the coupling constant discussed in sub-section 1b. 
In addition, the functions $\{u_{\nu}(\mathbf{r}),v_{\nu}(\mathbf{r})\}$ obey the orthonormality condition:
\begin{equation}
\int \! d\mathbf{r} \, \left[ u_{\nu}(\mathbf{r})^{*} u_{\nu'}(\mathbf{r})
\, + \, v_{\nu}(\mathbf{r})^{*} v_{\nu'}(\mathbf{r}) \right] \, = \, \delta_{\nu\nu'}  
                                                                                                  \label{normalization-condition}                       
\end{equation}

\noindent
where the Kronecker delta on the right-hand side is readily generalized to the eigenvalues of a continuous set.

The equations (\ref{BdG-equations}) are formally analogous to the Schr{\"o}dinger equation for a two-component wave function, in which $\Delta(\mathbf{r})$ plays the role of an off-diagonal potential.
We will argue that the fact that this off-diagonal potential has to be determined self-consistently is of crucial importance for sustaining the Josephson effect, especially on the BEC side of the crossover.
Note further that in Eqs.(\ref{BdG-equations}) we have not explicitly included a (Hartree-type) diagonal potential (which would also need, in principle, to be determined self-consistently \cite{BdG}), since it vanishes owing to the regularization we have adopted for the contact potential.

In the following, we shall restict to consider only the positive eigenvalues $\epsilon_{\nu}$ of 
Eq.(\ref{BdG-equations}), so that $f_{F}(\epsilon_{\nu})$ in Eq.(\ref{self-consistency}) can be taken to vanish in the zero-temperature limit.
One can show, in fact, that positive eigenvalues only are required to describe the stationary current flow below the critical current at given coupling.
More generally, out of the two ``branches'' of eigenvalues of Eq.(\ref{BdG-equations}) one can limit to consider only the one that reaches large positive energies, with the argument that it suffices to identify a complete set of wave functions. 

Physical quantities of special concern will be the number density and current associated with the solutions of Eq.(\ref{BdG-equations}).
They are given, respectively, by the expressions:
\begin{equation}
n(\mathbf{r}) \, = \, 2 \, \sum_{\nu} \left[ f_{F}(\epsilon_{\nu}) \, |u_{\nu}(\mathbf{r})|^{2}  
\, + \, \left( 1 - f_{F}(\epsilon_{\nu}) \right) \, |v_{\nu}(\mathbf{r})|^{2} \right]       \label{density}
\end{equation}
\begin{eqnarray}
\mathbf{j}(\mathbf{r}) & = & \frac{1}{i m} \, \sum_{\nu} \left\{
f_{F}(\epsilon_{\nu}) \, \left[ u_{\nu}(\mathbf{r})^{*} \nabla u_{\nu}(\mathbf{r})  \, - \,
(\nabla u_{\nu}(\mathbf{r})^{*}) u_{\nu}(\mathbf{r}) \right]   \right.                      \nonumber \\
& + & \left. \left( 1 - f_{F}(\epsilon_{\nu}) \right)
\left[ v_{\nu}(\mathbf{r}) \nabla v_{\nu}(\mathbf{r})^{*}  \, - \,
(\nabla v_{\nu}(\mathbf{r})) v_{\nu}(\mathbf{r})^{*} \right]  \right\}  \,\, .              \label{current}
\end{eqnarray}

\noindent
Again, only positive eigenvalues will be considered for the sums in Eqs.(\ref{density}) and (\ref{current}), 
so that $f_{F}(\epsilon_{\nu})$ can be set to vanish in these equations.

The crucial role played by the self-consistent condition (\ref{self-consistency}) emerges when considering 
the continuity equation associated with the current (\ref{current}).
With the help of the BdG equations (\ref{BdG-equations}) one, in fact, arrives at the result:
\begin{equation}
\nabla \cdot \mathbf{j}(\mathbf{r}) \, = \, \sum_{\nu} 
\left( 2 f_{F}(\epsilon_{\nu}) - 1 \right) \, Q_{\nu}(\mathbf{r})         \label{continuity-equation}    
\end{equation}

\noindent
where in the (apparently) ``source'' term on the right-hand side we have set
\begin{equation}
Q_{\nu}(\mathbf{r}) \, = \, 4 \, \mathrm{Im} \left\{ \Delta(\mathbf{r}) u_{\nu}(\mathbf{r})^{*} 
v_{\nu}(\mathbf{r}) \right\}   \,\, .                                                              \label{Q-definition}
\end{equation}

\noindent
The self-consistent condition (\ref{self-consistency}), however, restores the validity of the
continuity equation $\nabla \cdot \mathbf{j}(\mathbf{r}) = 0$, since the right-hand side of
Eq.(\ref{continuity-equation}) vanishes when entering the expression (\ref{self-consistency}) therein.

The strength $- g$ of the local fermionic attraction is suitably eliminated in favor of the fermionic scattering length $a_{F}$, by adapting to the inhomogeneous case a procedure
introduced for the homogeneous case \cite{PS-2000} (see also sub-section 1b).
In that case, one arrived at the following \emph{regularized} gap equation (that is, for which no ultraviolet cutoff is required):

\begin{equation}
\int \! \frac{d\mathbf{k}}{(2 \pi)^{3}} \, \left[ (1 - 2 f_{F}(\mathbf{k})) \, 
\frac{u(\mathbf{k})v(\mathbf{k})^{*} }{\Delta_{0}} \, - \, \frac{m}{\mathbf{k}^{2}} \right]
\, = \, - \, \frac{m}{4 \, \pi \, a_{F}}   \,\, ;                      \label{regularized-gap-homogeneous}
\end{equation}

\noindent
in the inhomogeneous case one obtains correspondingly for the same coupling:
\begin{equation}
\sum_{\nu} (1 - 2 f_{F}(\epsilon_{\nu})) \, u_{\nu}(\mathbf{r}) v_{\nu}(\mathbf{r})^{*} \, - \,
\Delta(\mathbf{r}) \int \! \frac{d\mathbf{k}}{(2 \pi)^{3}} \, \frac{m}{\mathbf{k}^{2}}
\, = \, - \, \Delta(\mathbf{r}) \, \frac{m}{4 \, \pi \, a_{F}}   \,\, .
                                                                              \label{regularized-gap-inhomogeneous}
\end{equation}

\noindent
Multiplying both sides of Eq.(\ref{regularized-gap-homogeneous}) by $\Delta(\mathbf{r})$ and
subtracting the resulting expression from Eq.(\ref{regularized-gap-inhomogeneous}) yields eventually:
\begin{eqnarray}
0 & = & \sum_{\nu} (1 - 2 f_{F}(\epsilon_{\nu}))\,  u_{\nu}(\mathbf{r}) v_{\nu}(\mathbf{r})^{*}  
                                                                                    \nonumber \\
& - & \frac{\Delta(\mathbf{r})}{\Delta_{0}} 
\int \! \frac{d\mathbf{k}}{(2 \pi)^{3}} \, (1 - 2 f_{F}(\mathbf{k})) \, u(\mathbf{k})v(\mathbf{k})^{*}
\,\,.                                                                                \label{final-regularized-gap-equation}
\end{eqnarray}

\noindent
This is the form of the regularized gap equation that we shall implement in the numerical calculations.
Roughly speaking, its ultraviolet convergence is ensured by the fact that, for a large enough value of their energy, the wave functions $\{u_{\nu}(\mathbf{r}),v_{\nu}(\mathbf{r})\}$ in the presence of the external potential reduce to those of the homogeneous case (with $\Delta(\mathbf{r})$ replacing $\Delta_{0}$).
The ultraviolet convergence of the expression (\ref{final-regularized-gap-equation}) will be further discussed in Appendix A.

We are interested in the solutions of the BdG equations (\ref{BdG-equations}) that are subject to the self-consistent condition (\ref{self-consistency}) and sustain a finite value of the current (\ref{current}).
To this end, let's first consider the corresponding solutions for the homogeneous case in the absence of an external potential.
In this case, one sets
\begin{equation}
u_{\nu}(\mathbf{r}) \, \rightarrow \, u_{\mathbf{q}}(\mathbf{k}) \, 
e^{i(\mathbf{k} + \mathbf{q}) \cdot \mathbf{r}}             \,\,\,\,\,\,\,\, , \,\,\,\,\,\,\,\,
v_{\nu}(\mathbf{r}) \, \rightarrow \, v_{\mathbf{q}}(\mathbf{k}) \, 
e^{i(\mathbf{k} - \mathbf{q}) \cdot \mathbf{r}}  \,\,\,\,\,\,\,\, ,            \label{u-v-homogeneous}
\end{equation}

\noindent
such that $\Delta(\mathbf{r}) = \Delta_{\mathbf{q}} \, e^{2 i \mathbf{q} \cdot \mathbf{r}}$ 
($\Delta_{\mathbf{q}}$ real) and
\begin{equation}
\epsilon_{\nu} \, \rightarrow  \, \frac{\mathbf{k} \cdot \mathbf{q}}{m} \, + \, 
\sqrt{ \left( \frac{\mathbf{k}^{2}}{2m} + \frac{\mathbf{q}^{2}}{2m} - \mu \right)^{2} \, + \,
\Delta_{\mathbf{q}}^{2}}                                          \label{eigeinvalue-homogeneous}
\end{equation}

\noindent 
where $\Delta_{\mathbf{q}}$ does not depend on $\mathbf{q}$ insofar as $\epsilon_{\nu}$ 
of Eq.(\ref{eigeinvalue-homogeneous}) remains positive (the term $\mathbf{q}^{2}/(2m)$ therein can, in fact, be reabsorbed by expressing $\mu = \mu_{0} + \mathbf{q}^{2}/(2m)$ in terms of the chemical potential $\mu_{0}$ in the absence of the current 
$\mathbf{j}= \mathbf{q} \, n/m$ where $n$ is the uniform density). 
Correspondingly, the eigenvectors can be expressed as follows:
\begin{eqnarray}
u_{\mathbf{q}}(\mathbf{k})^{2} & = & 
\frac{(E + \xi_{\mathbf{k}-\mathbf{q}})^{2}}
{(E + \xi_{\mathbf{k}-\mathbf{q}})^{2} + \Delta_{\mathbf{q}}^{2} }    \nonumber \\                                            
v_{\mathbf{q}}(\mathbf{k})^{2} & = & \frac{\Delta_{\mathbf{q}}^{2}}
{(E + \xi_{\mathbf{k}-\mathbf{q}})^{2} + \Delta_{\mathbf{q}}^{2} }    \label{u2-v2-homogeneous}                                                  
\end{eqnarray}

\noindent
where $E$ stands for the right-hand side of Eq.(\ref{eigeinvalue-homogeneous}) and 
$\xi_{\mathbf{k}-\mathbf{q}} = (\mathbf{k}-\mathbf{q})^{2}/(2m) - \mu$, such that $E + \xi_{\mathbf{k}-\mathbf{q}}$, too, does
not depend on $\mathbf{q}$.
The positive square root is eventually taken of the expressions (\ref{u2-v2-homogeneous}).

The expressions (\ref{u-v-homogeneous}) and (\ref{eigeinvalue-homogeneous}) could have been obtained directly from the requirement of Galilean invariance, by connecting the solutions of the BdG equations in the lab frame (in which the current has the value $\mathbf{j}= \mathbf{q} \, n/m$) and in the frame moving with relative velocity $\mathbf{V} = \mathbf{q}/m$ (in which the superfluid is at rest).
Note that, with respect to the moving frame where the current vanishes, in the lab frame the eigenvectors 
(\ref{u-v-homogeneous}) acquire the phase $\varphi(\mathbf{r}) = \mathbf{q} \cdot \mathbf{r}$ which is linearly varying with 
$\mathbf{r}$.
The presence of a barrier breaks Galilean invariance and adds to $\varphi(\mathbf{r})$ an additional phase 
$\phi(\mathbf{r})$ (such that 
$\varphi(\mathbf{r}) = \mathbf{q} \cdot \mathbf{r} + \phi(\mathbf{r})$) that \emph{varies sharply about the barrier\/} so as to keep the current uniform. 
We pass now to discuss in detail the corresponding solutions for a particular geometry.


\vspace{0.3cm}
\begin{center}
\begin{large}
{\bf 2b. Solutions for a barrier with slab geometry}
\end{large}
\end{center}
\addcontentsline{toc}{subsection}{2b. Solutions for a barrier with slab geometry}
\vspace{0.1cm}

To be specific, we realize a Josephson link with a \emph{slab} geometry, whereby a potential barrier $V(x)$ is embedded in a homogeneous superconductor which extends to infinity on both sides of the barrier.
Although the profile of the barrier is one-dimensional (along the $x$ direction), the slab is meant to be fully three-dimensional since it extends along the two ($y$ and $z$) orthogonal directions.
This is because on the BEC side of the crossover the formation of composite bosons in terms of their fermionic constituents requires one to include wave vectors with components in all three dimensions. 
In this respect, we depart from previous treatments of the Josephson effect both for fermions \cite{Wendin-1996} and point-like bosons
\cite{Smerzi-1999} for which the formation of composite bosons was not an issue, even though taking into account the two orthogonal directions renders the present treatment considerably more involved.
For the same reasons, the solution of the BdG equations in a fully three-dimensional geometry was also considered in
Ref.\cite{Antezza-2007} while studying the evolution of a dark soliton throughout the BCS-BEC crossover (albeit for real wave
functions only).

\begin{center}
\begin{figure}[htc]
\hspace{3.5cm}\includegraphics[angle=0,width=3.5in]{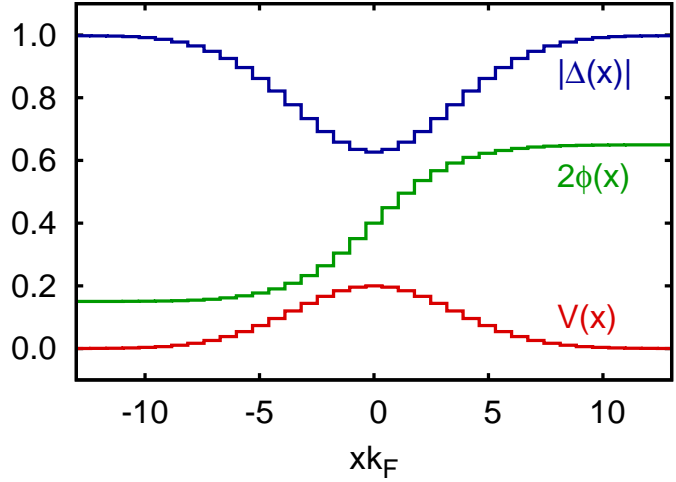}
\caption{Step-like representation of $V(x)$, $|\Delta(x)|$, and $2\phi(x)$ over a finite mesh of intervals about the
              position of a (Gaussian) barrier centered in $x=0$. Here, $|\Delta(x)|$ is normalized to its bulk value $\Delta_{0}$ 
              and $2\phi(x)$ has been shifted upward for clarity.}
\label{fig3}
\end{figure}
\end{center} 

The barrier we shall consider is of the SsS type, whereby the fermionic attraction extends unmodified under the barrier region (this is a typical situation encountered with ultracold Fermi atoms).
The more conventional SnS barrier, with a normal region embedded in an otherwise infinite superconductor, will not be considered in the present treatment.
Although the potential barrier $V(x)$ can in principle take a generic shape, in practice numerical calculations will be performed either with a rectangular barrier $V(x)$ of width $L$ and height $V_{0}$ or with a Gaussian barrier of the form $V(x) = V_{0} \, e^{-x^{2}/(2 \sigma^{2})}$ (where $V_{0} > 0$).

Operatively, we proceed to the self-consistent solution of the BdG equations (\ref{BdG-equations}) by extending to arbitrary values of the fermionic attraction the numerical approach to the scattering problem introduced for weak coupling in Ref.\cite{Bagwell-1996}. 
By this approach, the profiles of $V(x)$, $|\Delta(x)|$, and $\phi(x)$ are made piecewise constant (i.e., step-like) over a number $M$ 
of intervals, such that
\begin{equation}
V(x) \, = \, V_{\ell}  \,\,\,\,\,\,\,\,\, \mathrm{and}  \,\,\,\,\,\,\,\,\, 
\Delta(x) \, = \, \Delta_{\ell} \, e^{2 i (q x + \phi_{\ell})}                         \label{piecewise-constant}
\end{equation}

\noindent
($ \Delta_{\ell}$ real) when $x_{\ell - 1} \leq x \leq x_{\ell}$ with $\ell = (1, \cdots, M)$.
[The points $\{ x_{\ell} \}$ need not be equally spaced, and one can take $x_{0} = - \infty$ and $x_{M} = + \infty$.]
A typical step-like representation of $V(x)$, $|\Delta(x)|$, and $2\phi(x)$ is reported in Fig.\ref{fig3}.

The advantage of this representation is that, within each $\ell$-th interval, the BdG equations can be solved by elementary methods since the problem reduces to that of a homogeneous system with a uniform potential level $V_{\ell}$, while the solutions in contiguous intervals are connected by standard continuity conditions.
In analogy to Eqs.(\ref{u-v-homogeneous}), we thus set for the generic solutions of the BdG equations in the $\ell$-th interval:
\begin{eqnarray}
u(\mathbf{r};q,E) & = & u_{\ell}(q,E) \, e^{i(\mathbf{k} + \mathbf{q}) \cdot \mathbf{r} +
                                                                i \phi_{\ell}}            \nonumber \\
v(\mathbf{r};q,E) & = & v_{\ell}(q,E) \, e^{i(\mathbf{k} - \mathbf{q}) \cdot \mathbf{r} -
                                                                i \phi_{\ell}}            \label{BdG-discretized-solutions}                                                             
\end{eqnarray}

\noindent
for given energy $E$, where $\mathbf{r}=(x,y,z)$, $\mathbf{k}=(k_{\ell},k_{y},k_{z})$, and $\mathbf{q}=(q,0,0)$.
In this interval, the BdG equations then reduce to:
\begin{eqnarray}
\left( \frac{(k_{\ell}+q)^{2}}{2m} - \tilde{\mu}_{\ell} \right) u_{\ell}(q,E) +
\Delta_{\ell} \, v_{\ell}(q,E) & = & E \, u_{\ell}(q,E)              \nonumber \\
- \left( \frac{(k_{\ell}-q)^{2}}{2m} - \tilde{\mu}_{\ell} \right) v_{\ell}(q,E) +
\Delta_{\ell} \, u_{\ell}(q,E) & = & E \, v_{\ell}(q,E)              \label{BdG-discretized-equations}
\end{eqnarray} 

\noindent
where we have introduced the notation
\begin{equation}
\tilde{\mu}_{\ell} \, = \, \mu \, - \, V_{\ell} \, - \, \frac{k_{y}^{2}+k_{z}^{2}}{2m} 
                                                                                                            \label{mu-tilde}
\end{equation}

\noindent
which identifies the (local) reduced chemical potential.
The solution for $E$ in Eqs.(\ref{BdG-discretized-equations}) is thus formally analogous to the expression (\ref{eigeinvalue-homogeneous}) with the local replacements $k_{x} \rightarrow k_{\ell}$, 
$\mu \rightarrow \tilde{\mu}_{\ell}$, and $\Delta_{\mathbf{q}} \rightarrow \Delta_{\ell}$:
\begin{equation}
E \, = \, \frac{k_{\ell} \, q}{m} \, + \, 
\sqrt{ \left( \frac{k_{\ell}^{2}}{2m} + \frac{q^{2}}{2m} - \tilde{\mu}_{\ell} \right)^{2} \, + \, \Delta_{\ell}^{2}} \,\, .  
                                                                                                                                                           \label{eigeinvalue-E}
\end{equation}

The reason for replacing the variable $k_{x}$ in Eq.(\ref{eigeinvalue-homogeneous}) by the local value $k_{\ell}$ is now apparent.
For given values of $\tilde{\mu}_{\ell}$ and $\Delta_{\ell}$ in the $\ell$-th interval, there are in fact four distinct (complex) values of 
$k_{\ell}$ that satisfy Eq.(\ref{eigeinvalue-E}) for fixed $E$.
In the absence of current (i.e., for $q=0$), they correspond to the electron-like and hole-like excitations discussed in Ref.\cite{BTK}.
The evolution of these values of $k_{\ell}$ in the complex $k$-plane when $E$ varies from $0$ to $+ \infty$ (depending also on the sign of $\tilde{\mu}_{\ell}$) will be discussed in detail below.

Correspondingly, the eigenvectors of Eqs.(\ref{BdG-discretized-equations}) can be cast in a form similar to the solutions (\ref{u2-v2-homogeneous}).
Introducing the notation $\tilde{\xi}_{k_{\ell}-q} = (k_{\ell}-q)^{2}/(2m) - \tilde{\mu}_{\ell}$, we arrive at the expressions:
\begin{eqnarray}
u_{\ell}(q,E)^{2} & = & 
\frac{(E + \tilde{\xi}_{k_{\ell}-q})^{2}}{(E + \tilde{\xi}_{k_{\ell}-q})^{2} + \Delta_{\ell}^{2} }    \nonumber \\                                            
v_{\ell}(q,E)^{2} & = & \frac{\Delta_{\ell}^{2}} {(E + \tilde{\xi}_{k_{\ell}-q})^{2} + \Delta_{\ell}^{2} }        
                                                                                                         \label{u2-v2-discretized}                                                 
\end{eqnarray}

\noindent
where again the positive square root has to be taken.
Note that we have conventionally imposed on the solutions (\ref{u2-v2-discretized}) of the homogeneous system (\ref{BdG-discretized-equations}) the condition $u_{\ell}(q,E)^{2} + v_{\ell}(q,E)^{2} = 1$, even though this is not directly related to the overall normalization condition (\ref{normalization-condition}) for the total wave function which extends over the whole set of intervals.

Quite generally, to each value of $k_{\ell}^{(n)}$ [with $n = (1,\cdots,4)$] that satisfies 
Eq.(\ref{eigeinvalue-E}), we can associate the two-component wave function $\Upsilon_{\ell}^{(n)}(x;q,E)$
\begin{equation}
\Upsilon_{\ell}^{(n)}(x;q,E) \, = \, \left( \begin{array}{l}
                                                    u_{\ell}^{(n)}(q,E) \, e^{i(qx+\phi_{\ell})}  \\  
                                                    v_{\ell}^{(n)}(q,E) \, e^{-i(qx+\phi_{\ell})}
                                                    \end{array}  \right) e^{i k_{\ell}^{(n)}  x}     \label{Upsilon}
\end{equation}

\noindent
for $x_{\ell - 1} \leq x \leq x_{\ell}$.
We use the following convention: 
(i) $n=1$ represents an electron-like excitation impinging from the left (of the barrier); 
(ii) $n=2$ an electron-like excitation impinging from the right; 
(iii) $n=3$ a hole-like excitation impinging from the right;
(iv) and $n=4$ a hole-like excitation impinging from the left 
(the inverted convention for the hole excitations corresponds to the inverted current flow for hole with respect to electron excitations).
[There will actually occur an exception to the correspondence of this convention with the actual physical situation, in the energy region when $k_{\ell}^{(1)} < 0$ for $\ell=1,M$.]
Note that the terminology ``electron'' and ``hole'' is mantained for historical reasons, even though the fermions here
considered are ultracold atoms.

In terms of the wave functions (\ref{Upsilon}), we implement the solutions of the BdG equations in the $\ell$-th interval as follows:
\begin{equation}
\Psi_{\ell}(x;q,E) = a_{\ell} \Upsilon_{\ell}^{(1)}(x;q,E) + b_{\ell} \Upsilon_{\ell}^{(2)}(x;q,E)
                                 + c_{\ell} \Upsilon_{\ell}^{(3)}(x;q,E) + d_{\ell} \Upsilon_{\ell}^{(4)}(x;q,E)  
                                                                                                                \label{Psi-vs-Upsilon}
\end{equation}

\noindent
which generalize the solutions (\ref{BdG-discretized-solutions}) when all possible values of $k_{\ell}$ are considered, 
apart from the common factor $\exp (k_{y} y + k_{z} z)$ which has been dropped out.
The coefficients $(a_{\ell},b_{\ell},c_{\ell},d_{\ell})$ are determined by applying the \emph{continuity conditions} to the wave functions (\ref{Psi-vs-Upsilon}) and their derivatives at the points $ \{ x_{\ell};  \ell = 1,\cdots,M-1 \}$ between the $\ell$-th and $(\ell + 1)$-th intervals, plus imposing appropriate \emph{boundary conditions} in the $1$-st and $M$-th intervals at the edges.

The continuity conditions at the point $x_{\ell}$ separating the $\ell$-th and $(\ell + 1)$-th intervals can be expressed in the compact form:
\begin{equation}
\mathbf{M}_{\ell}(x=x_{\ell}) \, W_{\ell} \, = \, \mathbf{M}_{\ell + 1}(x=x_{\ell}) \, W_{\ell + 1} 
                                                                                                       \label{continuity-conditions}
\end{equation}

\noindent
where $W_{\ell}$ is the column vector formed by the coefficients of Eq.(\ref{Psi-vs-Upsilon})
\begin{equation}
W_{\ell} \, = \, \left( \begin{array}{c}
a_{\ell} \\
b_{\ell} \\
c_{\ell} \\
d_{\ell} 
\end{array} \right)                                       \label{column-vector-Wl}
\end{equation}

\noindent
and $\mathbf{M}_{\ell}(x)$ is the $4 \times 4$ matrix composed by the following column vectors $(n=1,2,3,4)$:
\begin{equation}
\left( \begin{array}{l}
M_{\ell}(x)_{1n} \\
M_{\ell}(x)_{2n} \\
M_{\ell}(x)_{3n} \\
M_{\ell}(x)_{4n}  \end{array} \right)
\, = \, \left( \begin{array}{l}
u_{\ell}^{(n)}(q,E) e^{i(qx + \phi_{\ell})} \\
v_{\ell}^{(n)}(q,E) e^{-i(qx + \phi_{\ell})} \\
i(k_{\ell}^{(n)}+q) u_{\ell}^{(n)}(q,E) e^{i(qx + \phi_{\ell})} \\
i(k_{\ell}^{(n)}- q) v_{\ell}^{(n)}(q,E) e^{-i(qx + \phi_{\ell})} \end{array}
\right) \,\, e^{i k_{\ell}^{(n)}x}                                  \label{column-vector-Mn}
\end{equation}

\noindent
(the dependence of $M_{\ell}(x)$ and $W_{\ell}$ on $(q,E)$ is understood throughout).
Since there are $(M-1)$ continuity conditions of the type (\ref{continuity-conditions}) (each of them corresponding to four conditions owing to Eqs.(\ref{column-vector-Wl}) and (\ref{column-vector-Mn})) and $4 M$ coefficients $(\{a_{\ell},b_{\ell},c_{\ell},d_{\ell} \}; \ell=1,\cdots,M$), the problem is apparently under-determined.
As we will see shorthly below, however, it turns out that in the $1$-st interval only the coefficients ($b_{1},c_{1}$) and in the last $M$-th interval only the coefficients ($a_{M},d_{M}$) need be determined, thus reducing the total number of coefficients to $4(M-1)$.
Grouping then together all continuity conditions in a single equation, one is left with solving the equation 
\begin{equation}
\mathbf{A} \, W \, = \, B                             \label{equation-AW=B}
\end{equation}

\noindent
where $W$ is now the column vector with $4(M-1)$ components
\begin{equation}
W  \, = \, \left( \begin{array}{c}
b_{1} \\
c_{1} \\
\cdots \\
a_{\ell} \\
b_{\ell} \\
c_{\ell} \\
d_{\ell} \\
\cdots \\
a_{M} \\
d_{M}
\end{array} \right)   \,\, ,                                    \label{column-vector-W}
\end{equation}

\noindent
and $\mathbf{A}$ is the $4(M-1) \times 4(M-1)$ matrix represented in the following block form:

\begin{tiny}
\begin{eqnarray*}
\mathbf{A} \, = \, \left( \begin{array}{cccccccc}
\tilde{\mathbf{M}}_{1}(x_{1}) & - \mathbf{M}_{2}(x_{1}) & 0 & 0 &  \cdots  & 0 & 0 & 0  \\
0 & \mathbf{M}_{2}(x_{2}) & - \mathbf{M}_{3}(x_{2}) & 0           &  \cdots  & 0 & 0 & 0  \\
0 & 0 & \mathbf{M}_{3}(x_{3}) & - \mathbf{M}_{4}(x_{3})           &  \cdots  & 0 & 0 & 0  \\
0 & 0 & \cdots & \cdots                                  &  \cdots  & 0 & 0 & 0  \\
0 & 0 & \cdots & \cdots                                  &  \cdots  & 0 & 0 & 0  \\
0 & 0 & \cdots & \cdots                                  &  \cdots  & \cdots & 0 & 0  \\
0 & 0 & 0 & 0 & 0 \cdots 0 & \mathbf{M}_{M-2}(x_{M-2}) & - \mathbf{M}_{M-1}(x_{M-2}) & 0 \\
0 & 0 & 0 & 0 & 0 \cdots 0 & 0 & \mathbf{M}_{M-1}(x_{M-1}) 
& - \tilde{\mathbf{M}}_{M}(x_{M-1}) 
\end{array} \right)                                                                                                                                                                                  
\end{eqnarray*} 
\end{tiny} 

\noindent
In the above expression, $\tilde{\mathbf{M}}_{1}$ and $\tilde{\mathbf{M}}_{M}$ are 
$4 \times 2$ matrices constructed from the pairs of column vectors (\ref{column-vector-Mn}) with $n=(1,4)$ and $n=(2,3)$, respectively.

\begin{center}
\begin{figure}[htc]
\includegraphics[angle=0,width=6.5in]{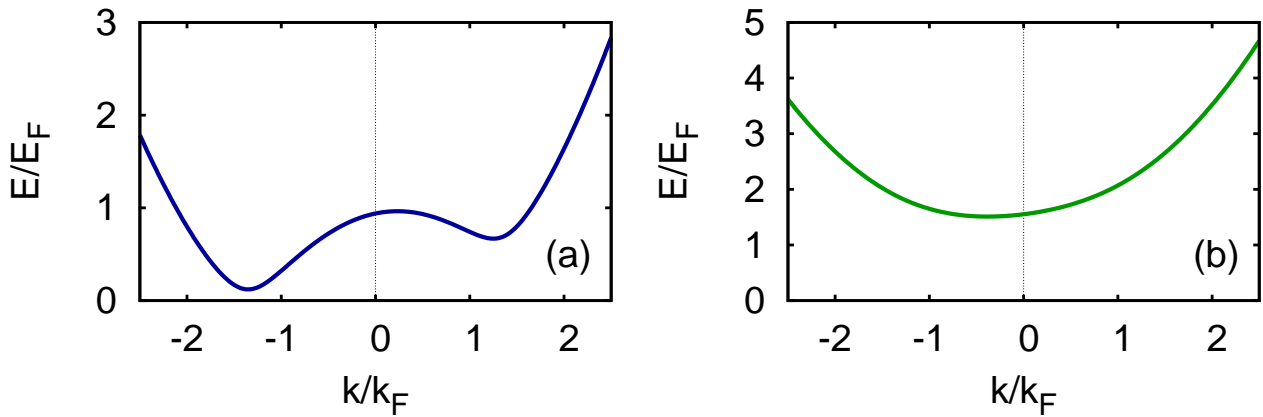}
\caption{Dispersion relation (\ref{eigeinvalue-E}) for $\ell=1$ or $\ell=M$ (whereby 
              $\Delta_{\ell} = \Delta_{0}$, $V_{\ell}=0$, and $k_{\ell}=k$), with (a) $\tilde{\mu} > 0$ 
              and (b) $\tilde{\mu} < 0$. These figures identify the seven energy ranges for which 
              the boundary conditions in the outermost (left and right) spatial intervals are specified.}
\label{fig4}
\end{figure}
\end{center}

The vector $B$ on the right-hand side of Eq.(\ref{equation-AW=B}) is specified by the boundary conditions in the 
$1$-st and $M$-th intervals, and takes accordingly different forms depending on the energy ranges identified by the expression (\ref{eigeinvalue-E}) whereby one sets $\Delta_{\ell}$ equal to the bulk value $\Delta_{0}$, $V_{\ell}=0$, and $k_{\ell}=k$.
The ensuing expression assumes different shapes depending on the sign of 
$\tilde{\mu} = \mu - (k_{y}^{2}+k_{z}^{2})/(2m)$.
In Fig.\ref{fig4} we plot this expression vs $k$ (with a value of $q>0$ for which it remains positive) for given values of 
$\Delta_{0}$ and $\tilde{\mu}$.
The two cases with $\tilde{\mu} > 0$ [Fig.\ref{fig4}(a)] and $\tilde{\mu} < 0$ [Fig.\ref{fig4}(b)] are considered separately. 
Note that negative values of $\tilde{\mu}$ result for large enough values of $(k_{y},k_{z})$ even when $\mu > 0$.

Four energy ranges are identified when $\tilde{\mu} > 0$ and three additional ones when 
$\tilde{\mu} < 0$, in the following way:
\vspace{0.1cm}

\noindent
- Range 1: \, $\tilde{\mu} > 0$ \, and \, $0 < E < E_{\mathrm{min}}^{\mathrm{left}}$

\noindent
- Range 2: \, $\tilde{\mu} > 0$ \, and \, $E_{\mathrm{min}}^{\mathrm{left}} < E < E_{\mathrm{min}}^{\mathrm{right}}$

\noindent
- Range 3: \, $\tilde{\mu} > 0$ \, and \, $E_{\mathrm{min}}^{\mathrm{right}} < E < E_{\mathrm{max}}$

\noindent
- Range 4: \, $\tilde{\mu} > 0$ \, and \, $E_{\mathrm{max}} < E$

\noindent
- Range 5: \, $\tilde{\mu} < 0$ \, and \, $0 < E < E_{\mathrm{min}}$

\noindent
- Range 6: \, $\tilde{\mu} < 0$ \, and \, $E_{\mathrm{min}} < E < E_{0}$

\noindent
- Range 7: \, $\tilde{\mu} < 0$ \, and \, $E_{0} < E$ .

\begin{center}
\begin{figure}[htc]
\includegraphics[angle=0,width=6.5in]{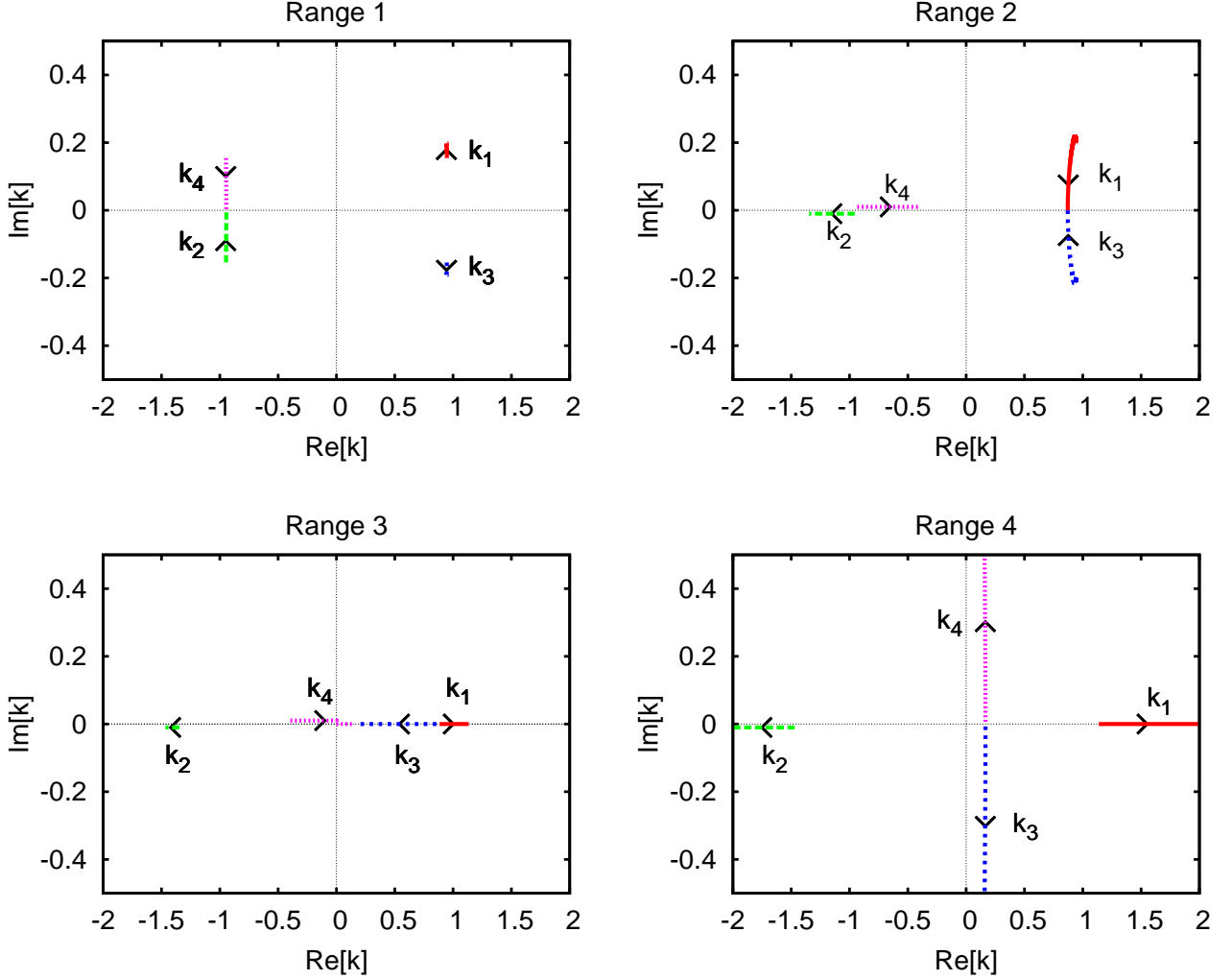}
\caption{Paths followed in the complex $k$-plane by the four solutions $k^{(n)}$ by increasing $E$ for $\tilde{\mu} > 0$.
              The panels correspond to ranges 1-4 of Table 1.}
\label{fig5}
\end{figure}
\end{center} 
\vspace{-0.1cm}

\noindent
With reference to the two different shapes of the dispersion relation reported in Fig.\ref{fig4}, 
$E_{\mathrm{min}}^{\mathrm{left}}$ stands for the absolute minimum on the left side of Fig.\ref{fig4}(a), 
$E_{\mathrm{min}}^{\mathrm{right}}$ for the local minimum on the right side of Fig.\ref{fig4}(a), $E_{\mathrm{max}}$ for the local maximum of Fig.\ref{fig4}(a), $E_{\mathrm{min}}$ for the absolute minimum of Fig.\ref{fig4}(b), and $E_{0}$ for the value corresponding to $k=0$ in Fig.\ref{fig4}(b).

\begin{center}
\begin{figure}[htc]
\includegraphics[angle=0,width=6.5in]{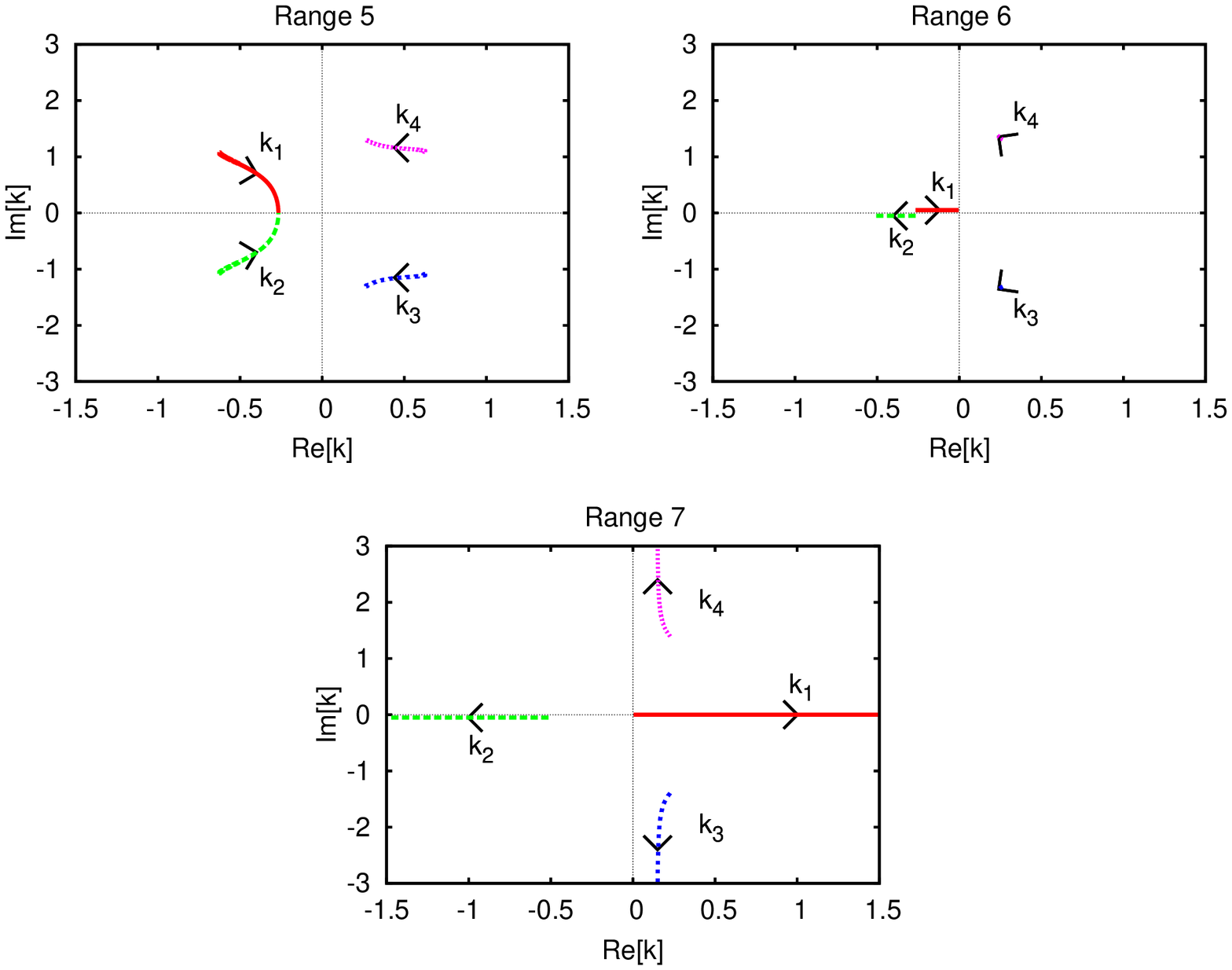}
\caption{Paths followed in the complex $k$-plane by the four solutions $k^{(n)}$ by increasing $E$ for $\tilde{\mu} < 0$.
              The panels correspond to ranges 5-7 of Table 1.}
\label{fig6}
\end{figure}
\end{center} 

In Figs.\ref{fig5} and \ref{fig6} we show how the different solutions $k^{(n)}$ of Eq.(\ref{eigeinvalue-E}) move in the complex $k$-plane by increasing $E$ in each of the above seven energy ranges.
From the form (\ref{Psi-vs-Upsilon}) of the wave function in the outermost left ($\ell=1$) and right ($\ell=M$) intervals, we conclude that the allowed (i.e., non divergent) components
(\ref{Upsilon}) must correspond to $\mathrm{Im}\{k^{(n)}\}<0$ when $\ell=1$ and 
$\mathrm{Im}\{k^{(n)}\}>0$ when $\ell=M$.
This leads us to exclude \emph{a priori} some of the components (\ref{Upsilon}) by setting to zero the corresponding coefficients in the wave function (\ref{Psi-vs-Upsilon}), as conveniently summarized in Table 1.

\begin{table}
\begin{large}
\begin{center} 
\begin{tabular}{|c|c|c|c|c|c|c|c|c|}  \hline \hline
\multicolumn{1}{|c|}{} & 
\multicolumn{4}{|c|}{outermost left interval} &
\multicolumn{4}{|c|}{outermost right interval}   \\ \hline
    & $k^{(1)}$ & $k^{(2)}$ & $k^{(3)}$ & $k^{(4)}$ & $k^{(1)}$ & $k^{(2)}$ & $k^{(3)}$ & $k^{(4)}$ \\  \hline
E &  $a_{1}$  & $b_{1}$   & $c_{1}$   & $d_{1}$   & $a_{M}$  & $b_{M}$  & $c_{M}$  & $d_{M}$  \\  \hline \hline
Range 1 & no  & yes & yes & no  & yes  & no  & no  & yes     \\ \hline 
Range 2 & no  & yes & yes & yes & yes & yes & no  & yes     \\ \hline
Range 3 & yes & yes & yes & yes & yes & yes & yes & yes    \\ \hline
Range 4 & yes & yes & yes & no  & yes & yes & no  & yes     \\ \hline
Range 5 & no  & yes & yes & no  & yes &  no  & no  & yes     \\ \hline
Range 6 & yes & yes & yes & no  & yes & yes & no  & yes     \\ \hline 
Range 7 & yes & yes & yes & no  & yes & yes & no  & yes     \\ \hline
\end{tabular}
\caption{Allowed form of the wave function (\ref{Psi-vs-Upsilon}) in the outermost left and right intervals (where the
             boundary conditions are enforced) depending on the value of the energy $E$.}
\end{center} 
\end{large}
\end{table}

Notice that, by allowing $k^{(n)}$ to span the complex $k$-plane when the energy $E$ varies from $0$ to $+\infty$, 
we avoid relying on the so-called Andreev approximation which is typical of a BCS treatment, whereby only wave vectors close to the minima of Fig.\ref{fig4}(a) are effectively retained (see, e.g., Ref.\cite{BT-97}).

To proceed further, we recall that our purpose here is to identify a complete set of energy eigenstates that enter the expressions (\ref{self-consistency}), (\ref{density}), and (\ref{current}) of the physical quantities of interest.
Apart from the presence of bound states which are localized about the barrier and exist only in the energy ranges 1 and 5, in the other energy ranges we are going to use ``\emph{outgoing boundary conditions\/}'' for waves impinging on the barrier from the left and right, in order to identify the contribution to the complete set from energies in the continuum.

We are now in a position to specify the vector $B$ of Eq.(\ref{equation-AW=B}) in each of the above seven energy ranges, which we treat separately as follows:
\vspace{0.1cm}

\noindent
(i) According to the first panel of Fig.\ref{fig5}, in {\bf range 1} there is no real solution to Eq.(\ref{eigeinvalue-E}). From Table 1 the wave functions in the outermost intervals are then given by:
\begin{equation}
\Psi_{1}(x;q,E) = b_{1} \Upsilon_{1}^{(2)}(x;q,E) + c_{1} \Upsilon_{1}^{(3)}(x;q,E)                                                 
                                                                                                       \label{Psi-1-range-1} 
\end{equation}

\noindent
for $x_{0} \leq x \leq x_{1}$, and 
\begin{equation}
\Psi_{M}(x;q,E) = a_{M} \Upsilon_{M}^{(1)}(x;q,E)  + d_{M} \Upsilon_{M}^{(4)}(x;q,E)  
                                                                                                      \label{Psi-M-range-1}                                                                                                             
\end{equation}
 
\noindent
for $x_{M-1} \leq x \leq x_{M}$.
In this case, the vector $B$ is identically zero and Eq.(\ref{equation-AW=B}) reduces to a homogeneous systems for the coefficients $W$, which admits nontrivial solutions only when the determinant of the matrix $\mathbf{A}$ vanishes. 
This occurs only for particular values of the energy $E$ in this range, which thus identify the bound-state energies.
[Note that there exists a continuous branch of bound states associated with different values of $(k_{y},k_{z})$.]
\vspace{0.1cm}

\noindent
(ii) In {\bf range 2}, $a_{1}=c_{M}=0$ according to Table 1, while according to the second panel of Fig.\ref{fig5} only the components (\ref{Upsilon}) corresponding to $k^{(2)}$ and $k^{(4)}$ propagate undamped.
The use of outgoing boundary conditions then results in two independent types of external wave functions, namely:
\vspace{0.05cm}

\noindent
- Hole-like excitation impinging from the left, such that in the outermost intervals
\begin{eqnarray}
\Psi_{1}(x;q,E) & = & b_{1} \Upsilon_{1}^{(2)}(x;q,E) + c_{1} \Upsilon_{1}^{(3)}(x;q,E) +
                                            \Upsilon_{1}^{(4)}(x;q,E)            \nonumber \\
\Psi_{M}(x;q,E) & = & a_{M} \Upsilon_{M}^{(1)}(x;q,E)  + d_{M} \Upsilon_{M}^{(4)}(x;q,E)  
                                                                                               \label{Psi-hole-range-2}                                            
\end{eqnarray}

\noindent
where the coefficient $d_{1}$ of $\Upsilon_{1}^{(4)}(x;q,E)$ has conventionally been set equal to unity (the overall normalization of the total wave function will be determined at a second stage).
According to the notation (\ref{column-vector-Mn}) and the convention (\ref{continuity-conditions}), in this case $B=(-M_{1}(x_{1})_{14}, -M_{1}(x_{1})_{24}, -M_{1}(x_{1})_{34},
 -M_{1}(x_{1})_{44}, 0,\cdots,0)^{T}$ (where $T$ stands for transpose).
\vspace{0.05cm}

\noindent
- Electron-like excitation impinging from the right, such that
\begin{eqnarray}
\Psi_{1}(x;q,E) & = & b_{1} \Upsilon_{1}^{(2)}(x;q,E) + c_{1} \Upsilon_{1}^{(3)}(x;q,E)              
                                                                                            \label{Psi-electron-range-2} \\                    
\Psi_{M}(x;q,E) & = & a_{M} \Upsilon_{M}^{(1)}(x;q,E) +  \Upsilon_{M}^{(2)}(x;q,E)  
                                + d_{M} \Upsilon_{M}^{(4)}(x;q,E)       \nonumber                                                     
\end{eqnarray}

\noindent
where now the coefficient $b_{M}$ of $\Upsilon_{M}^{(2)}(x;q,E)$ equals unity.
In this case, $B=(0,\cdots,0, M_{M}(x_{M-1})_{12},$ $ M_{M}(x_{M-1})_{22},
 M_{M}(x_{M-1})_{32}, M_{M}(x_{M-1})_{42})^{T}$.
\vspace{0.1cm}

\noindent
(iii) In {\bf range 3}, all components (\ref{Upsilon}) propagate undamped [cf. the third panel of Fig.\ref{fig5}] and the four coefficients $(a_{1},d_{1},b_{M},c_{M})$ are in principle all different from zero [cf. Table 1].
These coefficients are fixed by the outgoing boundary conditions that we have chosen to determine a complete set, which select four independent types of external wave functions as follows:
\vspace{0.05cm}

\noindent
- Electron-like excitation impinging from the left, such that 
$(a_{1}=1,d_{1}=0,b_{M}=0,c_{M}=0)$,
\begin{eqnarray}
\Psi_{1}(x;q,E) & = & \Upsilon_{1}^{(1)}(x;q,E) + b_{1} \Upsilon_{1}^{(2)}(x;q,E) 
                           + c_{1} \Upsilon_{1}^{(3)}(x;q,E)         \nonumber  \\           
\Psi_{M}(x;q,E) & = & a_{M} \Upsilon_{M}^{(1)}(x;q,E) + d_{M} \Upsilon_{M}^{(4)}(x;q,E) \,\, ,          
                                                                                       \label{Psi-electron-left-range-3}                                                                                      
\end{eqnarray}

\noindent
and $B=(-M_{1}(x_{1})_{11}, -M_{1}(x_{1})_{21},
 -M_{1}(x_{1})_{31}, -M_{1}(x_{1})_{41}, 0,\cdots,0)^{T}$.
\vspace{0.05cm}

\noindent
- Electron-like excitation impinging from the right, such that 
$(a_{1}=0,d_{1}=0,b_{M}=1,c_{M}=0)$,
\begin{eqnarray}
\Psi_{1}(x;q,E) & = & b_{1} \Upsilon_{1}^{(2)}(x;q,E) + c_{1} \Upsilon_{1}^{(3)}(x;q,E)  
                                                                                            \label{Psi-electron-right-range-3} \\                                   
\Psi_{M}(x;q,E) & = & a_{M} \Upsilon_{M}^{(1)}(x;q,E) + \Upsilon_{M}^{(2)}(x;q,E) 
                            + d_{M} \Upsilon_{M}^{(4)}(x;q,E) \,\, ,     \nonumber                                             
\end{eqnarray}

\noindent
and $B=(0,\cdots,0, M_{M}(x_{M-1})_{12}, M_{M}(x_{M-1})_{22},
 M_{M}(x_{M-1})_{32}, M_{M}(x_{M-1})_{42})^{T}$. 
 
\noindent
- Hole-like excitation impinging from the left, such that 
$(a_{1}=0,d_{1}=1,b_{M}=0,c_{M}=0)$,
\begin{eqnarray}
\Psi_{1}(x;q,E) & = & b_{1} \Upsilon_{1}^{(2)}(x;q,E) + c_{1} \Upsilon_{1}^{(3)}(x;q,E) 
                           + \Upsilon_{1}^{(4)}(x;q,E)                   \nonumber \\
\Psi_{M}(x;q,E) & = & a_{M} \Upsilon_{M}^{(1)}(x;q,E) + d_{M} \Upsilon_{M}^{(4)}(x;q,E) \,\, ,          
                                                                                                 \label{Psi-hole-left-range-3}                                            
\end{eqnarray}

\noindent
and $B=(-M_{1}(x_{1})_{14}, -M_{1}(x_{1})_{24},
 -M_{1}(x_{1})_{34}, -M_{1}(x_{1})_{44}, 0,\cdots,0)^{T}$.
\vspace{0.05cm}

\noindent
- Hole-like excitation impinging from the right, such that 
$(a_{1}=0,d_{1}=0,b_{M}=0,c_{M}=1)$,
\begin{eqnarray}
\Psi_{1}(x;q,E) & = & b_{1} \Upsilon_{1}^{(2)}(x;q,E) + c_{1} \Upsilon_{1}^{(3)}(x;q,E)  
                                                                                                 \label{Psi-hole-right-range-3} \\                                      
\Psi_{M}(x;q,E) & = & a_{M} \Upsilon_{M}^{(1)}(x;q,E) + \Upsilon_{M}^{(3)}(x;q,E) 
                            + d_{M} \Upsilon_{M}^{(4)}(x;q,E) \,\, ,         \nonumber                                    
\end{eqnarray}

\noindent
and $B=(0,\cdots,0, M_{M}(x_{M-1})_{13}, M_{M}(x_{M-1})_{23},
 M_{M}(x_{M-1})_{33}, M_{M}(x_{M-1})_{43})^{T}$. 

In each of the above four cases, one may identify the different processes as being alternatively the normal reflection, normal transmission, Andreev reflection, and inverted transmission.
[The last two processes are characteristic of the superconducting state \cite{Andreev-states}, and require one to consider the hole-like wave vectors $(k^{(3)},k^{(4)})$ besides the usual electron-like wave vectors $(k^{(1)},k^{(2)})$ characteristic of the normal state.]
\vspace{0.1cm}

\noindent
(iv) In {\bf range 4}, only the components (\ref{Upsilon}) corresponding to $k^{(1)}$ and 
$k^{(2)}$ propagate undamped [cf. the forth panel of Fig.\ref{fig5}] and $d_{1}=c_{M}=0$
[cf. Table 1].
It follows that the only allowed excitations are electron-like impinging either from the left 
[with $(a_{1}=1,b_{M}=0)$, cf. Eq.(\ref{Psi-electron-left-range-3})] or from the right 
[with $(a_{1}=0,b_{M}=1)$, cf. Eq.(\ref{Psi-electron-right-range-3})], whereby the vector $B$
has the same form as in those cases. 
\vspace{0.1cm}

\noindent
(v) The situation in {\bf range 5} is equivalent to that in range 1 [cf. the first panel of Fig.\ref{fig6} and Table 1].
There is no real solution to Eq.(\ref{eigeinvalue-E}), the wave functions in the outermost intervals are again given by Eqs.(\ref{Psi-1-range-1}) and (\ref{Psi-M-range-1}), the vector $B$ vanishes identically, and one is looking for bound-state solutions localized about the barrier for particular values of $E$.
\vspace{0.1cm}

\noindent
(vi) The situation in {\bf range 6} is similar to that in range 2 [cf. the second panel of Fig.\ref{fig6} and Table 1].
This is because, when the wave vector $k^{(1)}$ is negative, it should be associated with a hole-like excitation impinging from the left rather than with an electron-like excitation impinging from the right.
The corresponding wave function in the outermost intervals is given by the expression (\ref{Psi-hole-range-2}), with $\Upsilon_{1}^{(4)}$ in the upper equation now replaced by 
$\Upsilon_{1}^{(1)}$, and $B=(-M_{1}(x_{1})_{11}, -M_{1}(x_{1})_{21}, -M_{1}(x_{1})_{31},
 -M_{1}(x_{1})_{41}, 0,\cdots,0)^{T}$.
The excitation associated with $k^{(2)}$, on the other hand, remains an electron-like excitation impinging from the right as in Eq.(\ref{Psi-electron-range-2}).
\vspace{0.1cm}

\noindent
(vii) The situation in {\bf range 7} is fully equivalent to that in range 4 [cf. the third panel of Fig.\ref{fig6} and Table 1].
Here, the excitation associated with $k^{(1)}>0$ represents an electron-like excitation impinging from the left.
\vspace{0.1cm}
                                                
There remains to specify how the ortho-normalization condition 
(\ref{normalization-condition}) for the above wave functions is implemented in the different energy ranges.
In ranges 1 and 5 where the wave functions are normalizable (being exponentially localized about the barrier), one breaks up the $x$-integration into $M$ intervals and uses the generic form (\ref{Psi-vs-Upsilon}) of the wave function in the internal intervals
$x_{\ell} < x < x_{\ell + 1}$ with $\ell = (1,\cdots,M-2)$ as well as the forms
(\ref{Psi-1-range-1}) and (\ref{Psi-M-range-1}) in the outermost intervals. 
In all other ranges where the energy $E$ is continuous, on the other hand, the normalization per unit range of the wave vector $k_{x}$ is obtained by replacing the unit coefficient of the impinging excitation by $1/\sqrt{2 \pi}$ and rescaling the other coefficients accordingly.
We have also explicitly verified that degenerate wave functions with given $E$ of the kinds discussed above are orthogonal to each other, as required when constructing a complete set.
Finally, all types of wave functions are eventually normalized per unit range of the wave vectors $k_{y}$ and $k_{z}$ by multiplying them by $1/(2 \pi)$.


\vspace{0.3cm}
\begin{center}
\begin{large}
{\bf 2c. Weak-coupling (BCS) limit and the delta-like barrier}
\end{large}
\end{center}
\addcontentsline{toc}{subsection}{2c. Weak-coupling (BCS) limit and the delta-like barrier}
\vspace{0.1cm}

The general computational scheme, which we have set up in detail in sub-section 2b to solve the BdG equations, holds quite generally for any coupling across the BCS-BEC crossover from the weak-coupling (BCS) to the strong-coupling (BEC) limits.
Implementation of this computational scheme rests unavoidably on sophisticated numerical procedures (as described in Section 3) and involves heavy numerical calculations, which require that the outcomes of the calculations need be judiciously assessed before extracting firm physical conclusions from them.
To this end, it is of definite importance to rely on independent \emph{benchmarks}, with which the outcomes of the calculations could be confronted at least in some relevant limits.
The crossover problem itself identifies these benchmarks on physical grounds, separately for the two limiting (BCS and BEC) situations.
We discuss here the BCS limit and defer the discussion of the BEC limit to sub-section 2d.

\begin{center}
\begin{figure}[htc]
\hspace{3.5cm}\includegraphics[angle=0,width=3.5in]{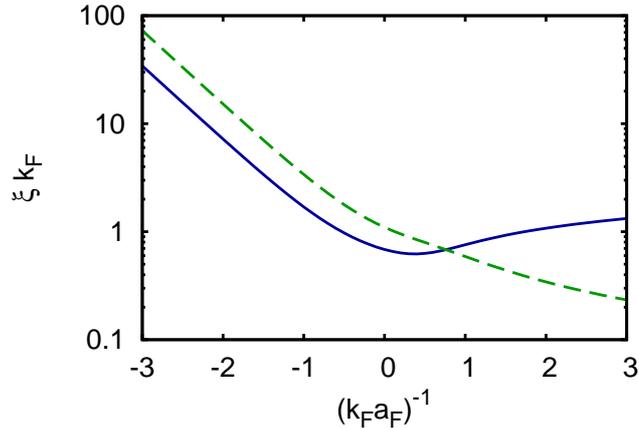}
\caption{Pair coherence length $\xi_{\mathrm{pair}}$ (dashed line) and phase coherence (healing) lenght $\xi_{\mathrm{phase}}$ (full line) 
              evaluated at zero temperature within the mean-field approximation according to their definitions as given respectively in 
              Refs.\cite{PS-1994} and \cite{PS-1996}, vs the coupling parameter $(k_{F} a_{F})^{-1}$ across the BCS-BEC crossover. 
              Both lengths are in units of the average interparticle distance $k_{F}^{-1}$. 
              Note that in weak coupling the two lenghts differ by an irrelevant numerical factor ($\xi_{\mathrm{pair}} \simeq (3/\sqrt{2}) \xi_{\mathrm{phase}}$) 
              owing to their independent definitions, so that in the plot the two curves result parallel to each other in this coupling regime.}
\label{fig7}
\end{figure}
\end{center} 

We begin by considering the homogeneous situation, whereby two length scales can be identified as being relevant to the BCS-BEC crossover problem at zero temperature \cite{PS-1996}.
The first one is the \emph{pair} coherence length $\xi_{\mathrm{pair}}$ for two-fermion correlation, which corresponds to the size of the Cooper pairs in the BCS limit and to the radius of the bound-electron pairs in the BEC limit.
The other one is the \emph{phase} coherence lenght $\xi_{\mathrm{phase}}$, which is associated with the spatial fluctuations of the order parameter (and is often referred to as the \emph{healing} length in the bosonic literature).
These two lengths coincide with each other (apart from a trivial numerical factor originating from their different definitions) in the BCS limit, where they also reduce to the Pippard coherence length characteristic of the superconductor electrodynamics \cite{FW}.
In the BEC limit, on the other hand, the two lengths differ considerably, representing the ``intra-'' and ``inter-'' boson correlation, respectively.

The two lengths are plotted in Fig.\ref{fig7} versus the coupling parameter $(k_{F} a_{F})^{-1}$ spanning the BCS-BEC crossover.
One sees from this figure that in the BCS limit $(k_{F} a_{F})^{-1} \ll -1$ both lengths grow without bound, exceeding eventually any other finite length scale that may enter the problem.
They will exceed, in particular, the characteristic width of a barrier (like that considered in sub-section 2b), so that the specific value of this width becomes irrelevant when calculating the scattering properties.

Under these circumstances, one expects on physical grounds that the barrier could be assimilated to a Dirac-delta barrier with potential $V(x) = \mathcal{Z} \delta(x)$, where the constant 
\begin{equation}
\mathcal{Z} \, = \, \int_{- \infty}^{+\infty} \! dx \, V(x)                    \label{total-area}
\end{equation}

\noindent
represents the total area of the potential. 
For instance, $\mathcal{Z} = V_{0} L$ for a rectangular barrier of width $L$ and height $V_{0}$, and $\mathcal{Z} = V_{0} \sigma \sqrt{2 \pi}$ for a Gaussian barrier with 
$V(x) = V_{0} \, e^{-x^{2}/(2 \sigma^{2})}$.
In practice, this expectation implies that alternative numerical calculations with different potential $V(x)$ that share the same area (\ref{total-area}) should produce the same results.
In the following, we shall use this kind of \emph{universality} as a stringent test on the correctness of the numerical calculations 
in the BCS limit of the crossover.

\begin{center}
\begin{figure}[htc]
\hspace{3.5cm}\includegraphics[angle=0,width=3.5in]{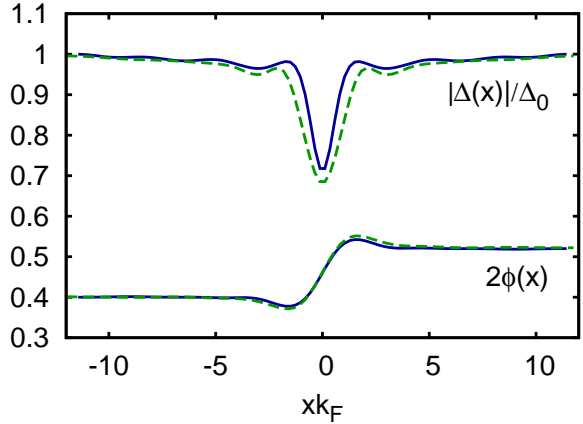}
\caption{Self-consistent profiles of $|\Delta(x)|$ and $2 \phi(x)$ in the weak-coupling limit ($(k_{F} a_{F})^{-1}=-1.5$)
              for two rectangular barriers with the same value of $\mathcal{Z}$. 
              Full lines corresponds to $(L k_{F} = 0.01,V_{0}/E_{F}= 30.0)$ and dashed lines to 
              $(L k_{F} = 0.5,V_{0}/E_{F}=0.6)$. 
              The function $2 \phi(x)$ has been shifted upward by 0.4 for clarity, and the supercurrent flows with 
              $q/k_{F}=0.045$.}
\label{fig8}
\end{figure}
\end{center} 

As an example, we report in Fig.\ref{fig8} the profiles of the magnitude $|\Delta(x)|$ (in units of its bulk value $\Delta_{0}$) and phase $2\phi(x)$ 
obtained by the self-consistent 
solution of the BdG equations for the coupling value $(k_{F} a_{F})^{-1}=-1.5$ and for two rectangular barriers of width $L$ and height $V_{0}$, in such a way that the product $\mathcal{Z} = L V_{0}$ is kept fixed.
Although the full line corresponds to a barrier more closely resembling a delta function (having $L / \xi_{\mathrm{phase}} \simeq 2.9 \times 10^{-3}$), one sees from the figure that there are actually only minor differences with respect to the dashed line corresponding to the larger barrier (having $L / \xi_{\mathrm{phase}} \simeq 0.15$), thus implying that the limit of a delta function is readily approached as soon as $L / \xi_{\mathrm{phase}} \ll 1$.
In Section 4 we shall return to this issue of how our numerical calculations approach this universal behavior.

An additional interesting issue, which has been discussed over the time in the literature when dealing with the BCS (weak-coupling) case, is the role played by self-consistency when solving the BdG equations in this limit.
Contrary to the BEC (strong-coupling) limit for which self-consistency always needs to be fully implemented, in the weak-coupling limit there are, in fact, situations for which the self-consistent process may not be necessarily activated in the barrier region.
These cases, when the solution of the scattering problem can be found analytically without requiring self-consistency, thus represent additional valuable tests for the fully numerical calculations.

Restricting to a Dirac-delta barrier as above, the non-self-consistent calculation is conveniently  \emph{defined} by taking $|\Delta(x)|=\Delta_{0}$ everywhere (except, possibly, at $x=0$ where we may set $\Delta(x)=0$), while $\phi(x)=0$ for $x<0$ and 
$2\phi(x)=\delta \phi$ for $x>0$.
In the non-self-consistent calculation this \emph{ansatz} is used to solve the BdG equations only once, without recalculating the profiles of $|\Delta(x)|$ and $\phi(x)$ from the solutions of these equations.
In such a case, it can be shown that only the bound-state solutions contribute to the current \cite{Beenakker-1991}.
These solutions have a form similar to Eqs.(\ref{Psi-1-range-1}) and (\ref{Psi-M-range-1}), namely,
\begin{equation}
\Psi_{L}(x) \, = \, b_{L} \left( \begin{array}{l} \bar{u} \\ \bar{v} \end{array} \right) e^{-i k^{(e)}  x} 
\, + \,  c_{L} \left( \begin{array}{l} \bar{v} \\ \bar{u} \end{array} \right) e^{i k^{(h)}  x}   
                                                                                               \label{Psi-left-non-self-consistent}
\end{equation}

\noindent
for $x<0$, and
\begin{equation}
\Psi_{R}(x) \, = \, a_{R} 
\left( \begin{array}{l} \bar{u} \, e^{i \delta \phi/2}  \\ \bar{v} \, e^{-i \delta \phi/2} \end{array} \right) e^{i k^{(e)}  x} 
\, + \,  d_{R} 
\left( \begin{array}{l} \bar{v} \, e^{i \delta \phi/2} \\ \bar{u} \, e^{-i \delta \phi/2} \end{array} \right) 
e^{-i k^{(h)}  x}                                                                      \label{Psi-right-non-self-consistent}
\end{equation}

\noindent
for $x>0$, where $k^{(e)}=\sqrt{2m ( \tilde{\mu} + i \sqrt{\Delta_{0}^{2} - E^{2}} )}$ and
$k^{(h)}=\sqrt{2m ( \tilde{\mu} - i \sqrt{\Delta_{0}^{2} - E^{2}} )}$ (with 
$\tilde{\mu}=\mu-\mathbf{k}_{\parallel}^{2}/(2m)>0$ and $E < \Delta_{0}$), while
\begin{eqnarray}
\bar{u} & = & \sqrt{ \frac{1}{2} \left( 1 + i \frac{\sqrt{\Delta_{0}^{2} - E^{2}}}{E} \right)}   \\     
                                                                                                                            \label{bar-u}
\bar{v} & = & \sqrt{ \frac{1}{2} \left( 1 - i \frac{\sqrt{\Delta_{0}^{2} - E^{2}}}{E} \right)} \,\, .
                                                                                                                            \label{bar-v}
\end{eqnarray}
By comparing with Eqs.(\ref{Psi-1-range-1}) and (\ref{Psi-M-range-1}), note that in the non-self-consistent solutions that we are considering we have set $q=0$, so that the (complex) wave vectors $(k^{(1)},k^{(2)},k^{(3)},k^{(4)})$ reduce in the order to 
$(k^{(e)},-k^{(e)},k^{(h)},-k^{(h)})$ as defined above.

In the weak-coupling limit, $\Delta_{0} \ll \mu \simeq E_{F}$ where $E_{F}=k_{F}^{2}/(2m)$ is the Fermi energy. We also anticipate that $E \approx \Delta_{0}$ for the bound-state solutions we are looking for, so that the real parts of $k^{(e)}$ and $k^{(h)}$ are approximately equal to
$k_{\tilde{\mu}} = \sqrt{2m \tilde{\mu}} \simeq \sqrt{k_{F}^{2} - \mathbf{k}_{\parallel}^{2}}$ in the weak-coupling limit we are considering, while their imaginary parts are much smaller than the real parts and will thus be neglected in the following calculation.
This corresponds to the so-called Andreev approximation, which has invariably been adopted in the solution of the BdG equation
in weak coupling (see, e.g. Ref.\cite{Deutscher-RMP-2005}).

Applying the appropriate boundary conditions at $x=0$, namely,
$\Psi_{L}(x=0^{-})=\Psi_{R}(x=0^{+})=\Psi(x=0)$ and $d\Psi_{L}(x)/dx|_{0^{-}} - d\Psi_{R}(x)/dx|_{0^{+}}= 2m \mathcal{Z} \Psi(x=0)$, we arrive at the following homogeneous system for the unknown coefficients $(b_{L},c_{L},a_{R},d_{R})$:
\begin{equation}
\left( \begin{array}{cccc}
\bar{u} & \bar{v} & -\bar{u}e^{i \delta \phi/2} & -\bar{v}e^{i \delta \phi/2} \\ 
\bar{v} & \bar{u} & -\bar{v}e^{-i \delta \phi/2} & -\bar{u}e^{-i \delta \phi/2} \\ 
\bar{u} & -\bar{v} & \bar{u}(1+2i\tilde{\mathcal{Z}}_{\tilde{\mu}})e^{i \delta \phi/2} &
\bar{v}(-1+2i\tilde{\mathcal{Z}}_{\tilde{\mu}})e^{i \delta \phi/2} \\ 
\bar{v} & -\bar{u} & \bar{v}(1+2i\tilde{\mathcal{Z}}_{\tilde{\mu}})e^{-i \delta \phi/2} & 
\bar{u}(-1+2i\tilde{\mathcal{Z}}_{\tilde{\mu}})e^{-i \delta \phi/2}  \end{array} \right) \, 
\left( \begin{array}{c} b_{L} \\ c_{L} \\ a_{R} \\ d_{R} \end{array} \right) \, = \, 
\left( \begin{array}{c} 0 \\ 0 \\ 0 \\ 0 \end{array} \right)               \label{eigen-equation-delta-barrier}
\end{equation}

\noindent
where we have set $\tilde{\mathcal{Z}}_{\tilde{\mu}}=m \mathcal{Z}/k_{\tilde{\mu}}$.
It can be verified that the only solution to Eq.(\ref{eigen-equation-delta-barrier}) has the form:
\begin{equation}
E_{0}(\delta \phi;k_{\tilde{\mu}}) \, = \Delta_{0} \, 
\sqrt{ \frac{1 \, + \, 2 \tilde{\mathcal{Z}}_{\tilde{\mu}}^{2} \, + \, \cos \delta \phi}
{2 \, (1 \, + \, \tilde{\mathcal{Z}}_{\tilde{\mu}}^{2})} }            \,\, .        \label{E-0-delta-barrier}
\end{equation}

This expression can be further manipulated by introducing the transmission coefficient 
$T_{k_{\tilde{\mu}}}$ for the normal case, which equals $(1 + \tilde{\mathcal{Z}}_{\tilde{\mu}}^{2})^{-1}$ for a Dirac-delta barrier. One obtains eventually:
\begin{equation}
E_{0}(\delta \phi;k_{\tilde{\mu}})\, = \Delta_{0} \, \sqrt{ 1 \, - \, T_{k_{\tilde{\mu}}} \, \sin^{2} \delta \phi/2}  \,\, .        
                                                                                                           \label{E-0-delta-barrier-T}
\end{equation}

\noindent
Under these circumstances, the Josephson (current vs $\delta \phi$) characteristic can be obtained \cite{SBW-1997} 
by taking the derivative of the above expression:
\begin{equation}
J(\delta \phi;k_{\tilde{\mu}}) \, = \, - 2 \, \frac{d E_{0}(\delta \phi;k_{\tilde{\mu}})}{d \delta \phi} 
\, = \, \frac{\Delta_{0} \, T_{k_{\tilde{\mu}}}}{2} \, \frac{ \sin \delta \phi}{ \sqrt{ 1 - T_{k_{\tilde{\mu}}} \sin^{2} \, \delta \phi/2} } \,\, .
                                                                                                                  \label{partial-current-delta-barrier}
\end{equation}

\noindent
Note that the energy eigenvalue (\ref{E-0-delta-barrier-T}) and the associated current
(\ref{partial-current-delta-barrier}) depend on the wave vector $\mathbf{k}_{\parallel}$ parallel to the surface of the barrier through $k_{\tilde{\mu}}$, so that the expression (\ref{partial-current-delta-barrier}) has to be integrated over $\mathbf{k}_{\parallel}$ to produce the total current. This integration will be done in a closed form in Appendix B.

The calculation simplifies in the limit of strong ($m \mathcal{Z}/k_{F} \gg 1$) and weak ($m \mathcal{Z}/k_{F} \ll 1$) barriers.
For a strong barrier, $T_{k_{\tilde{\mu}}} \simeq \tilde{\mathcal{Z}}_{\tilde{\mu}}^{-2} \ll 1$ and
\begin{equation}
J(\delta \phi;k_{\tilde{\mu}}) \, \simeq \, \frac{\Delta_{0}}{2 \, \tilde{\mathcal{Z}}_{\tilde{\mu}}^{2}} \, 
\sin \delta \phi \, = \, \frac{\Delta_{0} \, k_{\tilde{\mu}}^{2}}{2 \, m^{2} \, \mathcal{Z}^{2}} \, \sin \delta \phi                                                                                                           
                                                                                  \label{partial-current-delta-barrier-strong}
\end{equation}

\noindent
such that
\begin{equation}
J(\delta \phi) \, = \, \frac{1}{2 \pi} \, \int_{0}^{k_{F}} \! d k_{\parallel} \, k_{\parallel} \, J(\delta ª\phi;k_{\tilde{\mu}})
\, \simeq \, \frac{\Delta_{0} \, k_{F}^{2}}{16 \pi \tilde{\mathcal{Z}}_{F}^{2}} \, \sin \delta \phi
                                                                                  \label{total-current-delta-barrier-strong}
\end{equation}

\noindent
where we have introduced the dimensionless quantity $\tilde{\mathcal{Z}}_{F} = m \mathcal{Z}/k_{F}$.
We recognize in this type of $J$ vs $\phi$ dependence the standard form of the dc-Josephson current \cite{Mahan}.

This form can also be reproduced numerically from the full self-consistent calculation, by considering a sequence of rectangular barriers of width $L$ (such that $L/\xi_{\mathrm{phase}} \ll1$) and height $V_{0}$ with progressively increasing values of $\tilde{\mathcal{Z}}_{F} = (V_{0}/E_{F}) (L k_{F})/2$ such that $\tilde{\mathcal{Z}}_{F} \gg 1$ eventually, and fixing  
to the value $\delta \phi$ of Eq.(\ref{total-current-delta-barrier-strong}) the asymptotic phase difference $2\left[\phi(x=+\infty)-\phi(x=-\infty)\right]$ accumulated by the gap parameter across the barrier.

\begin{center}
\begin{figure}[htc]
\includegraphics[angle=0,width=6.5in]{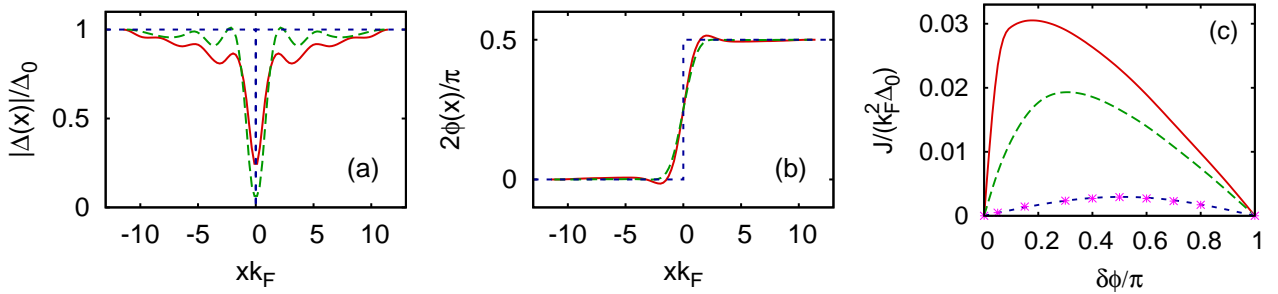}
\caption{Self-consistent profiles of (a) $|\Delta(x)|$ and (b) $2 \phi(x)$, and (c) Josephson characteristics $J(\delta \phi)$ 
              for the coupling $(k_{F} a_{F})^{-1}=-1.5$, obtained for a rectangular barrier with $\tilde{\mathcal{Z}}_{F}=0.25$
              (full lines) and 0.5 (dashed lines). 
              In the three panels, the dotted line represents for comparison the result of the non-self-consistent calculation.
              The asterisks in the lower curve of panel (c) further represent the results of the self-consistent calculation
              with $\tilde{\mathcal{Z}}_{F}=2.5$ for which the limit $1 \ll \tilde{\mathcal{Z}}_{F}$ has effectively been reached, 
              so that no deviation can be appreciated from the non-self-consistent calculation (dotted line).}
\label{fig9}
\end{figure}
\end{center} 

A comparison between self-consistent and non-self-consistent calculations is reported in Fig.\ref{fig9}.
We see from this figure that in the limit $\tilde{\mathcal{Z}}_{F} \gg 1$, not only the Josephson characteristics $J(\phi)$ obtained by the fully self-consistent calculation approach the result (\ref{total-current-delta-barrier-strong}) obtained analytically by the non-self-consistent approximation, but also the profiles of $|\Delta(x)|$ and $2\phi(x)$ approach the corresponding non-self-consistent profiles (barring, of course, the details manifesting at the microscopic length scale $k_{F}^{-1}$
that are associated with the quantum nature of the fermion gas).
The reason is that, when $\tilde{\mathcal{Z}}_{F} \gg 1$, the self-consistent profile of $|\Delta(x)|$ is strongly suppressed locally in the barrier region over a length scale of the order of $\mathcal{L}_{V_{0}} = (2 m V_{0})^{-1/2}$ or $k_{F}^{-1}$ (depending which is the largest one), whereas outside the barrier $|\Delta(x)|$ approaches its asymptotic value $\Delta_{0}$ over the larger length scale $\xi_{\mathrm{phase}}$.
At the same time, when $|\Delta(x)|$ gets close to zero for $x=0$, the phase $\phi(x)$ is allowed to vary sharply about $x=0$ without making the complex number $|\Delta(x)|e^{i 2\phi(x)}$ to move appreciably in the complex plane.
The result (\ref{total-current-delta-barrier-strong}) is thus universal in the weak-coupling limit, since it is obtained when both conditions $L/\xi_{\mathrm{phase}} \ll 1$ \emph{and\/}
$\mathcal{L}_{V_{0}}/\xi_{\mathrm{phase}} \ll 1$ are satisfied, leaving $\xi_{\mathrm{phase}}$ as the only relevant length scale in the problem (again, apart from the microscopic length scale  $k_{F}^{-1}$).

Returning to the general expression (\ref{partial-current-delta-barrier}), we now consider its limit for a weak barrier when $\tilde{\mathcal{Z}}_{F} \ll 1$.
In this case we approximate $T_{k_{\tilde{\mu}}} \simeq 1$ in Eq.(\ref{partial-current-delta-barrier}), such that 
$J(\delta\phi;k_{\tilde{\mu}}) \, \simeq \, \Delta_{0} \sin \delta \phi/2$ and 
\begin{equation}
J(\delta \phi) \, = \, \frac{1}{2 \pi} \, \int_{0}^{k_{F}} \! d k_{\parallel} \, k_{\parallel} \, J(\delta \phi;k_{\tilde{\mu}})
 \, \simeq \, \frac{\Delta_{0} \, k_{F}^{2}}{4 \pi} \, \sin \delta \phi/2 \,\, .
                                                                                  \label{total-current-delta-barrier-weak}
\end{equation}

\noindent
This result can be obtained more thoroughly (as shown in Appendix B), by first integrating over $k_{\parallel}$ the generic expression (\ref{partial-current-delta-barrier}) which is valid for any barrier strength, and then taking the limit of the result for a weak barrier.
Contrary to the case of a strong barrier discussed before, however, for which we have shown that the self-consistent and non-self-consistent calculations lead essentially to the same results, in the case of a weak barrier the non-self-consistent result (\ref{total-current-delta-barrier-weak}) is completely unreliable because it has the wrong dependence on $\delta \phi$ and too large a prefactor as well.
It will, in fact, be shown in Section 4 that for a generic weak barrier (which is not necessarily of the Dirac-delta type) the Josephson characteristics are proportional to $\cos \delta \phi/2$ and \emph{not\/} to 
$\sin \delta \phi/2$, and further that their prefactor cannot exceed the one dictated by the Landau criterion, which equals $\Delta_{0} k_{F}^{2} /(3 \pi^{2})$ in the weak-coupling limit we are considering in this Section. 


\vspace{0.3cm}
\begin{center}
\begin{large}
{\bf 2d. Strong-coupling (BEC) limit and the Gross-Pitaevskii equation}
\end{large}
\end{center}
\addcontentsline{toc}{subsection}{2d. Strong-coupling (BEC) limit and the Gross-Pitaevskii equation}
\vspace{0.1cm}

In the previous sub-section we have emphasized the importance of relying on definite benchmarks, with which the outcomes 
of the numerical calculations can be confronted in some relevant limits at least.
We have also pointed out that for the BCS-BEC crossover these benchmarks have to be looked for in the two limiting BCS 
and BEC situations.
In particular, in the BCS (weak-coupling) limit we have argued that the results of the fully numerical calculation are bound to depend only on the total area of the potential associated with the barrier and not on its specific shape, and we have consequently identified 
the reaching of this kind of universality (which is \emph{internal} to the numerical solution of the BdG equations) with the desired benchmark in this limit. 

In the BEC (strong-coupling) limit that we pass now to discuss, it turns out that one can do 
even better than this and identify a benchmark which is independent from the numerical 
solution of the BdG equations in the presence of a potential barrier.
We may again start our reasoning from the homogeneous situation where, as discussed in the Introduction, a description in terms of composite bosons naturally emerges in the strong-coupling limit of a fully fermionic (zero-temperature) mean-field approach provided the chemical potential is suitably renormalized.
In the presence of inhomogeneities (such as a spatially varying external potential) which induce spatial variations also in the gap parameter, the fermionic approach is embodied by the BdG equations (\ref{BdG-equations}) that generalize the homogeneous 
BCS approach.
In the strong-coupling limit of these equations, we still expect on physical grounds a description in terms of composite bosons to emerge even in the presence of an external potential, possibly with some limitations on its spatial variation.
In addition, to the extent the internal structure of the composite bosons is irrelevant that they can be treated as point-like bosons, one expects these bosons to be described by the Gross-Pitaevskii equation \cite{Gross,Pitaevskii} in the presence of a spatially varying external potential acting on the bosons.  

In this context, it was shown in Ref.\cite{PS-2003} that, in the strong-coupling limit $1 \ll (k_{F} a_{F})^{-1}$ 
of the fermionic attraction, the fermionic BdG equations (\ref{BdG-equations}) can be mapped onto the GP equation   
\begin{equation}
- \frac{\nabla^{2}}{4 \, m} \, \Phi(\mathbf{r}) + 2 \, V(\mathbf{r}) \, \Phi(\mathbf{r}) +
\frac{8 \, \pi \, a_{F}}{2 \, m} \, |\Phi(\mathbf{r})|^{2} \, \Phi(\mathbf{r}) 
\, = \, \mu_{B} \, \Phi(\mathbf{r})                                     \label{Gross-Pitaevskii-equation}                  
\end{equation}

\noindent
for the condensate wave function $\Phi(\mathbf{r})$ of composite bosons with mass $2m$.
In the above equation, $V(\mathbf{r})$ is the same potential of Eqs.(\ref{BdG-equations}) and
$\mu_{B} = 2 \mu + \varepsilon_{0}$ is the chemical potential for composite bosons (where 
$\varepsilon_{0} = (m a_{F}^{2})^{-1}$ is the two-fermion binding energy, with $\mu_{B} \ll \varepsilon_{0}$).
The two functions $\Phi(\mathbf{r})$ of Eq.(\ref{Gross-Pitaevskii-equation}) and $\Delta(\mathbf{r})$ of Eq.(\ref{self-consistency}) are related via the expression
\begin{equation}
\Phi(\mathbf{r}) \, = \, \Delta(\mathbf{r}) \,\, \sqrt{\frac{m^{2} \, a_{F}}{8 \, \pi}} \,\, .
                                                                                                  \label{Phi-vs-Delta}
\end{equation}

\noindent
We emphasize that this mapping is established at low-enough temperatures, so that all composite bosons are condensed and the mean-field approach is appropriate to describe (at least qualitatively) the whole BCS-BEC crossover.  

The derivation of the GP equation (\ref{Gross-Pitaevskii-equation}) from the 
BdG equations (\ref{BdG-equations}) was achieved in Ref.\cite{PS-2003}, by transforming first the differential BdG equations into the associated coupled integral equations for the normal and anomalous single-particle Green's functions, and by expanding then the latter equations in terms of the small parameter $\Delta / |\mu|$, in such a way that only integrals containing the non-interacting Green's function subject to the same external potential remain to be evaluated. 
In this process, a crucial step was represented by the introduction of a local-density approximation for the non-interacting Green's function, which plays an analogous role to the eikonal approximation introduced by Gorkov to derive the Ginzburg-Landau equation 
near the critical temperature in the opposite BCS (weak-coupling) limit \cite{FW-2}, whereby the presence of an external magnetic field is responsible for the spatial inhomogeneity.
The analogy between the two procedures also explains the formal similarity between the GP equation for condensed composite bosons at zero temperature and the Ginzburg-Landau equation for the superconducting order parameter near the critical temperature.

Note that the boson-boson scattering length $a_{B}$ that can be introduced  in Eq.(\ref{Gross-Pitaevskii-equation}) equals $2 a_{F}$, a value that corresponds to the Born approximation for the scattering between composite bosons.
This value is borne out by any mean-field treatment of the BEC limit of the BCS-BEC crossover.
Improvement on the treatment of the low-energy scattering between composite bosons results in different values for the ratio 
$a_{B} / a_{F}$ \cite{PS-2000}.
In particular, an exact treatment \cite{PSS-2004,Brodsky-2005} yields the value $a_{B} / a_{F} = 0.6$.
Although it is possible to include the fermionic scattering processes responsible for this reduced value of $a_{B} / a_{F}$ in the derivation of the GP equation for the composite bosons  \cite{PS-2006} (which thus would contain the correct value $0.6 a_{F}$ of $a_{B}$ instead of $2 a_{F}$), we regard this quantitative refinement in the BEC limit to be beyond the scope of the present treatment, where we concentrate instead in solving the BdG equations (\ref{BdG-equations}) throughout the whole BCS-BEC crossover.

Note further the essential role played by self-consistency in the derivation of the GP equation, since for this equation self-consistency can by no means be dismissed.
This is true in spite of the fact that the BdG equations (\ref{BdG-equations}) are apparently linear equations, while self-consistency 
is introduced by the condition (\ref{self-consistency}) on the local gap parameter $\Delta(\mathbf{r})$.

The derivation of Eq.(\ref{Gross-Pitaevskii-equation}) from the BdG equations provided in Ref.\cite{PS-2003} relies on the smallness of the parameter $\Delta / |\mu|$ when $1 \ll (k_{F} a_{F})^{-1}$, as well on the smoothness condition to be satisfied by the 
potential $V(\mathbf{r})$.
In practice, however, even for a square-well potential with sharp edges we have favorably compared the results of 
our numerical solution of the BdG equations (\ref{BdG-equations}) for large enough values of 
the coupling parameter $(k_{F} a_{F})^{-1}$ with the \emph{independent} numerical solution of the GP equation (\ref{Gross-Pitaevskii-equation}), obtained with the \emph{same} value of $a_{F}$ and with the value of the bosonic chemical potential $\mu_{B} = 2 \mu + (m a_{F}^{2})^{-1}$ expressed in terms of the corresponding fermionic chemical potential $\mu$.

\begin{center}
\begin{figure}[htc]
\includegraphics[angle=0,width=6.5in]{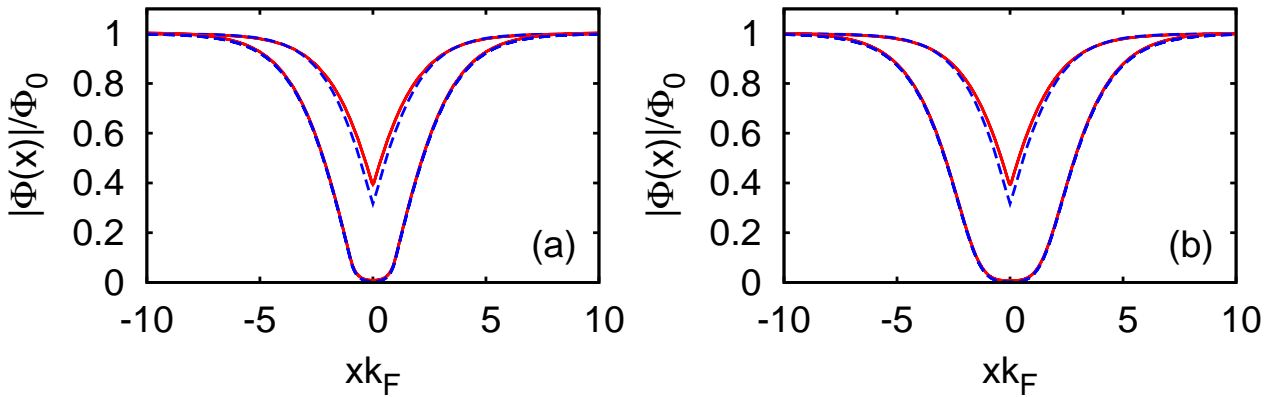}
\caption{Comparison of the solutions of the BdG equations (dashed lines) and GP equation (full lines) for 
              $(k_{F} a_{F})^{-1}=+3.0$. 
              The magnitude $|\Phi(x)|$ is here reported for $q=0$ in units of its bulk value $\Phi_{0}$.
              Two (a) rectangular and (b) Gaussian barriers of widths $L k_{F} = \sigma k_{F} \sqrt{2 \pi} = (0.1,2.0)$ 
              and height $V_{0}/E_{F} = 4.0$ are considered [the upper (lower) curves correspond to the narrow (wide)
              barriers].
              The barrier widths need be confronted with the size $\xi_{\mathrm{pair}} k_{F} = (3 \sqrt{2})^{-1} = 0.23$ of 
              the composite bosons for this coupling.}
\label{fig10}
\end{figure}
\end{center}

An example of this comparison is shown in Fig.\ref{fig10} when $(k_{F} a_{F})^{-1}=+3.0$ for rectangular and Gaussian barriers of comparable heights and widths.
Two alternative physical situations are further considered in this figure, namely, when the size of the composite bosons (given by $\xi_{\mathrm{pair}} = a_{F}/\sqrt{2}$ in this limit \cite{PS-1994}) is smaller or larger than the barrier width.
Only in the latter case, in fact, we expect deviations to occur between the solutions of the BdG and GP equations, since the composite nature of the bosons mostly manifests itself when they cannot fit into the barrier region. Moderate deviations along these lines are evident from Fig.\ref{fig10}.
We shall return to this interesting issue in Section 4 while providing an overall discussion of the results of the numerical calculations.

Besides being an important benchmark for the numerical solution of the BdG equations in the BEC limit, the independent solution of the GP equation plays in this case an additional role as it permits one to explore parameter ranges (concerning, for instance, the barrier) for which a direct numerical solution of the BdG equations becomes exceedingly difficult and/or time consuming.
In these cases, a preliminary solution of the GP equation may serve as an exploratory mean for focusing on the physical effects one is after. 
In the specific context of the Josephson effect, the solution of the GP equation may also serve for evidencing an alternative physical meaning of the effect itself, as discussed in Appendix C.

Still regarding the Josephson effect, the above mapping between the BdG and GP equations in the BEC limit allows one to establish a direct link with the literature concerning coherence effects in Bose-Einstein condensates, which are made up of point-like bosons and are themselves invariably described by the GP equation \cite{DGPS-rmp-1999}.

Finally, we refer the reader to Appendix D for some useful manipulations one needs to perform on the GP equation (\ref{Gross-Pitaevskii-equation}) in order to ease its numerical solution in the presence of a barrier with a slab geometry and a finite current.

\vspace{0.5cm}  
\section{Numerical strategies and procedures}

We pass now to discuss a number of procedures that prove necessary for an efficient numerical implementation of the method discussed in Section 2, whereby the solution of the BdG equations (\ref{BdG-equations}) in the presence of a barrier with a slab geometry was set up operatively.
In particular, we will discuss the numerical strategies that are required to speed up the cycles of self-consistency for Eq.(\ref{final-regularized-gap-equation}), a process that proves especially important when approaching the strong-coupling (BEC) limit.

To reconcile an accurate solution of the scattering problem based on the BdG equations with a feasible effort for attaining self-consistency, we have considered at the ouset two different spatial grids, one grid at the positions $\{x_{\ell}; \ell=1,\cdots,M-1\}$ where the continuity equations for the scattering problem are specified, and the other grid at the positions $\{\bar{x}_{\bar{\ell}}; \bar{\ell}=1,\cdots,\bar{M}-1\}$ where the local gap parameter $\Delta(x)$ is taken as a variable in the cycles of self-consistency.
In practice, the grid $\{\bar{x}_{\bar{\ell}}\}$ can be taken to be a coarse subset of the grid $\{x_{\ell}\}$ (such that $\bar{M} \ll M$), while the values of $\Delta(x_{\ell})$ between two successive points $\bar{x}_{\bar{\ell}}$ and $\bar{x}_{\bar{\ell} + 1}$ (where $\Delta(x)$ is specified in the cycles of self-consistency) are obtained by suitable interpolation (which is necessary because in the scattering problem the values of $\Delta(x)$ need be specified over the denser grid $\{x_{\ell}\}$).
Typically, we have taken $M \simeq 100$ and $\bar{M} \simeq 20$.
Furthermore, a judicious choice of the points $\{\bar{x}_{\bar{\ell}}\}$ of the coarse grid proves necessary, so as to avoid
obtaining sharp variations of $\Delta(x)$ where the barrier itself has jump discontinuities (this shortcoming is especially evident in the BEC limit when comparison with the results of the GP equation is possible).
To this end, it proves sufficient not to include in this grid the points where these discontinuities occur.

Equation (\ref{equation-AW=B}) is then solved separately in each of the seven energy ranges introduced in sub-section 2b.
Special care is necessary in ranges 1 and 5 where bound-state solutions occur, as these may sometimes escape even a quite refined search.
As a matter of fact, precise inclusion of these bound-state solutions turns out to be essential for an accurate determination of the physical quantities to be calculated, and also for reaching the self-consistency condition itself of the BdG equations.
In addition, in range 2 resonant (quasi-bound) states may occur which need also to be treated with sufficient accuracy.

Integration over $\mathbf{k} = (k_{\perp}, \mathbf{k}_{\parallel})$ is next performed in the self-consistency equation (\ref{final-regularized-gap-equation}) as well as in the equations (\ref{density}) for the density and (\ref{current}) for the current.
As discussed in Appendix A, the integration over $k_{\perp}$ need be performed first (and contains also a discrete sum
over the bound states that are defined for each given value of $\mathbf{k}_{\parallel}$).
In this integration (which extends in principle from $-\infty$ to $+\infty$), the variable $k_{\perp}$ coincides (in a piecewise continuos fashion) alternatively with one of the (real) wave vectors $k^{(n)}$ reported in Figs.\ref{fig5} and \ref{fig6}, depending on the energy ranges of Fig.\ref{fig4}.

Both integrals (over $k_{\perp}$ and $\mathbf{k}_{\parallel}$) are further split into two ranges. For each integral, a cutoff $k^{c}$ is introduced
(which may, of course, differ for the $\perp$ and $\parallel$ directions), such that for $k < k^{c}$
the integration is done numerically while for $k^{c} < k < + \infty$ one first extracts the appropriate power-law decay of the integrand together with the relative coefficient and then perform the integration analytically.
Since Eqs.(\ref{density}), (\ref{current}), and (\ref{final-regularized-gap-equation}) are all local relations, also $k^{c}$ depends parametrically on $x$, the largest values of $k^{c}(x)$ occurring at the position ($x=0$) of the barrier. 

\begin{center}
\begin{figure}[t]
\hspace{3cm}\includegraphics[angle=0,width=4in]{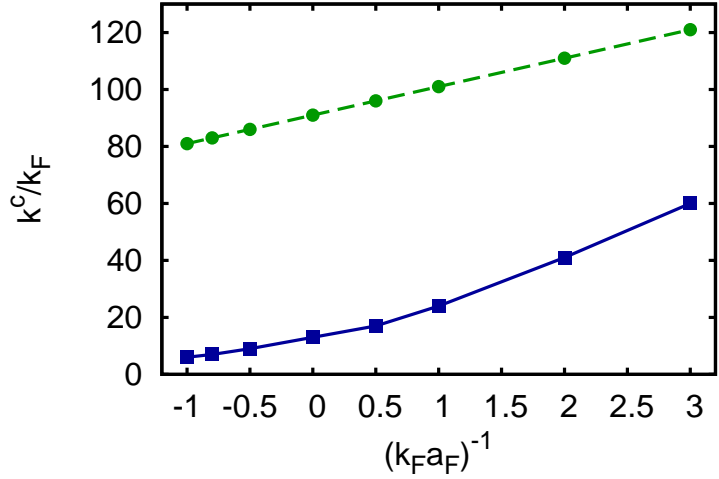}
\caption{Cutoffs $k^{c}_{\parallel}(x=0)$ (full line) and $k^{c}_{\perp}(x=0)$ (dashed line) in units of $k_{F}$, which 
              are used in the numerical integrations, vs the coupling $(k_{F} a_{F})^{-1}$ for a typical rectangular barrier 
              with $L k_{F} = 4.0$ and $V_{0}/E_{F} = 0.1$.}
\label{fig11}
\end{figure}
\end{center}

With these considerations in mind, we plot in Fig.\ref{fig11} the values of $k^{c}_{\parallel}(x=0)$ (full line) and $k^{c}_{\perp}(x=0)$ (dashed line) vs the coupling parameter $(k_{F} a_{F})^{-1}$ for a typical SsS barrier we have considered (we have found that these plots
depend only mildly on the chosen barrier).
Since $k^{c}_{\perp}(x=0)$, in turn, depends on the value of $k_{\parallel}(x=0)$, we have here considered $k^{c}_{\perp}(x=0)$ in
the extreme case when $k_{\parallel}(x=0)=k^{c}_{\parallel}(x=0)$.
Note that both quantities increase rapidly when approaching the BEC limit of the crossover
(where $1 \ll (k_{F} a_{F})^{-1}$), thus evidencing the increasing numerical effort one has 
to sustain on the BEC side with respect to the BCS side of the crossover.

Even though Eq.(\ref{final-regularized-gap-equation}) was originally introduced as the condition for enforcing self-consistency, a ``mixed choice'' is actually to be preferred for the purpose.
It turns out, in fact, that to the real and imaginary parts of Eq.(\ref{final-regularized-gap-equation}) there correspond different sensitivities to the self-consistent parameters $\{\Delta(x_{\bar{\ell}})\}$ and $\{\phi(x_{\bar{\ell}})\}$ as well as to the current 
wave vector $q$.
Specifically, the imaginary part of Eq.(\ref{final-regularized-gap-equation}) proves to be quite insensitive to the values of these variables, so that it cannot be the appropriate candidate for their self-consistent determination. 
The real part of Eq.(\ref{final-regularized-gap-equation}), on the other hand, is suitable to the purpose.
We have then replaced the imaginary part of Eq.(\ref{final-regularized-gap-equation}) by the current equation (\ref{current}), with its left-hand side set equal to $j = q n_{0} / m$ where 
$n_{0}$ is the asymptotic (bulk) value of the density far from the barrier (the current being uniform).
The choice of Eq.(\ref{current}) is also suggested by the fact that, in the strong-coupling (BEC) limit, the continuity equation (and thus the constancy of the current) results from taking the imaginary part of the GP equation [cf. Appendix D].

\begin{center}
\begin{figure}[htc]
\includegraphics[angle=0,width=6.5in]{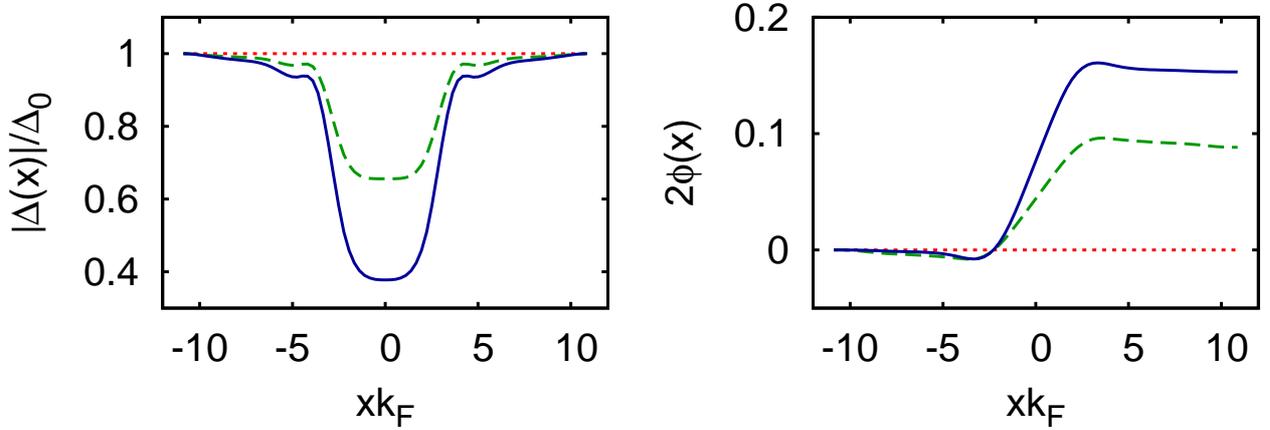}
\caption{Evolution toward self-consistency of $|\Delta(x)|$ and $2 \phi(x)$ for the coupling value 
              $(k_{F} a_{F})^{-1} = -0.5$ with a rectangular barrier of width $L k_{F} = 5.0$ and height $V_{0}/E_{F} = 0.5$.
              Dotted, dashed, and full lines stand for initial, intermediate, and final values of the functions, in the order.}
\label{fig12}
\end{figure}
\end{center} 

With the two above equations, the approach to self-consistency has been achieved via the multi-dimensional Newton 
method, which proves to be more efficient than the iterative method (used, for instance, in 
Ref.\cite{Bagwell-1996}) in reducing the number of cycles.
By the present method, self-consistency was invariably reached within at most five cycles.  
As an example, we plot in Fig.\ref{fig12} a typical evolution toward self-consistency of the profiles of 
$|\Delta(x)|$ and $2 \phi(x)$.
                                                                                
The numerical calculations have been performed by imposing the uniform value $qn_{0}/m$ of the current, with the provision that 
this value should be self-consistently sustained.
We will, in fact, find that to each coupling $(k_{F} a_{F})^{-1}$ there corresponds an upper (critical) value $q_{c}$ that can be self-consistently sustained, which we will associate to the Landau criterion for superfluidity as discussed in sub-section 4f.
By this procedure, a value $\delta\phi$ for the asymptotic phase difference $2[\phi(x = + \infty) - \phi(x = - \infty)]$ between the two sides of the barrier results for given $q$.
Alternatively, the numerical calculations can be performed by fixing the value of $\phi(x = + \infty) - \phi(x = - \infty)$ and letting $q$ be an independent variable to be detemined self-consistently. 
We have actually adopted either one of these two procedures depending on the circumstances.
Specifically, when the value of $q$ is selected in advance, we have found that the numerical solution invariably converges
to that corresponding to the ``left'' branch of the Josephson characteristics (see sub-section 4b).
To access also the solutions corresponding to the ``right'' branch of the Josephson characteristics, therefore, the value
of $\delta\phi$ need be selected in advance.
The corresponding stability of the solutions of the left and right branches of the Josephson characteristics will be
addressed in detail in sub-section 5b.

As a matter of fact, we have found it more convenient to use the derivative $\phi'(x)$ of the phase (and \emph{not\/} the phase $\phi(x)$ itself) taken at the coarse set of points $\{\bar{x}_{\bar{\ell}}\}$, as the independent variables to be inserted in the self-consistency conditions.
That this is a more appropriate choice of variables is also evident from the solution of the GP equation
[cf. Appendix D], although it appears less evident when solving the BdG equations troughout the BCS-BEC crossover 
since in this case $\phi'(x)$ occurs explicitly only in the current equation.
Operatively, we begin by an initial guess of $\phi'_{\bar{\ell}}$ at the points $\{\bar{x}_{\bar{\ell}}\}$, from 
which we obtain a continuous function $\phi'(x)$ by (a cubic spline) interpolation, and then obtain the required values
$\phi_{\bar{\ell}}$ for each $\bar{\ell}$ as follows:
\begin{equation}
\phi_{\bar{\ell}} \, = \,  \int_{- \infty}^{\bar{x}_{\bar{\ell}}} \! \, dx' \,\, \phi'(x')        \label{phi-from-phi'}  \,\, .
\end{equation}

\noindent
Numerically, the advantage of this procedure over the more obvious choice of $\phi(x)$ as the independent variable (from which $\phi'(x)$ could be obtained by numerical differentiation when necessary), is that the integration introduces a smaller numerical error than the differentiation, so that one can consider a more limited number ($\bar{M} - 1$ ) of points in the coarse grid. 

An additional numerical difficulty we have encountered when solving the equation (\ref{equation-AW=B}) is that some matrix elements (\ref{column-vector-Mn}) may become exceedingly large owing to the presence of the factor $\exp \{i k_{\ell}^{(n)} x \}$ therein.
To better appreciate this point, let's refer, for instance, to an electron-like excitation impinging from the 
left in range 4. 
In this case one sees from the forth panel of Fig.\ref{fig5} that $k^{(3)}$ possesses a negative imaginary part, in such a way that $\exp \{i k^{(3)} x \}$ grows without bound when $x$ becomes large to the right of the barrier.
This was precisely the reason for excluding the corresponding wave function (\ref{Upsilon}) from the linear superposition (\ref{Psi-vs-Upsilon}) in the outermost left interval with $\ell = M$ [cf. Table 1], thus setting $c_{M}=0$ at the ouset.
The same assignement cannot \emph{a priori} be done, however, with the coefficient $c_{M-1}$ 
of the next interval to the right, a coefficient that should turn out from the solution of Eq.(\ref{equation-AW=B}) to be exceedingly small in order to compensate for the large factor $\exp \{i k_{M-1}^{(n)} x \}$.
By this mechanism, one would inconveniently multiply quite large with quite small numbers, a procedure that may lead to substantial numerical errors.
To reduce these errors, one may absorb the large factors $\exp \{i k_{\ell}^{(n)} x \}$ (whenever they occur) in the associated coefficients of the expressions (\ref{Psi-vs-Upsilon}), by redefining in each interval:
\begin{eqnarray}
\Psi_{\ell}(x;q,E) & = & \tilde{a}_{\ell} \, \Upsilon_{\ell}^{(1)}(x - x_{\ell - 1};q,E) 
                                  + \tilde{b}_{\ell} \, \Upsilon_{\ell}^{(2)}(x - x_{\ell - 1};q,E)   \nonumber \\
                           & + & \tilde{c}_{\ell} \, \Upsilon_{\ell}^{(3)}(x - x_{\ell - 1};q,E)
                                  + \tilde{d}_{\ell} \, \Upsilon_{\ell}^{(4)}(x - x_{\ell - 1};q,E)  
                                                                                                        \label{Psi-vs-Upsilon-rescaled}
\end{eqnarray}

\noindent
where $\tilde{a}_{\ell} = a_{\ell} \exp \{i k_{\ell}^{(1)} x_{\ell - 1} \}$,
$\tilde{b}_{\ell} = b_{\ell} \exp \{i k_{\ell}^{(2)} x_{\ell - 1} \}$,
$\tilde{c}_{\ell} = c_{\ell} \exp \{i k_{\ell}^{(3)} x_{\ell - 1} \}$, and
$\tilde{d}_{\ell} = d_{\ell} \exp \{i k_{\ell}^{(4)} x_{\ell - 1} \}$.
In this way, the continuity conditions (\ref{continuity-conditions}) are replaced by
\begin{equation}
\mathbf{M}_{\ell}(x=x_{\ell} - x_{\ell - 1}) \, \tilde{W}_{\ell} \, = \, \mathbf{M}_{\ell + 1}(x=0) \, 
\tilde{W}_{\ell + 1}                                                        \label{continuity-conditions-rescaled}
\end{equation}

\noindent
where $\tilde{W}_{\ell}^{T} = (\tilde{a}_{\ell},\tilde{b}_{\ell},\tilde{c}_{\ell},\tilde{d}_{\ell})$  
[cf. Eq.(\ref{column-vector-Wl})].
The expressions entering Eq.(\ref{equation-AW=B}) are consequently modified.
By this expedient, we have effectively enlarged the range of the energy $E$ which is required 
for accurate numerical calculations (especially in the BEC limit of the crossover).
This is because the phase factors $\exp \{i k_{\ell}^{(n)} x \}$ entering the matrix elements  
(\ref{column-vector-Mn}) can blow up because either $\mathrm{Im}\{k_{\ell}^{(n)}\}$ or $|x|$ get 
large.
With the transformation (\ref{Psi-vs-Upsilon-rescaled}), $|x|$ becomes at most limited by the width of the largest interval $\Delta x_{\ell} = x_{\ell} - x_{\ell - 1}$.
Typically, $\Delta x_{\ell} \, \mathrm{Im}\{k\}_{\mathrm{max}} \lapprox 500$ in order to avoid numerical under- or over-floating.
Here, one can choose $\Delta x_{\ell} \approx \xi_{\mathrm{phase}} / 10$ in order to obtain the profile of the gap with sufficient accuracy (apart from the occurrence of the Friedel oscillations over a scale $k_{F}^{-1}$ in the BCS limit of the crossover, where  smaller values of $\mathrm{Im}\{k\}_{\mathrm{max}}$ need be considered).
This implies that $\xi_{\mathrm{phase}} \, \mathrm{Im}\{k\}_{\mathrm{max}} \lapprox 5000$.
Since $(k_{F} \, \xi_{\mathrm{phase}})_{\mathrm{max}} \lapprox 10$ in the relevant range of the crossover (cf. Fig.\ref{fig7}), one 
obtains $\mathrm{Im}\{k\}_{\mathrm{max}} / k_{F} \lapprox 500$ as a conservative estimate for the upper bound on $\mathrm{Im}\{k\}_{\mathrm{max}}$ while
$\Delta x_{\ell} \, k_{F} \lapprox 1$, values that turn out to be totally sufficient for all practical purposes.

Note further that the present method does not put too severe a limitation on the maximun 
number of zones $M$ one has to rely on for an accurate description of the problem (the major 
limitation originating, in practice, by the increasing computational time when $M$ increases).
This contrasts with the transfer-matrix method described in Appendix E, where the number of zones is severely limited by the numerical under- or over-floating problems one encounters 
when $M$ exceeds 40.
\vspace{0.3cm}

\vspace{0.5cm}  
\section{Results for an SsS barrier}

We pass now to give a detailed account of the results that we have obtained by solving numerically the BdG equations (\ref{BdG-equations}) subject to the self-consistent condition (\ref{self-consistency}) in the presence of a stationary superfluid flow impinging on a barrier, according to the methods described in Sections 2 and 3 which enable us to scan the whole BCS-BEC crossover at zero temperature \cite{Spuntarelli-07}.
Our systematic calculations will not only provide a bridge between the treatments of the stationary Josephson effect which were so far considered separately for fermionic and bosonic systems in the literature, but they will also uncover several novel and interesting features about the Josephson and related effects that could not emerge when addressing these (BCS and BEC) limits separately.


\vspace{0.3cm}
\begin{center}
\begin{large}
{\bf 4a. Spatial profiles of the fermionic wave functions and the complex gap parameter}
\end{large}
\end{center}
\addcontentsline{toc}{subsection}{4a. Spatial profiles of the fermionic wave functions and the complex gap parameter}
\vspace{0.1cm} 

Knowledge of the solutions $u_{\nu}(\mathbf{r})$ and $v_{\nu}(\mathbf{r})$ of the BdG equations (\ref{BdG-equations}) is required to calculate the gap function (\ref{self-consistency}) as well as the number density (\ref{density}) and current (\ref{current}) (or the total energy - cf. Eq.(\ref{thermal-average-T=0-regularized}) of sub-section 5a).
Even though these wave functions have no individual meaning, it is nevertheless instructive to examine their spatial behavior along $\mathbf{r}=(x,0,0)$ in some characteristic cases to get a physical feeling about their overall role.

\begin{center}
\begin{figure}[htc]
\includegraphics[angle=0,width=6.5in]{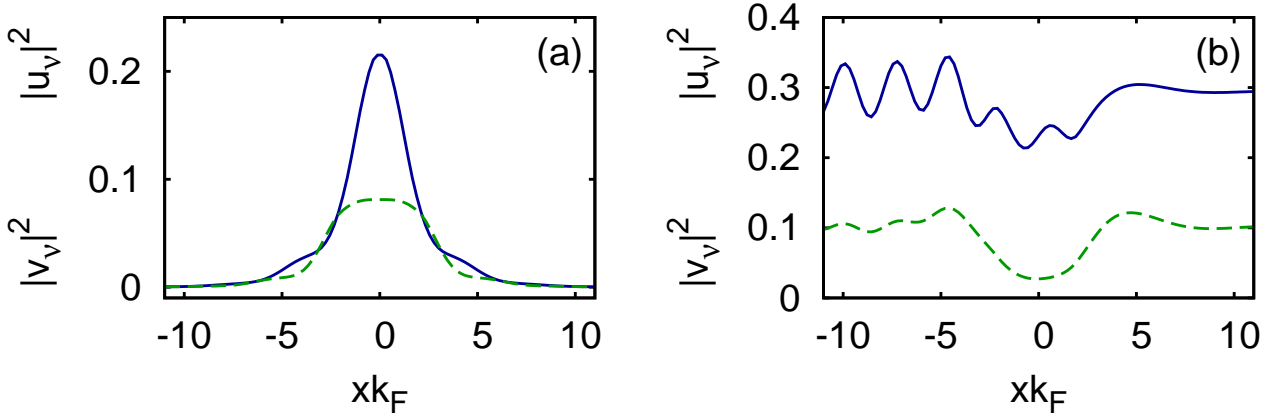}
\caption{Plot of $|u_{\nu}(x)|^{2}$ (full lines) and $|v_{\nu}(x)|^{2}$ (dashed lines) vs $x$ at unitarity for (a) a bound-state solution 
             and (b) an electron-like excitation in the continuum, for a rectangular barrier in the presence of a finite current. 
              The values of parameters are specified in the text.}
\label{fig13}
\end{figure} 
\end{center}

As an example, in Fig.\ref{fig13} we plot $|u_{\nu}(x)|^{2}$ and $|v_{\nu}(x)|^{2}$ vs $x$ at self-consistency for the 
following cases that correspond to two of the four ranges 

\noindent
identified in Fig.\ref{fig4}(a): 
(a) A bound-state solution occurring in range 1; 
(b) An electron-like excitation whose energy $E$ is comprised in range 3.
These states are calculated for a rectangular barrier of width $L k_{F}=4.0$ and height $V_{0}/E_{F}=0.2$,
with $\mathbf{k}_{\parallel}=0$ and $q/k_{F}=0.128$, and for the coupling value $(k_{F} a_{F})^{-1}=0$ at unitarity.
Both the continuum and the bound-state solutions are affected by the depression of the magnitude 
$|\Delta(x)|$ of the gap parameter about $x=0$, which originates from the presence of the repulsive barrier.
[In particular, the bound state would not exist were not for the depression occurring in $|\Delta(x)|$.]  

\begin{center}
\begin{figure}[htc]
\hspace{3cm}\includegraphics[angle=-90,width=4in]{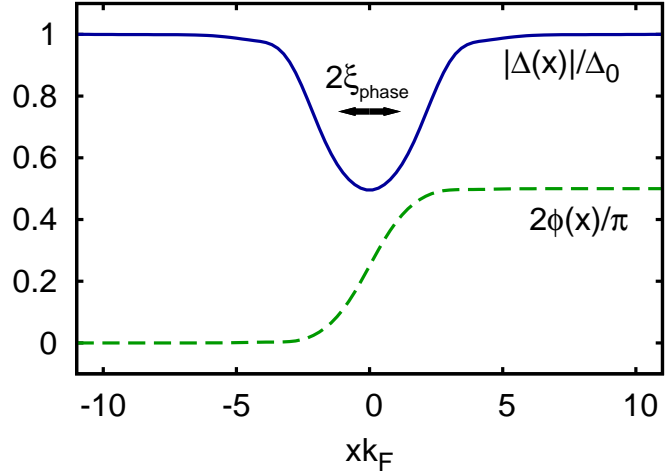}
\caption{The profiles of $|\Delta(x)|$ and $2 \phi(x)$ vs $x$ are shown for the same parameters of Fig.\ref{fig13}. 
             The arrow identifies the extension of the length $\xi_{\mathrm{phase}}$.}
\label{fig14}
\end{figure}
\end{center} 

The corresponding profiles of the magnitude $|\Delta(x)|$ and phase $2\phi(x)$ are reported in Fig.\ref{fig14} for the same parameter values. 
[Recall that, by our definition Eq.(\ref{piecewise-constant}), the phase of the local gap parameter is $2\phi(x)$ and not $\phi(x)$.] 
Note how the spatial variations of these quantities occur over the scale $\xi_{\mathrm{phase}}$ introduced for the homogeneous case (as identified by the arrow in the figure), and that these variations are mutually related by the condition of the current being uniform.

\begin{center}
\begin{figure}[htc]
\includegraphics[angle=0,width=6.5in]{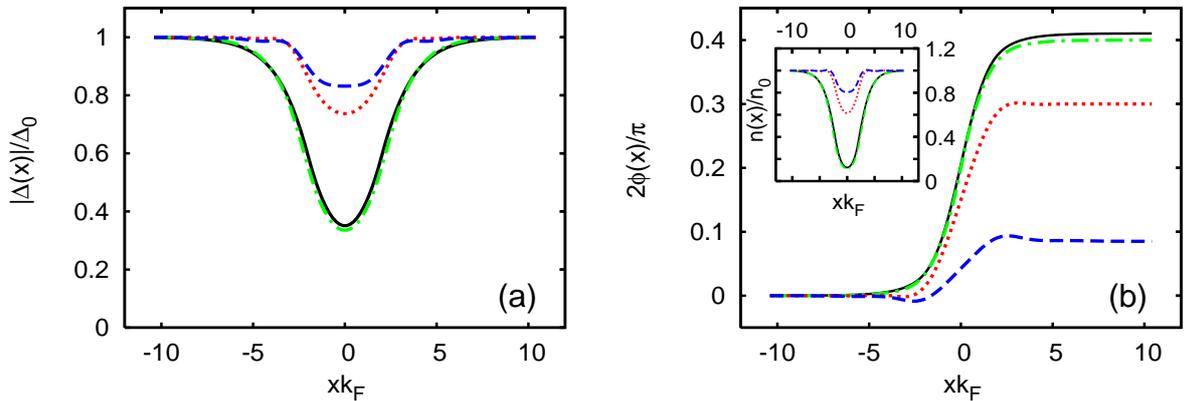}
\caption{Evolution of the profiles of $|\Delta(x)|$ and $2 \phi(x)$ from weak to strong coupling, obtained for a rectangular 
              barrier with $L k_{F}=4.0$ and $V_{0}/E_{F}=0.1$ (the values of the coupling are specified in the text). 
              The inset shows the corresponding density profiles $n(x)$. 
              The solution of the GP equation for the coupling value on the BEC side is shown for comparison (full line).}
\label{fig15}
\end{figure}
\end{center} 

We will devote sub-section 4e below to a detailed analysis of the bound-state solutions (the so-called Andreev-Saint James states), since they play a special role in the theory of the Josephson effect.
We examine instead here the way the profiles of $|\Delta(x)|$ and $2 \phi(x)$ for given barrier and $q$ evolve with increasing coupling, by considering in Fig.\ref{fig15} three characteristic coupling values $(k_{F} a_{F})^{-1}=-0.8$ (dashed lines),
$0.0$ (dotted lines), and $+3.0$ (dashed-dotted lines), for which $k_{F} \xi_{\mathrm{phase}} =(1.33,0.68,1.32)$ from weak to strong coupling, in the order.
In each case, the value of $q$ corresponds to the maximum sustainable value of the current for the given barrier and coupling.
The profiles obtained by the independent solution of the GP equation (\ref{Gross-Pitaevskii-equation}) are also
reported for comparison (full lines) when $(k_{F} a_{F})^{-1}=+3.0$, showing a remarkable agreement between the 
two independent (BdG and GP) calculations for this case.

Note in Fig.\ref{fig15} the progressive depression of $|\Delta(x)|$ at the center of the barrier as the coupling is increased, with an associated steepening of $ 2\phi(x)$ that results in an increase of  the asymptotic phase difference 
$\delta \phi = 2[\phi(x = + \infty) - \phi(x = - \infty)]$ between the two sides of the barrier.
The depression of $|\Delta(x)|$ is also reflected in the corresponding density profiles shown in the inset (where $n_{0}$ is the bulk fermionic density). In particular,  $|\Delta(x)|^{2}$ and $n(x)$ are proportional to each other in the BEC limit (cf. Eq.(\ref{Phi-vs-Delta})).
These features are independent from the shape of the barrier, as we have explicitly verified by repeating the calculation 
using a Gaussian barrier with the same area for definiteness  (that is, taking $\sigma = L / \sqrt{2 \pi} $).


\vspace{0.3cm}
\begin{center}
\begin{large}
{\bf 4b. Josephson characteristics: Current versus phase relations}
\end{large}
\end{center}
\addcontentsline{toc}{subsection}{4b. Josephson characteristics: Current versus phase relations}
\vspace{0.1cm} 

The above calculations, which relate the value of the asymptotic phase difference $\delta \phi$ to the value $q$ of the current wave vector, can be performed in a systematic way so as to obtain the function $q(\delta \phi)$ which identifies the characteristic Josephson current-phase relation $J(\delta \phi)$ where $J = q n_{0}/m$.
In this context, one actually finds that two different values of $\delta \phi$ (corresponding to two different
spatial profiles $\phi(x)$) are associated with a given value of $q$.
These two solutions give rise to \emph{two\/} different \emph{branches\/} of the $J$ vs $\phi$ relation, 
which we shall conventionally call the ``left'' and ``right'' branch and which will be shown in Section 5 to 
possess different stability properties against the introduction of fluctuations.
Setting apart for the moment the issue of their stability, we will represent here the two branches as a single continuous curve and discuss their overall properties across the BCS-BEC crossover. 

\begin{center}
\begin{figure}[htc]
\includegraphics[angle=0,width=6.5in]{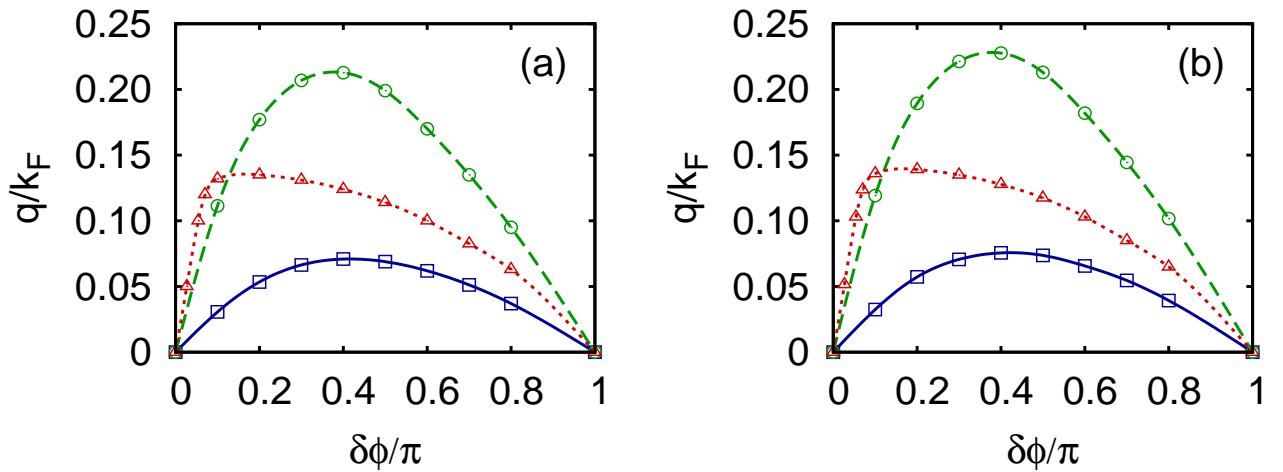}
\caption{Josephson characteristics $q(\delta \phi)/k_{F} = J(\delta \phi) m /(k_{F} n_{0})$ for (a) a rectangular barrier with $L k_{F}=4.0$ 
              and (b) a Gaussian barrier with $(\sigma k_{F})=1.6$ ($V_{0}/E_{F}=0.1$ in both cases).
              The three curves correspond to $(k_{F} a_{F})^{-1}=-1.0$ (dotted lines), 0.0 (dashed lines), and +1.5 (full lines).}
\label{fig16}
\end{figure}
\end{center}

In Fig.\ref{fig16} we report the behavior of $J$ vs $\delta \phi$ for given barrier and different coupling values spanning the crossover region.
A rectangular [Fig.\ref{fig16}(a)] and a Gaussian [Fig.\ref{fig16}(b)] barrier are considered with the same value of the 
(dimensionless) area $\tilde{\mathcal{Z}}_{F}=m \mathcal{Z}/k_{F}$ (which equals $(L k_{F}) (V_{0}/E_{F}) /2$ for a rectangular and 
$(\sigma k_{F}) (V_{0}/E_{F}) \sqrt{\pi/2}$ for a Gaussian barrier - cf. Eq.(\ref{total-area})).

\noindent

Two features appear evident from these plots. 
In both cases, the Josephson current is considerably enhanced at unitarity with respect to the BCS and BEC sides
(a phenomenon to which we will return more extensively in the following).
In addition, the curves are seen to stretch from being proportional to $\sin (\delta \phi)$ to being (almost) proportional 
to $\cos (\delta \phi/2)$ following the evolution from strong to weak coupling.

The proportionality to $\sin (\delta \phi)$ when approaching the BEC side with given barrier height represents 
a standard result for the Josephson characteristics, and reflects the fact that in this limit the barrier height $V_{0}$
becomes larger than the chemical potential $\mu_{B}$ of the composite bosons, which is the only other relevant 
energy scale of the problem.

The proportionality to $\cos (\delta \phi/2)$ when approaching the BCS side, on the other hand, results from having achieved full self-consistency in the calculation, whereas the non-self-consistent calculation discussed in sub-section 2c 
(cf. Eq.(\ref{total-current-delta-barrier-weak})) would yield a $\sin (\delta \phi /2)$ proportionality for a weak barrier with 
$\tilde{\mathcal{Z}}_{F} \ll 1$ ($\tilde{\mathcal{Z}}_{F} = 0.2$ for the barriers of Fig.\ref{fig16}).

\begin{center}
\begin{figure}[htc]
\includegraphics[angle=0,width=6.5in]{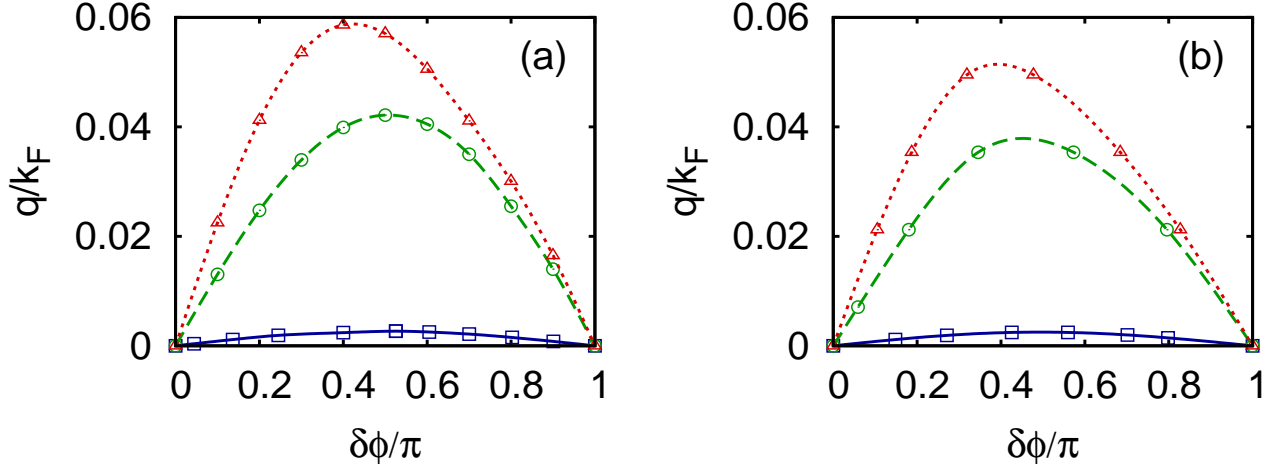}
\caption{Josephson characteristics $q(\delta \phi)/k_{F} = J(\delta \phi) m /(k_{F} n_{0})$ for (a) a rectangular and (b) a Gaussian 
              barrier with the same widths of Fig.\ref{fig16} but with the larger height $V_{0}/E_{F}=0.4$. 
              The coupling values are the same of Fig.\ref{fig16}.}
\label{fig17}
\end{figure}
\end{center} 

In Fig.\ref{fig16}, a rather small value was considered for the ratio $V_{0}/E_{F}$. 
[Note, however, that such a value for the height of the barrier becomes soon comparable with the chemical potential 
$\mu_{B}$ of the composite bosons when approaching the BEC side of the crossover, and in this sense it is not small.]
In Fig.\ref{fig17} we report for comparison the corresponding behavior of $J$ vs $\delta \phi$ for the larger value $V_{0}/E_{F}=0.4$, 
the other parameters remaining the same as in Fig.\ref{fig16}.
In this case, all curves are much suppressed with respect to those of Fig.\ref{fig16} and approach the $\sin (\delta \phi)$ dependence 
that characterizes a strong barrier not only on the BEC but also on the BCS side of the crossover.

\begin{center}
\begin{figure}[htc]
\hspace{3cm}\includegraphics[angle=0,width=4in]{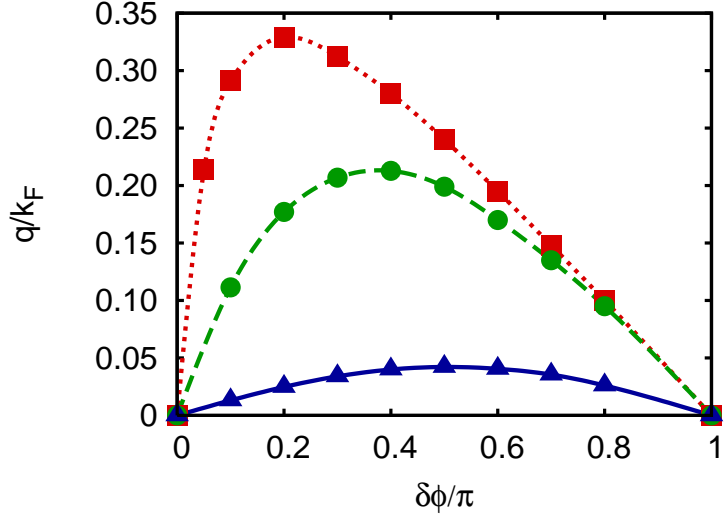}
\caption{Evolution of the Josephson characteristics $q(\delta \phi)/k_{F} = J(\delta \phi) m /(k_{F} n_{0})$ at unitarity for varying barrier height.
              Three different values of $V_{0}/E_{F}=0.025$ (dotted line), $0.1$ (dashed line), and $0.4$ (full line), 
              are considered for a rectangular barrier of width $L k_{F}=4.0$.}
\label{fig18}
\end{figure}
\end{center}

Conversely, in Fig.\ref{fig18} we show how the Josephson characteristics evolve when varying the strength of 
a rectangular barrier while keeping the coupling fixed at unitarity.
The example shown is for a rectangular barrier with height $V_{0}/E_{F}$ spanning a wide range.
One observes again the progressive evolution, from a $\sin (\delta \phi)$ dependence for strong barriers 
to an approximate $\cos (\delta \phi/2)$ dependence for weak barriers \cite{Sols-1994}.
We have found that this behavior is not restricted to unitarity, but is rather generally reproduced for any coupling.

Before discussing the additional features of the Josephson characteristics over the \emph{whole\/} BCS-BEC crossover, 
we restrict ourselves for a moment to  
the peculiar effects that can be identified from the numerical calculations in the two extreme BCS and BEC limits.

\noindent

\vspace{0.3cm}
\begin{center}
\begin{large}
{\bf 4c. The weak-coupling (BCS) limit and the occurrence of Friedel oscillations}
\end{large}
\end{center}
\addcontentsline{toc}{subsection}{4c. The weak-coupling (BCS) limit and the occurrence of Friedel oscillations}
\vspace{0.1cm}

We have already mentioned that for given coupling the value of $\xi_{\mathrm{phase}}$ sets the scale of the upper limit for 
the spatial range that need be explored by our numerical calculations about the position of the barrier, because only 
past this distance the magnitude $|\Delta(x)|$ of the gap parameter recovers its bulk value $\Delta_{0}$.
In Section 3, we have accordingly estimated the value $10$ as a resonable upper bound for the product $k_{F} \xi_{\mathrm{phase}}$, which in turn implies from Fig.\ref{fig7} that the coupling value $(k_{F} a_{F})^{-1} \simeq - 2$ represents, in practice, the limit as far as one can progress toward weak coupling.

\begin{center}
\begin{figure}[htc]
\hspace{3cm}\includegraphics[angle=0,width=4in]{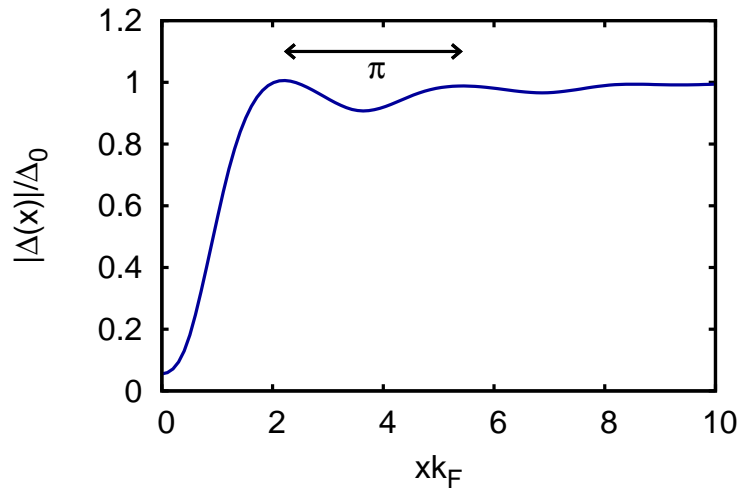}
\caption{Profiles of $|\Delta(x)|$ when $(k_{F} a_{F})^{-1} = -1.5$ and $q/k_{F}=0.01$, for a rectangular barrier of width
              $L k_{F}=3.4\times 10^{-2}$ and height $V_{0}/E_{F}=88.24$.}
\label{fig19}
\end{figure}
\end{center}

In Fig.\ref{fig19} we report a typical profile of $|\Delta(x)|$ in weak coupling, obtained for $(k_{F} a_{F})^{-1} = -1.5$ 
when $k_{F}\xi_{\mathrm{phase}}=3.4$.
In this case, a supercurrent with $q/k_{F}=0.01$ impinges on a rectangular barrier of width $L/\xi_{\mathrm{phase}}=0.01$ and height $V_{0}/E_{F}= 3/(L k_{F})$ that resembles a delta function.

Note from this figure the presence of \emph{Friedel oscillations\/} with characteristic wave vector $2 k_{F}$, which originate 
from the sharpness of the Fermi surface in wave-vector space and distintictly modulate the profile of $|\Delta(x)|$.
Apart from these oscillations, note also the presence of two length scales in the overall profile of $|\Delta(x)|$, as discussed
in sub-section 2c:
A first scale (given by $k_{F}^{-1}$ in this case) within which $|\Delta(x)|$ is strongly suppressed locally by the barrier,
and the second scale $\xi_{\mathrm{phase}}$ over which $|\Delta(x)|$ smoothly approaches its bulk value $\Delta_{0}$.
[The additional lenght scale $\mathcal{L}_{V_{0}}=(2mV_{0})^{-1/2}$ associated with the barrier does not affect  
the profile of $|\Delta(x)|$ in the present case, since $k_{F} \mathcal{L}_{V_{0}} \simeq 0.106$.]

In weak coupling, we have thus far considered barriers resembling a delta function for which $L/\xi_{\mathrm{phase}} \ll 1$.
It is, however, interesting to consider also barriers whose width $L$ is comparable with (or possibly even larger than) $\xi_{\mathrm{phase}}$.

\begin{center}
\begin{figure}[htc]
\includegraphics[angle=0,width=6.5in]{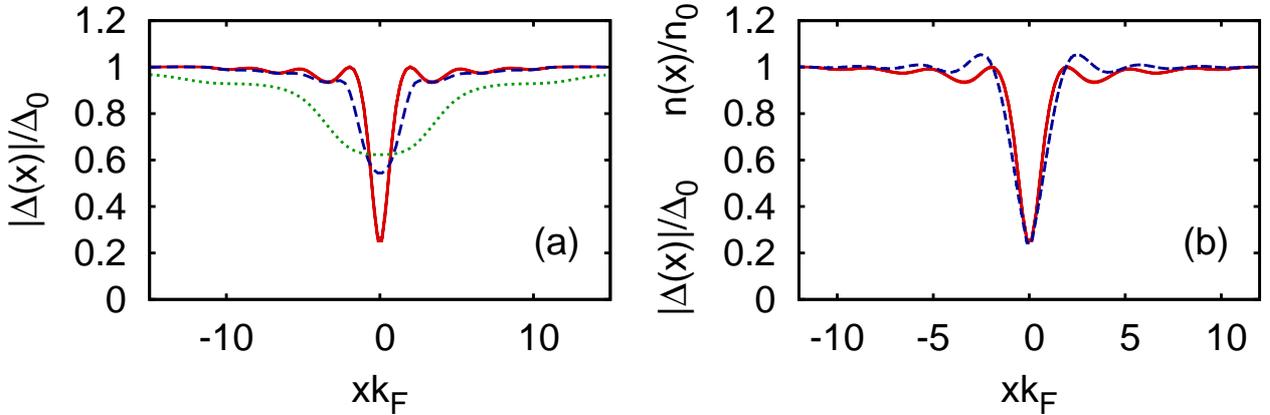}
\caption{(a) Profiles of $|\Delta(x)|$ for $(k_{F} a_{F})^{-1} = -1.5$ and $q/k_{F}=0.01$, obtained with three 
              rectangular barriers having $Lk_{F}=0.01$ (full line), $Lk_{F}=2.6$ (dashed line), and $Lk_{F}=5.3$ 
              (dotted line), while $\tilde{\mathcal{Z}} = 0.525$ in all cases.
              (b) Comparison of the profiles of $|\Delta(x)|$ (full line) and $n(x)$ (dashed line) for the narrowest barrier 
              considered in (a).}
\label{fig20}
\end{figure}
\end{center} 

The profiles of $|\Delta(x)|$ obtained for the coupling $(k_{F} a_{F})^{-1} = -1.5$ and current $q/k_{F}=0.01$, and
corresponding to three different values of $L$ which are respectively smaller than, comparable with, and larger than 
$\xi_{\mathrm{phase}}$, are shown in Fig.\ref{fig20}(a).
This figure makes it evident that there are definite differences between narrow and wide barriers, and that the length scale 
$\xi_{\mathrm{phase}}$ is responsible for these differences.

The profile of $|\Delta(x)|$ for the narrowest barrier of Fig.\ref{fig20}(a) is further compared in Fig.\ref{fig20}(b) 
with the number density $n(x)$, showing a characteristic de-phasing of the Friedel oscillations affecting these two quantities.
This implies that $|\Delta(x)|$ and $n(x)$ are somewhat correlated with each other even in the weak-coupling limit
(while in strong coupling we know from the relation (\ref{Phi-vs-Delta}) that $|\Delta(x)|^{2}$ is strictly proportional to the 
bosonic density).


\vspace{0.3cm}
\begin{center}
\begin{large}
{\bf 4d. The strong-coupling (BEC) limit and the internal structure of the composite bosons}
\end{large}
\end{center}
\addcontentsline{toc}{subsection}{4d. The strong-coupling (BEC) limit and the internal structure of the composite bosons}
\vspace{0.1cm} 

We have already discussed at some length in sub-section 2d the role played by the bosonic GP equation 
when the fermionic BdG equations are solved in the BEC limit of the BCS-BEC crossover,
that is, when fermion pairs of opposite spin bind together so as to form composite bosons whose size is 
much smaller than the average interparticle distance.

Even in this limiting case, however, the presence of a barrier may affect the comparison between the results 
of the corresponding solutions of the GP and BdG equations, when the size of the composite bosons is 
comparable with the width of the barrier. 
Under these circumstances, the composite nature of the bosons plays a role and the fermionic 
nature of their constituents emerges, when a detailed comparison is made with the solution of the GP equation 
which completely ignores the internal structure of the bosons.

\begin{center}
\begin{figure}[htc]
\includegraphics[angle=0,width=6.5in]{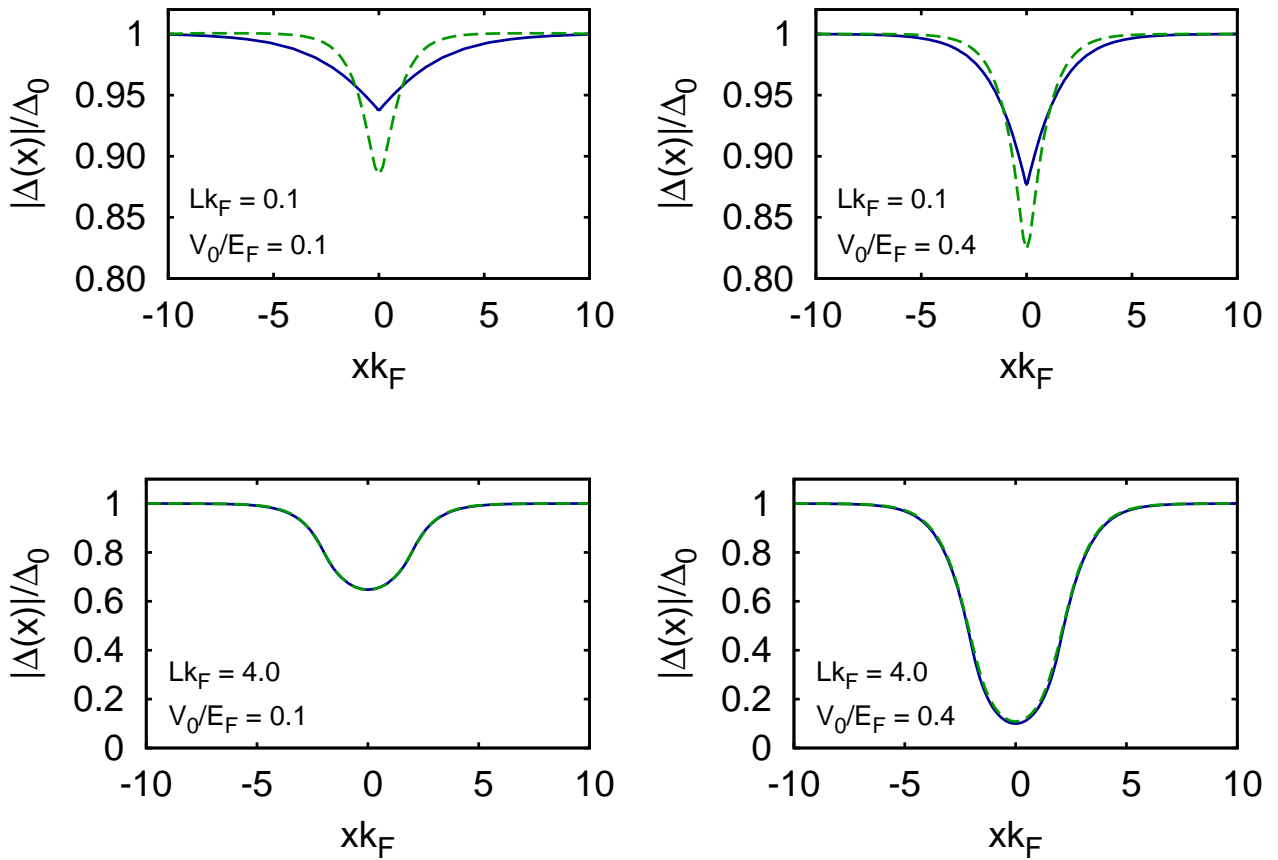}
\caption{The profiles $|\Delta(x)|$ when $(k_{F} a_{F})^{-1}=+1.5$ (dashed lines) are compared with the corresponding
             solutions of the GP equation for the same coupling (full lines), for four rectangular barriers with 
             $L k_{F} = (0.1,4.0)$ and $V_{0}/E_{F} = (0.1,0.4)$. In all cases here considered, we have fixed the value
             $\delta \phi/\pi = 0.1$.}
\label{fig21}
\end{figure}
\end{center}

To spot these kinds of effects with more evidence than in Fig.\ref{fig10}, we report in Fig.\ref{fig21} the self-consistent profiles 
$|\Delta(x)|$ for the lesser extreme coupling value $(k_{F} a_{F})^{-1} = +1.5$ and for a rectangular barrier with
two different widths $L k_{F} = (0.1,4.0)$ and heights $V_{0}/E_{F} = (0.1,0.4)$.
For this coupling value, the size of the composite bosons equals $k_{F} \xi_{\mathrm{pair}}=0.44$, and it is respectively
larger (upper panels) and smaller (lower panels) than the barrier width for the cases here considered.
In each case, the solution of the GP equation is also shown for comparison.

The comparison evidences definite deviations between the two (BdG and GP) calculations in the region close to the barrier when the size of the composite bosons is larger than the barrier width, such that the internal wave function of the fermion pair is considerably affected by the presence of the barrier.
Specifically, taking into account the composite nature of the bosons as in the BdG equations leads to a noticeable depression 
$|\Delta(x)|$ in the vicinity of the barrier with respect to point-like bosons.
The panels of Fig.\ref{fig21} make it further evident that the height of the barrier has little effect on the degree of discrepancy between the BdG and GP calculations.

 
\vspace{0.3cm}
\begin{center}
\begin{large}
{\bf 4e. The role of the Andreev-Saint-James bound states}
\end{large}
\end{center}
\addcontentsline{toc}{subsection}{4e. The role of the Andreev-Saint-James bound states}
\vspace{0.1cm} 

The Andreev-Saint-James reflection was originally conceived as occurring at an interface between a superconductor and a normal material \cite{Andreev-states}.
By this process, an electron incident on the interface from the normal material (with an energy smaller than the superconducting gap) is retro-reflected as a hole of opposite spin and momentum, while a Cooper pair forms on the superconductor side. 
This occurs because an electron impinging from the normal side at energies smaller than superconducting gap cannot propagate inside the superconductor, while Cooper pairs can do so. 
At the same time, the reflected hole ensures current conservation. 

This process is actually a key feature for the solutions of the BdG equations.
As such, it also occurs in the different situation considered in this paper when a barrier separates two superconductors. 
In this case, the multiple Andreev-Saint-James reflections occurring at the barrier sides may interfere constructively to give rise to  Andreev bound states localized at the barrier.
These bound states, which emerge when solving the BdG eigenvalue problem even with a repulsive potential barrier, originate from the lowering of the gap profile at the potential barrier, which acts as an effective binding potential in the off-diagonal terms of the BdG equations. 

The importance of these bound states with subgap energies stems from the fact that, following the pioneering work of Kulik \cite{Kulik-1970}, they are believed to provide the main channel (pictorially, a bridge) for Cooper-pair transfer via Andreev-Saint-James scattering between the two superconductors separated by a barrier
 \cite{SBW-1997,Kummel-1994}.
As a consequence, the process of Andreev-Saint-James scattering is intimately related to the dc Josephson effect when a pair of correlated fermions is transferred from one to another superconductor \cite{GKI-RMP-2004}.
It is also worth mentioning that the spectroscopy of these levels constitutes a quite active field of research both in conventional as well as in cuprate superconductors \cite{Deutscher-RMP-2005}.

The above considerations, about the connection between the Andreev-Saint-James reflections and the dc Josephson effect, have historically been applied to weak-coupling situations where the presence of the Cooper pairs play a major role.
With the present calculation based on the self-consistent solution of the BdG equations, whereby the fermionic attraction can be varied at will from weak to strong coupling, we are in a position to address the connection between the Andreev bound states and the flow of the Jospehson current within a more general context, up to the point when the Cooper pairs give the way to the presence of composite bosons whose underlying fermionic degrees of freedom become progressively more irrelevant for increasing fermionic attraction.
In addition, our need for taking into account the fermionic wave vector transverse to the current flow (in order to be able to form the composite bosons in the strong-coupling limit) brings out the issue of the dispersion of the Andreev bound states as the magnitude of this wave vector is varied.

\begin{center}
\begin{figure}[htc]
\hspace{2.5cm}\includegraphics[angle=-90,width=4.5in]{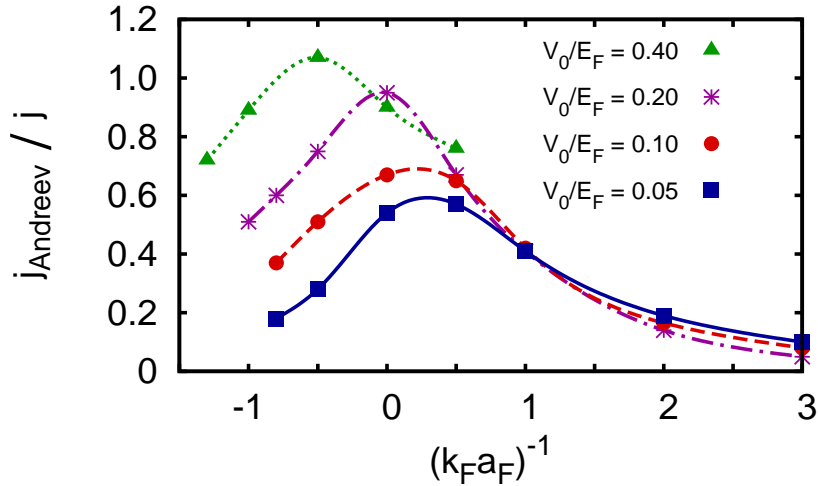}
\caption{Fraction of the superfluid current carried by the Andreev bound states vs $(k_{F} a_{F})^{-1}$ at the position 
              of a barrier with $Lk_{F} = 4.0$ and $V_{0}/E_{F} = (0.05, 0.10, 0.20, 0.40)$ from bottom to top.}
\label{fig22}
\end{figure}
\end{center} 

We have already discussed in sub-section 2c that, in the weak-coupling limit, a non-self-consistent calculation suffices to obtain a reliable description of the Josephson effect for a Dirac-delta barrier whose strength is large enough, and that in this case a single Andreev bound state accounts for the whole Josephson current at the position of the barrier (a result that was well-known in the literature - cf. Ref.\cite{Bagwell-1992}).
The point is here to understand to what an extent the above statement survives to (i) the introduction of self-consistency, (ii) the presence of a finite barrier width, and (iii) the evolution toward strong coupling.

To this end, we plot in Fig.\ref{fig22} the ratio of the superfluid current $j_{\mathrm{Andreev}}$ carried at the position of the barrier through the Andreev bound states with respect to the total current $j$, across the BCS-BEC crossover.
The results are shown for several barriers and include the contribution of the whole Andreev band spanned by the
wave vector $\mathbf{k}_{\parallel}$ transverse to the current flow.
Note how the contribution of the Andreev bound states may sometimes exceed $100\%$ of the whole current; in this case, the continuum levels contribute a current flowing in the opposite direction (this feature was already pointed out in Ref.\cite{Wendin-1996}
for the weak-coupling limit).

\begin{center}
\begin{figure}[htc]
\hspace{3cm}\includegraphics[angle=0,width=4in]{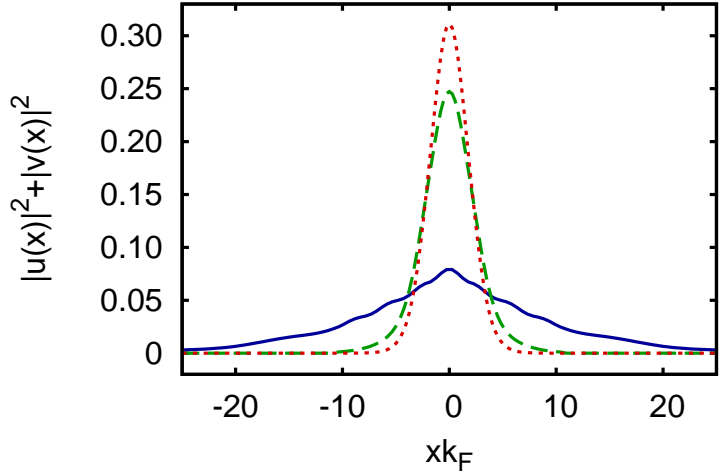}
\caption{Square magnitude of the Andreev wave function with $\mathbf{k}_{\parallel} = 0$ vs $xk_{F}$ across the position 
              of a barrier with $L k_{F} = 4.0$ and $V_{0}/E_{F} = 0.10$, for three coupling values: $(k_{F} a_{F})^{-1}=-1.0$ 
              (full line), $0.0$ (dashed line), $+1.0$ (dotted line). In each case, the superfluid current is taken at its 
              maximum sustainable value.}
\label{fig23}
\end{figure}
\end{center} 

In addition, Fig.\ref{fig23} shows the degree of localization of an Andreev bound state with $\mathbf{k}_{\parallel} = 0$,
by plotting the square magnitude $|u_{\nu}(x)|^{2} + |v_{\nu}(x)|^{2}$ of the corresponding wave function vs $x$ 
for three characteristic coupling values across the BCS-BEC crossover.
One sees from this figure that the spatial extension of the Andreev bound state decreases for increasing coupling 
from the BCS to the BEC limits.

We shall return in sub-section 4g to the special role played by the Andreev bound states on the BCS side of the crossover with
respect to the stability of the superfluid flow.


\vspace{0.3cm}
\begin{center}
\begin{large}
{\bf 4f. Critical Josephson current and its relation with the Landau criterion for superfluidity}
\end{large}
\end{center}
\addcontentsline{toc}{subsection}{4f. Critical Josephson current and its relation with the Landau criterion for superfluidity}
\vspace{0.1cm} 

We have seen in sub-section 4b that the overall shape of the Josephson characteristics varies considerably by changing the width $L$ and height $V_{0}$ of the barrier as well as the value of the coupling throughout the BCS-BEC crossover.
In particular, these variations affect the maximum Josephson current, which we have seen to increase as the 
ratio $V_{0}/E_{F}$ is decreased.
Since the maximum Josephson current is probably the most important feature of the Josephson characteristics for practical purpose, it is relevant to address the question whether there exists an intrinsic  \emph{maximum} allowed value of the Josephson current for any given coupling. 
In other words, we shall address the question whether adopting the limiting procedure of progressively lowering the barrier height leads to an intrinsic upper value of $q_{c}/m$ for the maximum velocity of the superfluid flow at any given coupling.

One may anticipate on physical grounds that a vanishingly small barrier acts as an impurity that probes the stability of the homogeneous superfluid flow.
In this sense, it plays a similar role to the walls of the container in the context of the \emph{Landau criterion for
superfluidity} \cite{AGD}. 
One thus expects that the velocity $q/m$ of the superfluid flow never exceeds the corresponding value of the
critical velocity as obtained by the Landau criterion. 
This critical velocity is determined by the onset of quasiparticle excitations which are of a different nature 
on the two sides of the crossover, namely, pair-breaking excitations on the BCS side and sound-mode
quanta on the BEC side.

Before showing in detail how our numerical self-consistent solutions of the BdG equations indeed capture this important physical feature on both sides of the crossover, we shall briefly consider the mechanism through which the BdG equations are able to recognize the occurrence of an instability when the velocity of the superfluid flow increases beyond a certain limit.

We begin with a preliminary discussion about the role of Galilean invariance in the BdG equations.
We thus consider how the wave functions $u_{\mathbf{k}}(\mathbf{r})$ and $v_{\mathbf{k}}(\mathbf{r})$, that are solutions of 
Eq.(\ref{BdG-equations}) in the absence of a barrier (\emph{homogeneous} case), are transformed when passing from the lab frame $K$ to a frame $K'$ which moves with velocity $\mathbf{V}$ with respect to $K$.
This transformation reads:
\begin{equation}
\left( \begin{array}{c} u_{\mathbf{k}}(\mathbf{r}) \\ v_{\mathbf{k}}(\mathbf{r})^{*} \end{array} \right) \, = \,
e^{i \mathbf{q}\cdot\mathbf{r}} \, 
\left( \begin{array}{c} u_{\mathbf{k'}}(\mathbf{r}) \\ v_{\mathbf{k'}}(\mathbf{r})^{*} \end{array} \right)  
                                                                                                       \label{Galilean-transformation-BdG-equations}
\end{equation}

\noindent
where $\mathbf{q} = m \mathbf{V}$, $\mathbf{k}=\mathbf{k'}+\mathbf{q}$, 
$u_{\mathbf{k'}}(\mathbf{r})=u_{\mathbf{k'}} \, e^{i \mathbf{k'}\cdot\mathbf{r}}$, and
$v_{\mathbf{k'}}(\mathbf{r})=v_{\mathbf{k'}} \, e^{i \mathbf{k'}\cdot\mathbf{r}}$ (with $u_{\mathbf{k'}}$ and 
$v_{\mathbf{k'}}$ real).
Solving for the BdG equations alternatively in the two frames then yields the following relations between the gap parameter, chemical potential, and energy eigenvalues in the two frames:
\begin{equation}
\Delta(\mathbf{r}) \, = \, e^{2 i \mathbf{q}\cdot\mathbf{r}} \, \Delta' \,\,\, , \,\,\,
\mu \, = \, \mu' \, + \, \frac{\mathbf{q}^{2}}{2m} \,\,\, , \,\,\, 
\epsilon_{\mathbf{k}} \, = \, \epsilon'_{\mathbf{k'}} \, + \, \frac{\mathbf{q}}{m} \cdot \mathbf{k'}  \label{relations-K-K'}
\end{equation}

\noindent
where $\epsilon'_{\mathbf{k'}}=\sqrt{(\mathbf{k'}^{2}/(2m) - \mu')^{2} + \Delta'^{2}}$.
These relations have been already utilized while discussing 
Eqs.(\ref{u-v-homogeneous})-(\ref{u2-v2-homogeneous}) near the end of sub-section 2a, although with a slightly different notation. 
Note that in the above relations, no factor appears that could limit the value of the relative velocity $\mathbf{V}$.
For a truly homogeneous superfluid, this feature would naivly reflect itself in the absence of an upper value for the current $j=n\mathbf{V}$ flowing in the frame $K$ at rest with respect to the superfluid, in contrast to the Landau criterion.

To appreciate the way the BdG equations recognize the occurrence of an instability when the velocity of the superfluid flow increases beyond a certain limit, we rewrite the gap equation (\ref{self-consistency}) for the homogeneous case in the form:
\begin{equation}
\frac{1}{g} \, = \, \int \! \frac{d\mathbf{k'}}{(2 \pi)^{3}} \, \frac{1}{\epsilon'_{\mathbf{k'}}}
\left[ 1 \, - \, 2 \, f_{F}\left(\epsilon'_{\mathbf{k'}} \, + \, \frac{\mathbf{q}}{m} \cdot \mathbf{k'} \right) \right]
                                                                                                             \label{gap-equation-homogeneous}
\end{equation}

\noindent
with reference to the newly introduced notation in the frame $K'$ where the superfluid is at rest.
Consider, in particular, \emph{an infinitesimally small} (albeit non-zero) temperature for which the presence of the Fermi function can be neglected whenever its argument is positive, resulting in a gap parameter $\Delta'$ that does not depend
on $\mathbf{q}$.
A different situation arises, however, as soon as $\epsilon'_{\mathbf{k'}} \, + \, \frac{\mathbf{q}}{m} \cdot \mathbf{k'}$ vanishes, a condition which leads to an instability of the system felt by the BdG equations.
To study this situation, we take $\mathbf{q}=(q,0,0)$ and set $\tilde{\mu}' = \mu' - [(k'_{y})^{2}+(k'_{z})^{2})]/(2m)$, so that the above condition becomes:
\begin{equation}
\sqrt{ \left( \frac{k'^{2}}{2m} - \tilde{\mu}' \right)^{2} \, + \, \Delta'^{2}} \, + \, k' \, \frac{q}{m} \, = \, 0     \label{condition}
\end{equation}

\noindent
where $k'=k_{x}'$.
Two real solutions exist for this equation provided $\tilde{\mu}'^{2} + \Delta'^{2} \le (\tilde{\mu}' + q^{2}/m)^{2}$.
The minimum value $q_{c}$ of $q$ for this to occur is given by:
\begin{equation}
\frac{q_{c}^{2}}{m} \, = \sqrt{\mu'^{2} + \Delta'^{2}} \, - \, \mu'  \,\, .               \label{minimum-condition-BCS}
\end{equation}

\noindent
This result is consistent with the Landau criterion for the collapse of the superfluid flow in the case when the relevant quasiparticles consist of pair-breaking excitations.

It is also interesting to monitor the collapse of the superfluid flow directly in terms of the absorption of energy 
by the system, which becomes possible when quasiparticles can be excited.
To this end, instead of considering the superfluid moving with velocity $\mathbf{V}$ with respect to
a fixed barrier, we pass to the frame where the superfluid is at rest and the barrier moves with velocity
$-\mathbf{V}$.
In this way, the barrier acts on the superfluid as an external \emph{time-dependent\/} potential of the
form 
\begin{equation}
V_{\mathrm{ext}}(\mathbf{r},t)=V(\mathbf{r} + \mathbf{V} t) \, F(t;T_{0}) \,\, .        \label{external-potential}
\end{equation}

\noindent
Here, $V(\mathbf{r})$ is the static potential entering the BdG equations (\ref{BdG-equations}) while
the function $F(t;T_{0})$ accounts for the gradual switching on and off of the barrier within the time
interval $(-T_{0},+T_{0})$, where $T_{0}$ may depend on the velocity $\mathbf{V}$.
We could, for instance, take $F(t;T_{0})$ of the Gaussian form $ \exp\{-t^{2}/T_{0}^{2}\}$, such that at time $t=0$ the full barrier
appears centered at $\mathbf{r}=0$ in an otherwise homogeneous superfluid.
Correspondingly, the time-dependent Hamiltonian
\begin{equation}
\hat{H}'(t) \, = \, \int \! d\mathbf{r} \, \hat{\rho}(\mathbf{r}) \, V_{\mathrm{ext}}(\mathbf{r},t)         
                                                                                                              \label{time-dependent-H}
\end{equation}

\noindent
acts on the system through the total-density operator $ \hat{\rho}(\mathbf{r})$.

We are interested in the rate at which the external agent (represented by the moving barrier)
does work on the system.
From standard expressions of linear-response theory \cite{CL}, the total work done by the moving barrier
is given by:
\begin{equation}
\Delta W \, = \, - \, \int \! \frac{d\mathbf{p}}{(2 \pi)^{3}} \, \int_{-\infty}^{+\infty} \! \, \frac{d\omega}{2 \pi} 
\,\, V_{\mathrm{ext}}(\mathbf{p},\omega)^{*} \, \omega \, \chi^{''}_{\rho\rho}(\mathbf{p},\omega) \,
V_{\mathrm{ext}}(\mathbf{p},\omega)       \,\, .                                          \label{total-work-done}
\end{equation}

\noindent
Here, $\chi^{''}_{\rho\rho}(\mathbf{p},\omega)$ is the imaginary part of the Fourier transform of
the density-density correlation function and $V_{\mathrm{ext}}(\mathbf{p},\omega)$ the Fourier transform of the external potential (\ref{external-potential}):
\begin{equation}
V_{\mathrm{ext}}(\mathbf{p},\omega) \, = \, V(\mathbf{p}) \, 
F(\omega + \mathbf{p} \cdot \mathbf{V};T_{0})   \,\, .                  \label{Fourier-transform-external-potential}
\end{equation}

\noindent
With $F(t;T_{0})$ of a Gaussian form, $F(\omega;T_{0}) = |T_{0}| \sqrt{\pi} 
\exp\{-\omega^{2}T_{0}^{2}/4\}$.
In particular, in the limit $T_{0} \rightarrow + \infty$ one has
$F(\omega + \mathbf{p} \cdot \mathbf{V};T_{0}) \rightarrow 
2 \pi \delta(\omega + \mathbf{p} \cdot \mathbf{V})$ in Eq.(\ref{Fourier-transform-external-potential}) for
the range of wave vectors $\mathbf{p}$ represented in $V(\mathbf{p})$.
This implies that the total work (\ref{total-work-done}) vanishes unless
$\chi^{''}_{\rho\rho}(\mathbf{p},\omega=-\mathbf{p} \cdot \mathbf{V})$ is different from zero for (at least)
one of those wave vectors. 

The simplest form for $\chi^{''}_{\rho\rho}(\mathbf{p},\omega)$ that can be studied in the broken-symmetry (superfluid) phase corresponds to the BCS approximation, which reads in the zero-temperature limit:
\begin{eqnarray}
\chi^{''}_{\rho\rho}(\mathbf{p},\omega)^{BCS} \, \rightarrow & \pi & \int \! \frac{d\mathbf{k}}{(2 \pi)^{3}} \,
\left( u(\mathbf{k+p}) v(\mathbf{k}) \, + \, v(\mathbf{k+p}) u(\mathbf{k}) \right)^{2}     
                                                                                                             \label{response-function-BCS-T-0} \\
& \times & \left[ \delta(E(\mathbf{k+p}) + E(\mathbf{k}) + \omega) \, - \,
        \delta(E(\mathbf{k+p}) + E(\mathbf{k}) - \omega) \right]               \nonumber
\end{eqnarray}

\noindent
where $u(\mathbf{k})$ and $v(\mathbf{k})$ are the coherence factors and $E(\mathbf{k})$ the eigenvalues of BCS theory in the frame where 
the superfluid is at rest \cite{FW-3} (which were written as $E_{\mathbf{k}}$ in Eqs.(\ref{gap-homogeneous}) and (\ref{density-homogeneous})).
Within this approximation, therefore, the superfluid at zero temperature is perturbed by the action of the moving barrier 
\emph{provided\/} $E(\mathbf{k+p}) + E(\mathbf{k}) = \pm \omega = 
\mp \mathbf{p} \cdot \mathbf{V}$ for the range of wave vectors $\mathbf{p}$ represented in 
$V(\mathbf{p})$.
This implies that
\begin{equation}
E(\mathbf{p-k}) \pm (\mathbf{p-k}) \cdot \mathbf{V} + E(\mathbf{k}) \pm \mathbf{k} \cdot \mathbf{V} 
\, = \, 0 \,\, ,                                                                              \label{Landau-criterion-BCS-bubble}
\end{equation}

\noindent
which is equivalent to the Landau criterion for BCS quasiparticle excitations, and requires that $\mathbf{p} = \pm 2\mathbf{k}$
for the wave vector $\mathbf{k}$ at which \emph{a pair} of these excitations begins to appear.

Additional types of excitations come into play when considering more refined forms of the density-density correlation function.
In particular, the BCS-RPA approximation adds the following RPA contribution \cite{PS-rc-2004} to the BCS approximation considered above for the density-density correlation function:
\begin{equation}
\chi_{\rho\rho}(p)^{RPA} \, = \, 
\left( C(p), C(-p) \right) \, \frac{\left(  \begin{array}{cc}
                                            - A(-p) & B(p) \\
                                              B(p)  & -A(p)
                                            \end{array}        \right)}{A(p) A(-p) \, - \, B(p)^{2}}
\left(  \begin{array}{c} C(p) \\ C(-p) \end{array}  \right)       \label{response-function-RPA-T-0}
\end{equation}

\noindent
where $p$ now stands for the four-vector $(\mathbf{p},\omega)$.
In the zero-temperature limit, the quantities $A(p)$, $B(p)$, and $C(p)$ in Eq.(\ref{response-function-RPA-T-0}) are given by the expressions \cite{MPS}:
\begin{eqnarray}
A(p) \, = \, - \frac{m}{4 \pi a_{F}} & + &  \int \! \frac{d\mathbf{k}}{(2 \pi)^{3}} \, \left[
\frac{u(\mathbf{k+p})^{2} u(\mathbf{k})^{2}}{\omega + i \eta - E(\mathbf{k+p}) - E(\mathbf{k})} \right.     \nonumber \\
& - &  \left. \frac{v(\mathbf{k+p})^{2} v(\mathbf{k})^{2}}{\omega + i \eta + E(\mathbf{k+p}) + E(\mathbf{k})} \, + \,
\frac{m}{\mathbf{k}^{2}} \right]                                                                                                                   \label{A-q}                                                                                                                                 
\end{eqnarray}

\noindent
where $\eta$ is a positive infinitesimal and the regularization (\ref{regularization}) has been used;
\begin{eqnarray}
B(p) & = &  - \, \int \! \frac{d\mathbf{k}}{(2 \pi)^{3}} \, 
\left[ \frac{1}{\omega + i \eta - E(\mathbf{k+p}) - E(\mathbf{k})}  \right.        \label{B-q} \\
& - & \left. \frac{1}{\omega + i \eta + E(\mathbf{k+p}) + E(\mathbf{k})} \right] 
\, u(\mathbf{k+p}) v(\mathbf{k+p}) u(\mathbf{k}) v(\mathbf{k}) \,\, ;              \nonumber
\end{eqnarray}

\noindent
and
\begin{eqnarray}
C(p) & = & \int \! \frac{d\mathbf{k}}{(2 \pi)^{3}} \, 
\left[ \frac{u(\mathbf{k+p})^{2}}{\omega + i \eta - E(\mathbf{k+p}) - E(\mathbf{k})}  \right.     \label{C-q} \\
& + & \left. \frac{v(\mathbf{k+p})^{2}}{\omega + i \eta + E(\mathbf{k+p}) + E(\mathbf{k})} \right] 
\, u(\mathbf{k}) v(\mathbf{k}) \,\, .                                                                                           \nonumber
\end{eqnarray}

Note that, when taken individually, the imaginary parts of the expressions (\ref{A-q}), (\ref{B-q}), and (\ref{C-q})
are nonvanishing whenever $E(\mathbf{k+p}) + E(\mathbf{k}) = \pm \omega$, thus leading again to the condition
(\ref{Landau-criterion-BCS-bubble}).
When considering the imaginary part of the expression (\ref{response-function-RPA-T-0}), on the other hand, a
novel contribution originates from the zeros of the denominator therein, which results in sound-mode quanta 
being the relevant quasiparticle excitations with dispersion $\omega = s |\mathbf{p}|$ where $s$ is the sound 
velocity.
In this case, the Landau criterion for the collapse of the superfluid yieds the well-known condition \cite{AGD}
\begin{equation}
\frac{q_{c}}{m} \, =  \, s                              \label{minimum-condition-BEC}
\end{equation}

\noindent
in the place of Eq.(\ref{minimum-condition-BCS}).
The sound velocity $s$ resulting from the zeros of the denominator in Eq.(\ref{response-function-RPA-T-0}) corresponds to the Bogoliubov-Anderson mode obtained within the BCS-RPA approximation, which can be calculated throughout the BCS-BEC crossover \cite{MPS}.

\begin{center}
\begin{figure}[htc]
\hspace{3cm}\includegraphics[angle=0,width=4in]{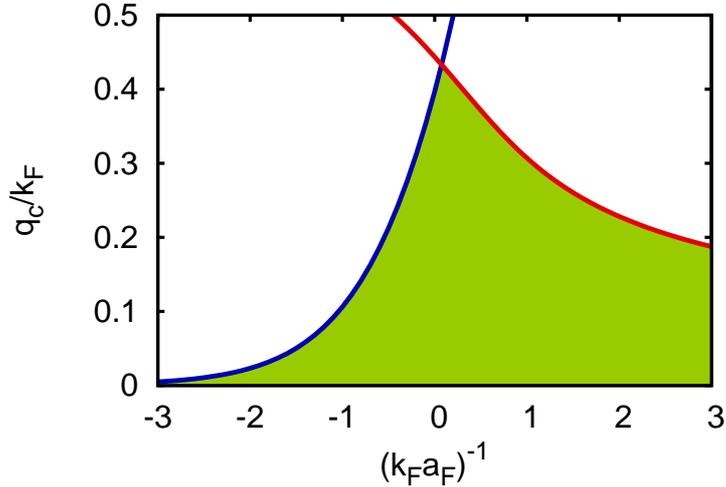}
\caption{The critical wave vector $q_{c}^{\mathrm{Landau}}$ at which the superfluid flow becomes unstable is shown 
              as a function of $(k_{F} a_{F})^{-1}$ throughout the BCS-BEC crossover.
              The ``left'' branch (blue curve) is associated with the appearance of pair-breaking excitations according to
              Eq.(\ref{minimum-condition-BCS}), while the ``right'' branch (red curve) is associated with the appearance of 
              sound-mode excitations according to Eq.(\ref{minimum-condition-BEC}).
              The shaded (green) area evidences the region of allowed superfluid flow, where the superfluid wave vector $q$ 
              lies below both curves.}
\label{fig24}
\end{figure}
\end{center}

The two curves showing the critical wave vector $q_{c}^{\mathrm{Landau}}$ vs $(k_{F} a_{F})^{-1}$, as obtained independently from the conditions (\ref{minimum-condition-BCS}) and (\ref{minimum-condition-BEC}), are plotted in Fig.\ref{fig24} from the BCS to the BEC limits.
In this figure, the ``left'' branch corresponds to the expression (\ref{minimum-condition-BCS}) appropriate to the
pair-breaking excitations characteristic of BCS theory, while the ``right'' branch corresponds to the expression 
(\ref{minimum-condition-BEC}) appropriate to the Bogoliubov-Anderson mode for sound-mode excitations.
In particular, in the BEC limit the value of $q_{c}$ of the ``right'' branch can be related to the chemical potential $\mu_{B}$  for composite bosons introduced in Eq.(\ref{Gross-Pitaevskii-equation}) via the relation $q_{c}^{2}/m = \mu_{B}/2$, in agreement with the Bogoliubov theory for point-like bosons.

Note in Fig.\ref{fig24} that the two Landau branches cross each other near the unitary limit when $(k_{F} a_{F})^{-1}=0$, 
where the maximum value of $q_{c}/m$ is then achieved.
That superfluidity is most robust against suppressing mechanisms in the crossover region was pointed out 
in Ref.\cite{APS-2003} (see also Refs.\cite{SRH-2006} and \cite{CKS-2006}).

The two branches drawn in Fig.\ref{fig24} provide an upper limit for the the velocity $\mathbf{V}$ of the superfluid flow, 
because its magnitude $|\mathbf{V}|$ can never exceed the \emph{smallest} value of the ratio $q_{c}/m$ corresponding 
to the two curves. 
The motion of the superfluid can thus occur, in principle, only within the shaded area of this figure.

The upper value for $q_{c}/m$ is \emph{intrinsic} in nature, in the sense that it results \emph{through a limiting procedure of a vanishing barrier}.
This is because an infinitesimal barrier acts as a ``seed'' for the occurrence of the superfluid collapse via nonlinear effects, while for a truly homogeneous system no upper limit for $|\mathbf{V}|$ would exist by Galilean invariance, as 
we have already emphasized.

The question that arises specifically when solving the BdG equations (\ref{BdG-equations}) up to self-consistency, is whether the curves for $q_{c}$ vs $(k_{F} a_{F})^{-1}$, which are obtained numerically from the maximum value $q_{c}$ of the Josephson characteristics in the presence of a finite barrier, tend \emph{from below} to the intrinsic boundary drawn in Fig.\ref{fig24} 
when the barrier height is progressively decreased for \emph{any} value of $(k_{F} a_{F})^{-1}$ throughout the BCS-BEC crossover.

That this could be a rather subtle point results by considering that, from the above procedure, one would 
naively expect to recover at most the ``left'' branch of Fig.\ref{fig24} corresponding to pair-breaking excitations, by arguing 
that the BdG equations correspond to the BCS mean field which explicitly contains only this type of excitations.
As we have discussed in sub-section 2d, however, the BdG equations are able to recover the GP equation in the 
BEC limit.
As a consequence, on the BEC side also their excitations should correctly correspond to the sound-mode excitations 
of the GP equation, since these excitations result within linear-response theory in terms of the BCS-RPA approximation
(\ref{response-function-RPA-T-0}) (cf. Ref.\cite{PS-rc-2004}), as  confirmed by the present calculation.
On physical grouds, this quite remarkable result has to be expected from the fact that, in the presence of a non-trivial geometry, the imprint of the excitation spectrum is found already in the ground-state wave function \cite{Gross}, 
which in the present context corresponds to the solutions of the BdG equations at zero temperature for given supercurrent.
In addition, reaching the upper value $q_{c}$ for a given barrier and coupling will be related in Section 5 to the onset of the
energy instability of the self-consistent solutions of the BdG equations against the inclusion of fluctuations.

\begin{center}
\begin{figure}[htc]
\includegraphics[angle=0,width=6.5in]{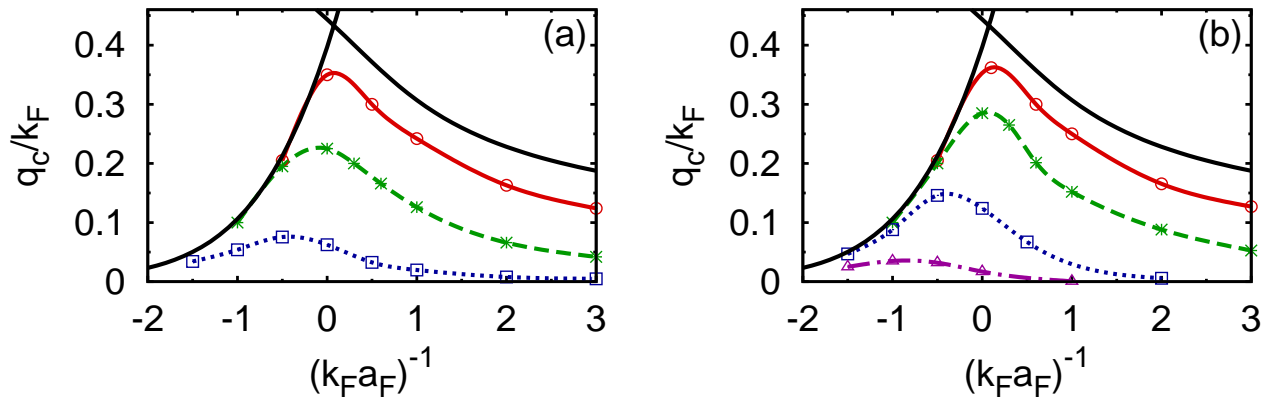}
\caption{The maximum velocity $q_{c}/m$ obtained from the maximum of the Josephson characteristics is 
              shown as a function of $(k_{F} a_{F})^{-1}$ throughout the BCS-BEC crossover for several rectangular 
              barriers, with:
             (a) $L k_{F} = 2.65$ and $V_{0}/E_{F} = (0.02,0.10,0.40)$ from top to bottom;
             (b) $L k_{F} = 5.30$ and $V_{0}/E_{F} = (0.01,0.05,0.20,0.50)$ from top to bottom.
             [The width $L k_{F} = 2.65$ has been conventionally chosen to correspond to the value of 
             $2 k_{F} \xi_{\mathrm{phase}}$ when $(k_{F} a_{F})^{-1} = +3.0$.]
             The two limiting branches of Fig.\ref{fig24} for vanishing barrier are also reported for comparison.}
\label{fig25}
\end{figure}
\end{center} 

Figure \ref{fig25} shows the maximum velocity $q_{c}/m$ vs $(k_{F} a_{F})^{-1}$ throughout the BCS-BEC crossover, 
as obtained from the maximum of the Josephson characteristics which result by solving the BdG equations 
(\ref{BdG-equations}) up to self-consistency for a given barrier.
Values of $q_{c}/m$ corresponding to several barriers are reported in the figure.
At a given coupling, these values are seen to approach \emph{from below\/} the corresponding maximum values of 
Fig.\ref{fig24} related to the Landau criterion as the barrier height is progressively decreased. 

Two additional features are evident from Fig.\ref{fig25}.
First, the maximum of each curve corresponding to a given barrier height shifts from about unitarity toward the BCS side
upon increasing the barrier height, thus implying that bosonic features get amplified by stronger barriers (this point will be
discussed further in the next sub-section).
Second, the rate of approach to the limiting Landau curves differs on the two sides of the crossover, being apparently
faster on the BCS side with respect to the BEC side.

\begin{center}
\begin{figure}[htc]
\hspace{3cm}\includegraphics[angle=0,width=4in]{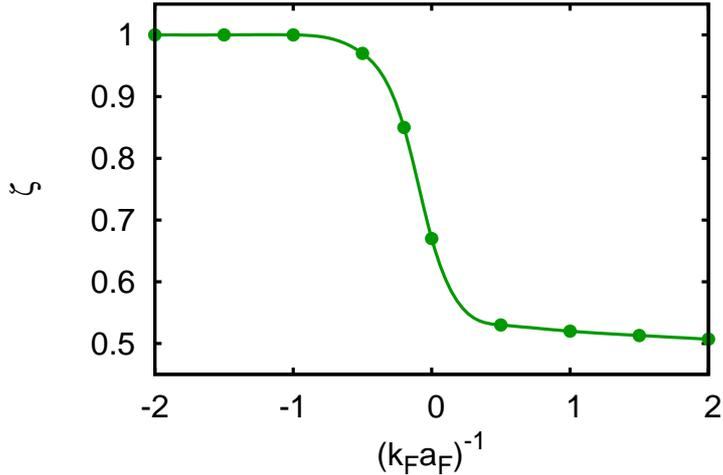}
\caption{Exponent $\zeta$ of the relation (\ref{exponent}) vs $(k_{F} a_{F})^{-1}$.}
\label{fig26}
\end{figure}
\end{center} 

To quantify these features, we report in Fig.\ref{fig26} the exponent $\zeta$ of the relation 
\begin{equation}
\frac{\left( q_{c}^{\mathrm{Landau}} \, - \, q_{c} \right)}{q_{c}^{\mathrm{Landau}}} \, = \, 
\Gamma\, \left( \frac{V_{0}}{E_{F}} \right)^{\, \zeta}                                                                \label{exponent}
\end{equation}

\noindent
which is determined over the limited range $0.005 \lapprox V_{0}/E_{F} \lapprox 0.05$, where $\Gamma$ is a prefactor (this range of
\emph{small} values of $V_{0}/E_{F}$ can also be explored experimentally, as done in Ref.\cite{Ketterle-2007}).
The exponent $\zeta$ turns out to change from the limiting value $1$ on the BCS side when $(k_{F} a_{F})^{-1} \lapprox - 1$, 
to the limiting value $1/2$ on the BEC side when $1 \lapprox (k_{F} a_{F})^{-1}$, through a rather quick variation across the 
unitary limit $(k_{F} a_{F})^{-1} = 0$.
These results imply that the present calculation, by extracting the values of $q_{c}$ from the self-consistent solutions of the BdG equations, is not only capable of recovering from below the limiting Landau curves reported in Fig.\ref{fig25}, but is also able to calculate the exponent that governs the approach to these limiting curves when the height of the barrier becomes progressively
smaller.

It is relevant to mention in this context that, after publication of Ref.\cite{Spuntarelli-07} where the theoretical predictions about the critical velocities throughout the BCS-BEC crossover (like those shown in Fig.\ref{fig25}) were reported for the first time, critical velocities of an ultracold superfluid Fermi gas throughout the BEC-BCS crossover have indeed been observed \cite {Ketterle-2007}, albeit by the use of a different geometry. 
These critical velocities were experimentally determined from an abrupt onset of dissipation when the velocity of a moving one-dimensional optical lattice was varied.
In spite of the different geometry, a pronounced peak of the critical velocity was indeed found at unitarity (cf. Fig.2 of Ref.\cite {Ketterle-2007}), thus confirming the physical prediction made in Ref.\cite{Spuntarelli-07} about superfluidity being most robust at unitarity.
In addition, the experimental shape of the critical velocity curve vs. $(k_{F} a_{F})^{-1}$ shows the behavior predicted theoretically,
of being sharper on the BCS side than on the BEC side of the crossover.

In Fig.4 of Ref.\cite{Ketterle-2007} the critical velocities at unitarity were further reported for varying lattice depth up to the point that $V_{0} \ll E_{F}$, 
a limiting situation for which differences in the specific geometry should become irrelevant.
From that figure, in the limit $V_{0} \ll E_{F}$ one can extract for $q_{c}/k_{F}$ the value $0.25$ and for the exponent $\zeta$ of Eq.(\ref{exponent}) the value $0.66$ (plus the value $1.23$ for the prefactor $\Gamma$).
The experimental value $0.25$ for $q_{c}/k_{F}$ is somewhat smaller than the theoretical prediction $0.397$ obtained
at unitarity for $q_{c}^{\mathrm{Landau}}/k_{F}$ (cf. Fig.\ref{fig24}).
The reason for this difference is due to the fact that this theoretical value has been obtained from the condition
(\ref{minimum-condition-BCS}) using the BCS mean-field values for the gap parameter and chemical potential at zero temperature (as reported in Fig.\ref{fig1}).
A more refined theoretical treatment would require the inclusions of fluctuations beyond mean field to calculate these
quantities. 
For instance, the inclusion of pairing fluctuations \cite{PPS-2004} yields at unitarity the values $0.455$ for $\mu/E_{F}$ 
and $0.530$ for $\Delta/E_{F}$ at zero temperature (in the absence of a superfluid current),
to be contrasted with the values $0.590$ and $0.686$ at the mean-field level, respectively.
Using the above improved values of $\mu/E_{F}$ and $\Delta/E_{F}$ in the expression (\ref{minimum-condition-BCS}),
one obtains $q_{c}/k_{F} = 0.349$ which is improved with respect to the mean-field value but still larger than the
experimental one.
On the other hand, the theoretical value $0.67$ of the exponent $\zeta$ obtained from Fig.\ref{fig26} at unitarity turns
out to be quite close to the experimental value $0.66$ (while the corresponding theoretical value for the prefactor 
$\Gamma$ is 2.5).
Note that the comparison between theory and experiment proves better for the exponent $\zeta$ than for the
prefactor $\Gamma$ and the value of $q_{c}^{\mathrm{Landau}}$ in Eq.(\ref{exponent}).
This reminds us of the known fact about scaling relations, for which only the critical exponents and not the prefactors are 
considered to be universal.

It should be mentioned in this context that the true ``critical'' exponent $\zeta_{c}$ of a relation of the kind (\ref{exponent}) where $V_{0} \rightarrow 0^{+}$ \emph{strictly} (meaning that the lower bound $0.005 \lapprox V_{0}/E_{F}$ considered for obtaining the exponent $\zeta$ reported in
Fig.\ref{fig26} is now removed) can be obtained analytically at unitarity as well as in the BCS and BEC limits by means of a local-density argument similar to that considered at the end of sub-section 4g.
Specifically, one obtains $\zeta_{c}=1$ in the BCS limit as well as at unitarity, and $\zeta_{c}=1/2$ in the BEC limit.
Comparison with Fig.\ref{fig26} then shows that $\zeta = \zeta_{c}$ in the BCS and BEC limit, while at unitarity the two
exponents differ from each other.
This difference is due to the fact that at unitarity the critical exponent $\zeta_{c}$ can actually be revealed only for extremely small values $V_{0}/E_{F} \lapprox 0.001$.
Approaching the BCS limit, on the other hand, the range of $V_{0}/E_{F}$ where the critical exponent $\zeta_{c}$ holds becomes
progressively wider, extending beyond the lower bound $V_{0}/E_{F} = 0.005$ considered for Fig.\ref{fig26}.
Along these lines we expect the value $\zeta_{c} = 1$ to be the imprint of the fermionic character of the system, which
is restricted to a progressively smaller range of $V_{0}/E_{F}$ upon approaching the BEC limit, when the bosonic
value $\zeta_{c} =1/2$ takes effectively over for all practical purposes.


\begin{center}
\begin{large}
{\bf 4g. Landau-type criterion for Andreev bound states}
\end{large}
\end{center}
\addcontentsline{toc}{subsection}{4g. Landau-type criterion for Andreev bound states}

\begin{center}
\begin{figure}[htc]
\hspace{3cm}\includegraphics[angle=0,width=4in]{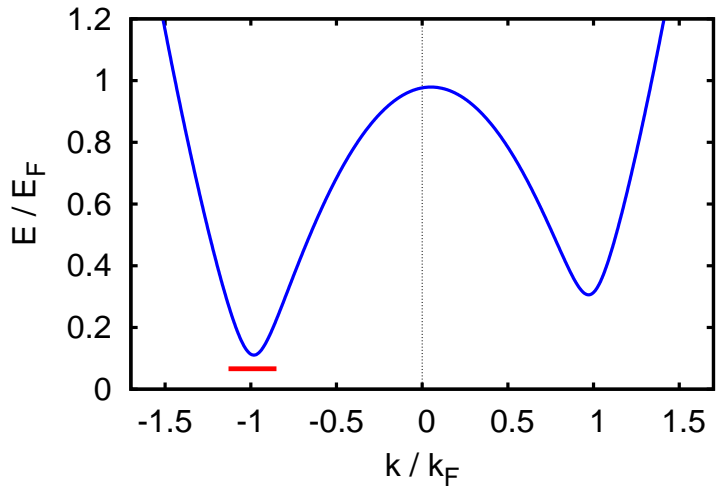}
\caption{An Andreev bound state with $\mathbf{k}_{\parallel}=0$
              (red line) lies below the threshold of the continuum (blue line) corresponding to Fig.\ref{fig4}(a).
              Here, a supercurrent with $q/k_{F}=0.05$ flows through a rectangular barrier of width $L k_{F} = 4.0$ 
              and height $V_{0}/E_{F}=0.4$ when the coupling is $(k_{F} a_{F})^{-1}=-1$.}
\label{fig27}
\end{figure}
\end{center}

In sub-section 4e we have emphasized the role played by the Andreev bound states on the BCS side 
of the crossover up to unitarity, in enabling the supercurrent to flow between the two sides of a barrier.
These bound states, localized about the position of the barrier, act as a ``bridge'' 
through which most of the local transport occurs.
The energy levels corresponding to these bound states lie \emph{below\/} the threshold of the continuum levels.
A typical Andreev level is shown in Fig.\ref{fig27}, with reference to the continuum branch of Fig.\ref{fig4}(a) identified 
by the left-hand side of Eq.(\ref{condition}).

In addition, in sub-section 4f we have discussed how the self-consistent solutions of the BdG equations (which we 
have obtained across the BCS-BEC crossover) are sensitive to the instability associated with the Landau criterion 
for superfluidity when the barrier is allowed to vanish ($V_{0} \rightarrow 0^{+}$).
In particular, on the BCS side of the crossover up to unitarity, we have recovered the onset of the instability due to the
spontaneous generation of pair-breaking excitations, whose \emph{energy becomes negative} as soon as the wave vector $q$ associated with the superfluid current 
exceeds the critical value $q_{c}$ given by Eq.(\ref{minimum-condition-BCS}).
This limiting situation corresponds to the vanishing of the threshold of the continuum levels (in Fig.\ref{fig27} this threshold
is represented by the bottom of the left dip of the curve).

\begin{center}
\begin{figure}[htc]
\hspace{2.5cm}\includegraphics[angle=0,width=4.5in]{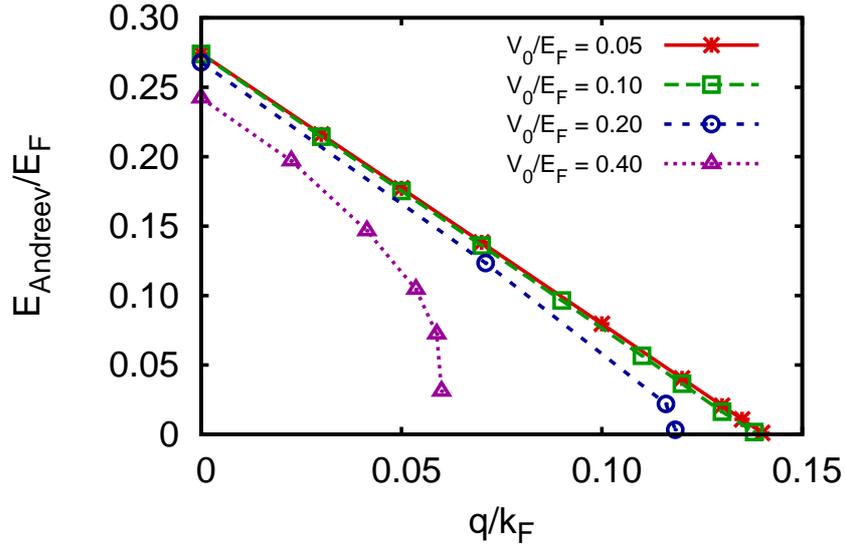}
\caption{Energy of the Andreev bound state (in units of $E_{F}$) with $\mathbf{k}_{\parallel}=0$ vs $q/k_{F}$
              when $(k_{F} a_{F})^{-1}=-0.8$, for rectangular barriers with given width $L k_{F} = 4.0$ and different heights.}
\label{fig28}
\end{figure}
\end{center}

In the presence of a finite barrier (with $V_{0} \ne 0$) when an Andreev bound state lies below the threshold 
of the continuum, upon increasing the value of $q$ it is clearly the energy of this bound state to reach the zero value  
\emph{before} the bottom of the continuum would do so.
In this situation, it is natural to wonder whether an analog of the Landau criterion could be
formulated even in the presence of a finite barrier. 

With reference to the BCS side of the crossover where the Andreev bound states are expected to play a crucial role, we thus attempt to formulate the
``conjecture'' that, in the presence of a finite barrier, the superfluid continues to flow until the energy of the lowest 
Andreev bound state (with $\mathbf{k}_{\parallel}=0$) becomes negative.

\begin{center}
\begin{figure}[tc]
\hspace{3cm}\includegraphics[angle=0,width=4in]{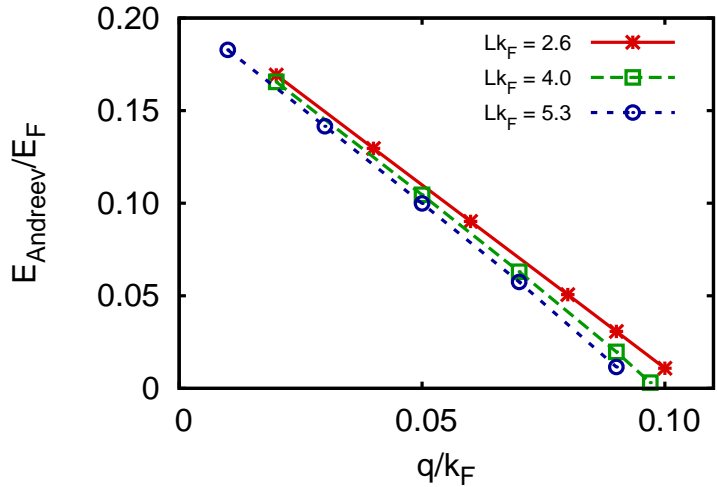}
\caption{Energy of the Andreev bound state (in units of $E_{F}$) with $\mathbf{k}_{\parallel}=0$ vs $q/k_{F}$
             when $(k_{F} a_{F})^{-1}=-1.0$, for rectangular barriers with the same height $V_{0}/E_{F}=0.2$ and 
             different widths.}
\label{fig29}
\end{figure}
\end{center}

To verify this conjecture, we follow the evolution of the energy $E_{\mathrm{Andreev}}$ of this bound state for a given barrier while increasing 
the value of the wave vector $q$ of the supercurrent, and determine accordingly whether there exists a critical value $q_{c}^{\mathrm{Andreev}}$ 
at which $E_{\mathrm{Andreev}}$ vanishes (within some numerical error).
To this end, in Fig.\ref{fig28} we report $E_{\mathrm{Andreev}}$ vs $q$ for several barriers (with the same width but different heights) for the coupling value $(k_{F} a_{F})^{-1}=-0.8$ on the weak-coupling side of the crossover, while in Fig.\ref{fig29} we do the same for barriers with the same height but different widths 
(and for the slightly different coupling value $(k_{F} a_{F})^{-1}=-1.0$).
The corresponding values of $q_{c}^{\mathrm{Andreev}}$ can be readily extracted from these plots.

\begin{table}
\begin{center}
\begin{large}
\begin{tabular}{||c||c|c|c||} \hline \hline
    $(k_{F} a_{F})^{-1}$ & $V_{0}/E_{F}$ & $Lk_{F}$ 
    & $q_{c}^{\mathrm{Josephson}}/q_{c}^{\mathrm{Andreev}}$ \\ \hline \hline
    $-1.0$ & $0.20$ & $2.6$ & \,\,\,$0.995$ \\ \hline
    $-1.0$ & $0.20$ & $4.0$ & $0.98$ \\ \hline
    $-1.0$ & $0.20$ & $5.3$ & $0.96$ \\ \hline
    $-0.8$ & $0.05$ & $4.0$ & \,\,\,$0.995$ \\ \hline
    $-0.8$ & $0.10$ & $4.0$ & $0.98$ \\ \hline
    $-0.8$ & $0.20$ & $4.0$ & $0.99$ \\ \hline
    $-0.8$ & $0.40$ & $4.0$ & $0.90$ \\ \hline 
\end{tabular}      
\caption{Ratio of the critical values of $q$, obtained from the maximum of the Josephson characteristics and
              the vanishing of the energy of the Andreev bound state with $\mathbf{k}_{\parallel}=0$, for given barrier and
              coupling on the BCS side.}
\end{large}              
\end{center} 
\end{table}

\noindent

The values of $q_{c}^{\mathrm{Andreev}}$ extracted from these plots are then compared in Table 2 with the critical values $q_{c}^{\mathrm{Josephson}}$ corresponding to the maximum of 
the associated Josephson characteristics for the same barrier.
From this comparison we conclude that these two values of $q_{c}$ are indeed in quite good agreement with each other
within some numerical error.
This result then verifies explicitly our conjecture about the role played by the Andreev bound states for the collapse 
of the superfluid flow, at least for the couplings here considered on the BCS side of the crossover.

\begin{center}
\begin{figure}[htc]
\hspace{3cm}\includegraphics[angle=0,width=4in]{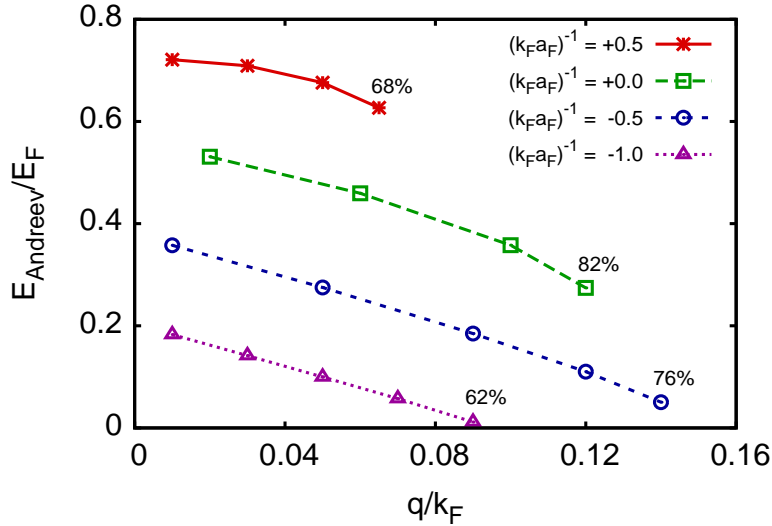}
\caption{Energy of the Andreev bound state (in units of $E_{F}$) with $\mathbf{k}_{\parallel}=0$ vs $q/k_{F}$, 
              for a given rectangular barrier with $L k_{F}= 5.30$ and $V_{0}/E_{F}= 0.20$, and several coupling values 
              across unitarity. The percentage of the current carried at the upper value of $q$ by the whole branch 
              of Andreev states at the barrier center is also reported.}
\label{fig30}
\end{figure}
\end{center}

This conjecture, however, starts failing as soon as the fermionic coupling becomes stronger and the unitary limit is approached.
This can be seen from Fig.\ref{fig30} where the energy $E_{\mathrm{Andreev}}$ is reported vs $q/k_{F}$ for a given barrier and
several coupling values.
It appears evident from the figure that $E_{\mathrm{Andreev}}$ is far from vanishing when the maximum
sustainable current is approached, as soon as the coupling $(k_{F} a_{F})^{-1}$ enters the crossover regime.
This occurs in spite of the fact that the fraction of the current carried by the Andreev states becomes maximum at unitarity.

\begin{center}
\begin{figure}[htc]
\hspace{3cm}\includegraphics[angle=0,width=4in]{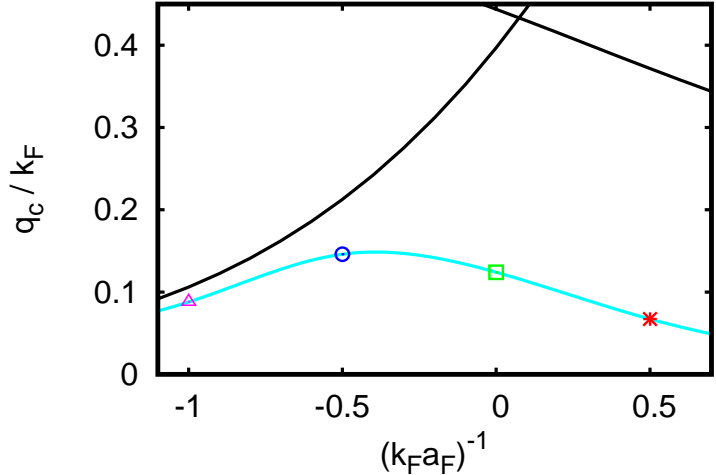}
\caption{The curve that interpolates the critical values $q_{c}$ vs $(k_{F} a_{F})^{-1}$ corresponding to the four cases 
               of Fig.\ref{fig30} is compared with the two Landau limiting branches of Fig.\ref{fig24}.}
\label{fig31}
\end{figure}
\end{center}

Nevertheless, the point we like to emphasize here is that, for a given barrier, the coupling at which the energy of the Andreev bound state no longer approaches zero for the maximum sustainable current has already effectively verged toward the BEC side of the crossover.
This can be clearly 
seen in Fig.\ref{fig31} where the critical values $q_{c}$ corresponding to the four curves of Fig.\ref{fig30} are plotted vs $(k_{F} a_{F})^{-1}$, 
and the curve that interpolates over these data points is compared with the two Landau limiting branches of Fig.\ref{fig24}.

\vspace{-0.5cm}
This comparison shows that the maximum of the curve corresponding to the finite barrier has moved toward weaker 
couplings with respect to the limiting Landau curve associated with a vanishing barrier.
In this way, features of the BEC side of the crossover are now effectively felt already for couplings $(k_{F} a_{F})^{-1} <0$,
so that the Andreev bound states (for which the internal structure of the Cooper pairs is an essential ingredient) loose the 
special role they have on the BCS side of the crossover.
\vspace{-1.0cm}
\vspace{-1.0cm}
\begin{center}
\begin{figure}[htc]
\hspace{3cm}\includegraphics[angle=0,width=4in]{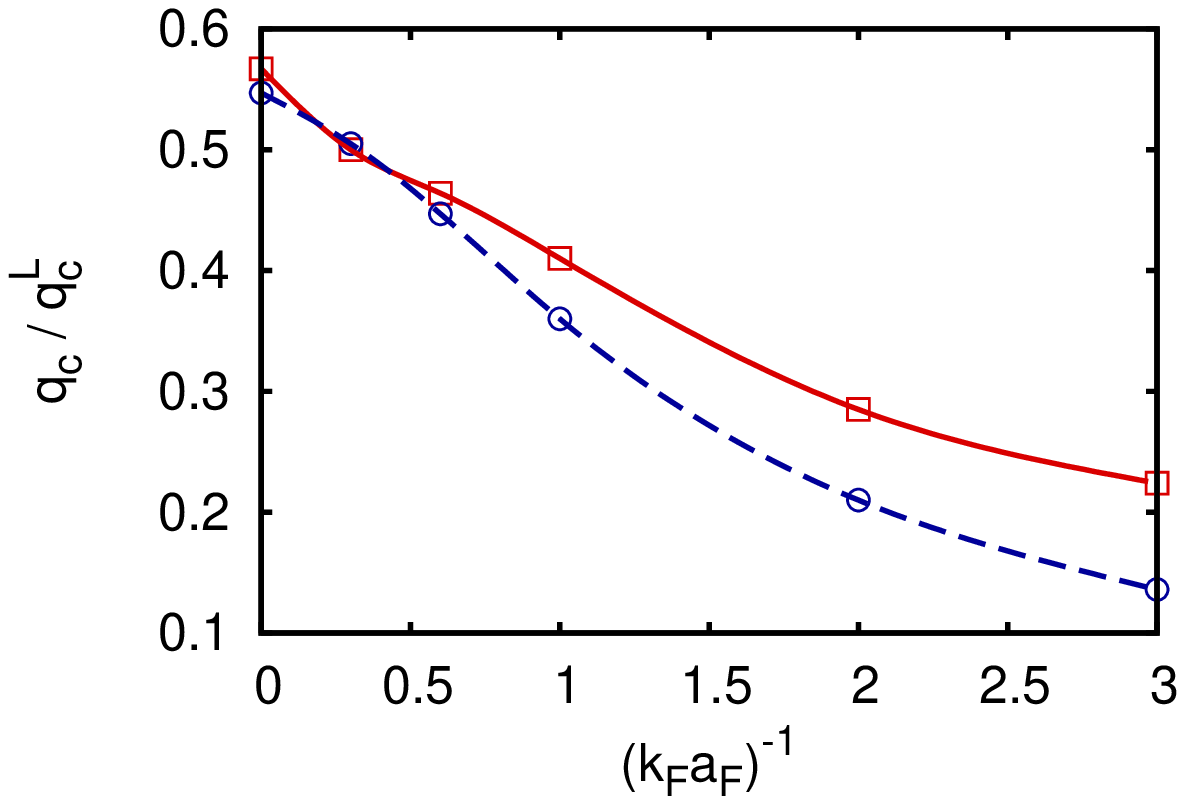}
\caption{Ratio $q_{c}/q_{c}^{\mathrm{Landau}}$ vs $(k_{F} a_{F})^{-1}$ on the BEC side of the crossover for a rectangular barrier of width $L k_{F} = 2.65$ and height $V_{0}/E_{F}= 0.10$, obtained from the solution of the BdG equations (full line) and the expression (\ref{local-Landau-criterion}) (dashed line).}
\label{fig32}
\end{figure}
\end{center}  
\vspace{-1.0cm}

It is worth mentioning in this context that, for coupling values such that the critical current $q_{c}$ in the presence of a barrier is no longer determined by
the occurrence of the Andreev bound states localized at the barrier, an alternative mechanisms is activated inside \emph{a wide enough barrier}, 
whereby it is the \emph{local} sound velocity to limit the amount of supercurrent that can flow across the barrier.
Suppose, in fact, that the barrier is wide enough for the relation $j = n(x) v(x)$ to occur between the local density $n(x)$ and velocity $v(x)$, in such a way that $j$ does not depend on $x$.
As a consequence, the local velocity $v(x=0) = j/n(x=0)$ at the center $x=0$ of the barrier can be quite considerably increased over its bulk value $j/n_{0}$ away from the barrier, owing to the local depression ($n(x=0) < n_{0}$) of the density at the center of the barrier, in such a way that $v(x=0)$ may as well exceed the local value $s(n(x=0))$ of the sound velocity.
Under these circumstances, the Landau criterion is activated locally and the superfluid current ceases to flow.
The following simple relation 

\begin{equation}
\frac{q_{c}}{q_{c}^{\mathrm{Landau}}} \, = \, \frac{n(x=0)}{n_{0}} \, \frac{s(n(x=0))}{s}
\label{local-Landau-criterion}
\end{equation}

\noindent
is then found to hold between the critical wave vector for a finite ($q_{c}$) and a vanishing ($q_{c}^{\mathrm{Landau}}$) barrier (in the BEC limit, the right-hand side of Eq.(\ref{local-Landau-criterion}) reduces to $|\Phi(x=0)|^{3}/|\Phi_{0}|^{3}$ where $\Phi(x)$ is the solution of the GP equation).

To show the (approximate) validity of this local criterion, we report in Fig.\ref{fig32} the ratio $q_{c}/q_{c}^{\mathrm{Landau}}$ for different coupling values $(k_{F} a_{F})^{-1}$ on the BEC side of the crossover for a wide enough barrier, as obtained alternatively from the maximum of the Josephson characteristics and from the right-hand side of Eq.(\ref{local-Landau-criterion}), where the local density $n(x=0)$ results from the solution of the BdG equations (\ref{BdG-equations}) and the local sound velocity $s(n(x=0))$ is obtained from Ref.\cite{MPS} for the given coupling and value of the local density.
Note that the two values of $q_{c}$ are comparable to each other, even though the agreement becomes worse when approaching
the strong-coupling limit. 
This is consistent with the observation we made in sub-section 4b, concerning the dominance of quantum tunneling when approaching the
BEC side for given barrier, since the barrier height $V_{0}$ eventually becomes larger than the bosonic chemical potential $\mu_{B}$.
In addition, when approaching the BEC side the healing length (cf. the quantity $\xi_{\mathrm{phase}}$ of Fig.\ref{fig7}) becomes
larger than the barrier width, thus further invalidating any local approximation.

We mention in this context that an analogous local Landau criterion was recently considered in Ref.\cite{Trento-LLC-2009} 
within an hydrodynamic approach based on a local-density approximation (see also Ref.\cite{Toigo-2009}).

\vspace{0.3cm}
\section{Stability of the self-consistent solutions}

In the previous Section, we have reported a number of numerical results that were obtained by solving the zero-temperature BdG equations throughout the BCS-BEC crossover until self-consistency was attained.
Even though our numerical procedure has proven capable of determining these self-consistent solutions rather unambiguously, the question naturally arises about the stability of these solutions when fluctuations beyond mean field would be introduced.

To a certain extent, this question can be addressed still remaining within our mean-field treatment, by exploiting the knowledge of the non-self-consistent solutions of the BdG equations about the self-consistent one. 
In a related fashion, one may also regard the two approaches (namely, of finding the mean-field solution via a self-consistent
calculation as we did so far or via an energy minimization as we are going to do) as complementing each other along the
lines of standard alternative mean-field approximations \cite{PB-1994}.

In the following, we shall first provide some general considerations concerning the expression for the energy related to the (not necessarily self-consistent) solutions of the BdG equations in the presence of a current, before embarking in its explicit numerical calculation for the cases of interest.
The BEC limit, for which an independent knowledge of the results obtained by the GP equation can be exploited, will again be especially emphasized.
In particular, we shall verify that the critical value $q_{c}$ obtained (like in Fig.\ref{fig25}) for given coupling and barrier corresponds also to the maximum value of the superfluid current for which the self-consistent solutions of the BdG equations are energetically stable.


\vspace{0.3cm}
\begin{center}
\begin{large}
{\bf 5a. General considerations about the energy stability}
\end{large}
\end{center}
\addcontentsline{toc}{subsection}{5a. General considerations on the energy stability}
\vspace{0.1cm} 

The quantity that needs to be estimated for the purpose is the total energy $\mathcal{E}$ of the system, as it changes when non-self-consistent solutions about the self-consistent one are considered.
Specifically, we have to compare the total energy of the system in the presence \emph{or\/} in the absence of a barrier when the respective solutions are subject to the same ``macroscopic boundary conditions'' of given superfluid current $j$ and overall phase difference $\Delta \phi$.
The reason for taking this energy difference is that, even to a strictly homogeneous solution in the presence of the superfluid current $j$, there corresponds an energy cost needed to set up the current and the finite phase difference $\Delta \phi = \phi(x_{B}) - \phi(x_{A})$ between two distant spatial points 
$x_{A}$ and $x_{B}$ with $(x_{B} - x_{A}) = D \gg \xi_{\mathrm{phase}}$ \cite{Anderson-1966}.
[Note that, for notational convenience, in the present Section the phase difference  $\Delta \phi$ corresponds to the quantity $\delta \phi/2$ of Section 4.]
This energy cost should be appropriately subtracted off from the corresponding value of the total energy in the presence of 
a barrier (while keeping the same values $j$ for the superfluid current and $\Delta \phi$ for the overall phase difference that accumulates accross the barrier), in order to probe the stability of the solution when a barrier is introduced into the system (say, in the middle of the interval $D$).

By this procedure, we mean that the reference solution of the homogeneous case in the absence of a barrier (nb) should not only carry the same value $j = q n_{0}/m$ of the superfluid current ($n_{0}$ being the bulk density) but should also contain an infinitesimal local phase gradient $\phi_{\mathrm{nb}}'(x)$, in such a way that the total accumulated phase difference $\phi_{\mathrm{nb}}(x_{B}) - \phi_{\mathrm{nb}}(x_{A})$ has the same value $\Delta \phi$ as for the corresponding solution in the presence of the barrier. 
The only difference is that the latter solution develops the total phase difference $\Delta \phi$ sharply in a rather narrow spatial range of the order 
$\xi_{\mathrm{phase}}$ about the position of the barrier, as we have explicitly verified by our numerical calculations.

In addition, since the value of the chemical potential $\mu$ is kept uniform throughout the system, rather than the total energy $\mathcal{E}$ we should actually consider the value of the grand-canonical energy $\mathcal{E} - \mu N$ in the frame where the barrier is at rest, and look for the minimum of $\mathcal{E} - \mu N$ to determine the most probable configuration of the system \cite{Baym-1967}.

To obtain the value of the total energy $\mathcal{E}$, one could be tempted to calculate it by using the following standard expression of many-particle theory \cite{FW-4} 
\begin{equation}
\mathcal{E} \, = \, \int \! d\mathbf{r} \, \lim_{\mathbf{r'} \rightarrow \mathbf{r}} \, \frac{1}{\beta} \, \sum_{n} \, 
e^{i \omega_{n} \eta} \, \left( - \frac{\nabla^{2}}{2m} \, + V(\mathbf{r}) \, + \mu \, + \, i \omega_{n} \right)
\, \mathcal{G}_{11}(\mathbf{r},\mathbf{r'};\omega_{n})                             \label{total-energy-many-body}
\end{equation}

\noindent 
which holds in the present form for a temperature $T$ below the critical temperature.
In this expression, $\beta = 1/(k_{B} T)$, $\omega_{n} = (2n+1)\pi/\beta$ ($n$ integer) a fermionic Matsubara frequency, $\eta$ a positive infinitesimal, and $\mathcal{G}_{11}$ the ``normal'' component of the single-particle Green's function in the broken-symmetry phase which admits the following eigenfunction expansion in terms of the solutions of the BdG equations \cite{PS-2003}:

\begin{equation}
\mathcal{G}_{11}(\mathbf{r},\mathbf{r'};\omega_{n}) \, = \, \sum_{n} \, 
\left( \frac{u_{\nu}(\mathbf{r}) \, u_{\nu}(\mathbf{r'})^{*}}{ i \omega_{n} \, - \, \epsilon_{\nu}} \, + \,
\frac{v_{\nu}(\mathbf{r})^{*} \, v_{\nu}(\mathbf{r'})}{ i \omega_{n} \, + \, \epsilon_{\nu}} \right) \,\, .
                                                                                                              \label{eigenfunction-expansion-G-11}
\end{equation}

\noindent
The problem with the expression (\ref{total-energy-many-body}) when using the solutions of the BdG equations in Eq.(\ref{eigenfunction-expansion-G-11}), however, is that Eq.(\ref{total-energy-many-body}) holds when the 
\emph{self-consistency} condition for these solutions has been achieved, while for the present purpose we need to know the value of $\mathcal{E}$ also away from self-consistency.
This important point results particularly evident when one attempts to derive the strong-coupling (BEC) limit of 
Eq.(\ref{total-energy-many-body}), which is seen to coincide with the expression resulting from the Gross-Pitaevskii theory \cite{PS-book-2003} only when the self-consistency condition represented by the GP equation (\ref{Gross-Pitaevskii-equation}) itself is applied. 
The correct derivation of the BEC limit for $\mathcal{E}$ also away from self-consistency will be considered in detail below.

To obtain the expression of the total energy that holds \emph{also away from self-consistency}, we have to resort to the general procedure discussed in Ref.\cite{BdG} and, besides the original (grand-canonical) Hamiltonian 
\begin{equation}
H \, - \mu N \, = \, \sum_{\sigma} \int \! d\mathbf{r} \, \Psi_{\sigma}^{\dagger}(\mathrm{r}) 
\mathcal{H}(\mathbf{r}) \Psi_{\sigma}(\mathrm{r}) \, - \, g
\int \! d\mathbf{r} \, \Psi_{\uparrow}^{\dagger}(\mathrm{r}) \Psi_{\downarrow}^{\dagger}(\mathrm{r})
                               \Psi_{\downarrow}(\mathrm{r}) \Psi_{\uparrow}(\mathrm{r})  \label{original-Hamiltonian}
\end{equation}

\noindent
expressed in terms of the field operator $\Psi_{\sigma}(\mathrm{r})$ ($\sigma=\uparrow,\downarrow$ being the spin component)
with a contact interaction, we have to introduce the \emph{effective} (grand-canonical) \emph{Hamiltonian}:
\begin{eqnarray}
H_{\mathrm{eff}} \, - \mu N & = & \sum_{\sigma} \int \! d\mathbf{r} \, \Psi_{\sigma}^{\dagger}(\mathrm{r}) \,
\mathcal{H}(\mathbf{r}) \, \Psi_{\sigma}(\mathrm{r})             \nonumber \\
& + & \int \! d\mathbf{r} \, \left[ 
\Delta(\mathrm{r}) \Psi_{\uparrow}^{\dagger}(\mathrm{r}) \Psi_{\downarrow}^{\dagger}(\mathrm{r}) \, + \,
\Delta(\mathrm{r})^{*} \Psi_{\downarrow}(\mathrm{r}) \Psi_{\uparrow}(\mathrm{r}) \right] \,\, .
                                                                                          \label{effective-Hamiltonian}
\end{eqnarray}

\noindent
Here, $\Delta(\mathbf{r})$ is the same arbitrary function entering the BdG equations (\ref{BdG-equations}) \emph{before} the self-consistency condition (\ref{self-consistency}) is eventually achieved.
We then evaluate the thermal average $\langle(H \, - \mu N)\rangle_{H_{\mathrm{eff}} \, - \mu N}$ with respect to the effective Hamiltonian (\ref{effective-Hamiltonian}), and obtain the following expression with the use of the BdG equations (\ref{BdG-equations}):
\begin{eqnarray}
\langle(H \, - \mu N)\rangle_{H_{\mathrm{eff}} \, - \mu N} & = & \int \! d\mathbf{r} \, \left\{
2 \sum_{\nu} \, \epsilon_{\nu}  \, \left[ |u_{\nu}(\mathbf{r})|^{2} f_{F}(\epsilon_{\nu}) -
                                                           |v_{\nu}(\mathbf{r})|^{2} \left(1 - f_{F}(\epsilon_{\nu})\right) \right]  \right.      
                                                                                                                                     \nonumber \\
& + & 2 \, \Delta(\mathbf{r}) \, \sum_{\nu} u_{\nu}(\mathbf{r})^{*} v_{\nu}(\mathbf{r}) 
\left(1 - 2 f_{F}(\epsilon_{\nu})\right)                                                                             \nonumber  \\   
& - & g \, \sum_{\nu} u_{\nu}(\mathbf{r})^{*} v_{\nu}(\mathbf{r}) \left(1 - 2 f_{F}(\epsilon_{\nu})\right) 
                                                                                                                                 \nonumber \\
& \times & \, \left. \sum_{\nu'} u_{\nu'}(\mathbf{r}) v_{\nu'}(\mathbf{r})^{*} \left(1 - 2 f_{F}(\epsilon_{\nu'})
\right) \right\} \,\, .                                                                                                           \label{thermal-average}
\end{eqnarray}

\noindent
This expression is formally a functional of the function $\Delta(\mathbf{r})$ that appears in the BdG equations (\ref{BdG-equations}), a function which can \emph{a priori\/} assume arbitrary profiles when the self-consistent condition (\ref{self-consistency}) is not satisfied.

In particular, in the zero-temperature limit (whereby $f_{F}(\epsilon)=0$ for $\epsilon >0$) which we restrict to in the present paper, Eq.(\ref{thermal-average}) simplifies as follows:
\begin{eqnarray}
\langle(H \, - \mu N)\rangle_{H_{\mathrm{eff}} \, - \mu N}^{(T=0)} & = & \int \! d\mathbf{r} \, \left\{
- 2 \sum_{\nu} \, \epsilon_{\nu} |v_{\nu}(\mathbf{r})|^{2} \, + \, 2 \, 
\Delta(\mathbf{r}) \, \sum_{\nu} u_{\nu}(\mathbf{r})^{*} v_{\nu}(\mathbf{r}) \right.     \nonumber  \\   
& - & \left.  g \, \sum_{\nu} u_{\nu}(\mathbf{r})^{*} v_{\nu}(\mathbf{r}) 
\sum_{\nu'} u_{\nu'}(\mathbf{r}) v_{\nu'}(\mathbf{r})^{*}  \right\} \,\, .                         \label{thermal-average-T=0}
\end{eqnarray}

The above expression need be regularized owing to the presence of the contact potential with strength $g$.
This can be done by assuming that the non-self-consistent solutions of the BdG equations share the same ultraviolet behavior 
that was identified in order to regularize the gap equation (\ref{regularized-gap-inhomogeneous}).
In the zero-temperature limit we then obtain in the place of the expression (\ref{thermal-average-T=0}):
\begin{eqnarray}
\langle(H \, - \mu N)\rangle_{H_{\mathrm{eff}} \, - \mu N}^{(T=0)} & = & \int \! d\mathbf{r} \, \left\{
- 2 \sum_{\nu} \, \epsilon_{\nu} |v_{\nu}(\mathbf{r})|^{2} \, + \, |\Delta(\mathbf{r})|^{2} \, 
\int \! \frac{d\mathbf{k}}{(2 \pi)^{3}} \, \frac{m}{\mathbf{k}^{2}} \right.      \nonumber  \\   
& - & \left. |\Delta(\mathbf{r})|^{2} \, \frac{m}{4 \pi a_{F}} \right\}       \label{thermal-average-T=0-regularized}
\end{eqnarray}

\noindent
where use has been made of the relation (\ref{regularization}) to eliminate $g$ in favor of $a_{F}$.

As we have already discussed at the beginning of this Section, what we have actually to compare are the values of the quantity $\mathcal{E} - \mu N = \langle(H \, - \mu N)\rangle$ calculated in the presence \emph{and\/} in the absence of a barrier, while keeping in the two cases the same values of the superfluid current $j$ flowing in and out of the system and of the overall phase difference $\Delta\phi$.
We thus write with reference to Eq.(\ref{thermal-average-T=0-regularized}):
\begin{eqnarray}
& & \left. \frac{(\mathcal{E} - \mu N )}{\mathcal{A}} \right|_{\mathrm{with \, barrier}}  \, - \, 
       \left. \frac{(\mathcal{E} - \mu N )}{\mathcal{A}} \right|_{\mathrm{no \, barrier}}             \nonumber  \\
& = &  \int_{- \infty}^{+ \infty} \! dx \, \left\{
2 \left( \int \! \frac{d\mathbf{k}}{(2 \pi)^{3}} \, E_{\mathbf{k}} |v_{\mathbf{k}}|^{2} \, - \,
           \sum_{\nu} \, \epsilon_{\nu} |v_{\nu}(x)|^{2} \right) \right.             \nonumber \\
& + & \left( |\Delta(x)|^{2} \, - \, \Delta_{0}^{2} \right)
\int \! \frac{d\mathbf{k}}{(2 \pi)^{3}} \, \frac{m}{\mathbf{k}^{2}}                             \nonumber  \\   
& - & \left. \left( |\Delta(x)|^{2} \, - \, \Delta_{0}^{2} \right) \frac{m}{4 \pi a_{F}} \right\}  
\, - \, j \, \Delta \phi                                                                              \label{quantity-to-be-calculated-fermions}                                                                                        
\end{eqnarray}                                                                                                     

\noindent
where $\mathcal{A}$ is the area transverse to the current flow (that we here consider finite for convenience),
$E_{\mathbf{k}}$ is defined after Eq.(\ref{density-homogeneous}), and $\Delta_{0}$ is the bulk value 
of the gap parameter far away from the barrier.

Note the presence in Eq.(\ref{quantity-to-be-calculated-fermions}) of the term $j \, \Delta \phi$, which represents quite generally
the energy cost required to set up the superfluid current $j$ 
\emph{and} the phase difference $\Delta \phi$ \cite{Anderson-1966}. 
In the present context, when superfluid fermions are described by the homogeneous BdG equations with no barrier, the origin of the term $j \, \Delta \phi$ can be understood as follows.

For superfluid fermions flowing with no barrier, we can use the results 
(\ref{u-v-homogeneous})-(\ref{u2-v2-homogeneous}) with $\mu = \mu_{0} + \mathbf{q}^{2}/(2m)$,
which imply that $\Delta_{\mathbf{q}}$, $u_{\mathbf{q}}(\mathbf{k})$, and $v_{\mathbf{q}}(\mathbf{k})$ do not depend on $\mathbf{q}$ as long as the superflow is stable.
As a consequence, the first term on the right-hand side of Eq.(\ref{eigeinvalue-homogeneous}) does not contribute to the right-hand side of Eq.(\ref{thermal-average-T=0-regularized}) when specified to the homogeneous case with the eigenvalues $\epsilon_{\nu}$ replaced by $E_{\mathbf{k}}$. 
The only dependence of $\mathcal{E} = \langle H \rangle$ on $\mathbf{q}$ thus originates from the explicit dependence on $\mathbf{q}$ of the chemical potential on the left-hand side of Eq.(\ref{thermal-average-T=0-regularized}). 
Writing the total particle number as $N = n \mathcal{V}$, where $n$ is the uniform fermionic density and
$\mathcal{V} = \mathcal{A} D$ is the volume of interest containing the system, we then obtain:
\begin{equation}
\frac{\partial \mathcal{E}}{\partial |\mathbf{q}|} \longrightarrow \frac{\partial \mu}{\partial |\mathbf{q}|} \,\, N \, = \, 
\frac{|\mathbf{q}|}{m} \, n \, \mathcal{A} \, D \,\, .                                                             \label{dE-dq}
\end{equation}                                                                               

\noindent
Suppose now to vary $|\mathbf{q}| \rightarrow \, |\mathbf{q}| + \lambda$ by an infinitesimal amount 
$\lambda$, in order to generate the overall phase difference $\Delta \phi$ while mantaining unchanged
the values $j$ of the local current and $\mu$ of the chemical potential.
Owing to Eq.(\ref{dE-dq}), this variation induces the following increase in the total energy:

\begin{equation}
\left. \Delta \mathcal{E} \right|_{\mathrm{no \, barrier}} \, = \, \frac{\partial \mathcal{E}}{\partial |\mathbf{q}|} \, \lambda \, = \,
\frac{|\mathbf{q}|}{m}  \, \lambda \,\, n \, \mathcal{A} \, D \, = \, \mathcal{A} \, j \, \Delta \phi
                                                                                                                    \label{Delta-E-fermions}
\end{equation}

\noindent
where $j = q \, n / m$ and
\begin{equation}
\Delta \phi \, = \, \lambda \, D  \, = \, \int_{x_{A}}^{x_{B}} \! dx \, \lambda \, = \,  
\int_{x_{A}}^{x_{B}} \! dx \, \frac{d \phi_{\mathrm{nb}}(x)}{dx}  \,\, .                    \label{Delta-phi-homogeneous-fermions}
\end{equation}

\noindent
Here, $\phi_{\mathrm{nb}}(x)$ represents the additional phase entering the fermionic wave function [cf. Eq.(\ref{BdG-discretized-solutions})]
\begin{equation}
v_{\nu}(\mathbf{r}) \rightarrow v_{\mathbf{q}}(\mathbf{k}) 
\exp \left\{i (\mathbf{k} - \mathbf{q}) \cdot \mathbf{r} - i \phi_{\mathrm{nb}}(x) \right\}   \label{additional-phase}
\end{equation}

\noindent
which is needed to produce the overall phase difference $\Delta \phi$ even with no barrier.
The result (\ref{Delta-E-fermions}) justifies eventually the presence of the last term on the right-hand side of 
Eq.(\ref{quantity-to-be-calculated-fermions}).

The expressions (\ref{thermal-average-T=0}) and (\ref{thermal-average-T=0-regularized}) are written in terms of fermionic quantities and are accordingly able to span the whole BCS-BEC crossover from the weak- to the strong-coupling limits.
However, the same conceptual and practical instances that have led us in sub-section 2d to consider the strong-coupling (BEC) limit of the BdG equations, by deriving from them the GP equation (\ref{Gross-Pitaevskii-equation}) for composite bosons, manifest here for the calculation of the total energy which we would like to express directly in terms of the condensate wave function $\Phi(\mathbf{r})$ of composite bosons [cf. Eq.(\ref{Phi-vs-Delta})].
We then need to extend the results of Ref.\cite{PS-2003} in order to provide a formal derivation of the expression of the total energy in terms of composite bosons, starting from the original expression in terms of the constituent fermions. We include here for completeness this missing derivation.

To this end, it is convenient to express from the outset the grand-canonical energy (\ref{thermal-average}) in terms of the ``normal'' ($\mathcal{G}_{11}$) and ``anomalous'' ($\mathcal{G}_{21}$) single-particle Green's functions that are compatible with the solutions of the BdG equations for an \emph{arbitrary\/} choice of the function $\Delta(\mathbf{r})$ (this is required by our demand of studying the stability of the self-consistent solutions of the BdG equations).
We have already reported in Eq.(\ref{eigenfunction-expansion-G-11}) the form of the ``normal'' component
$\mathcal{G}_{11}$, which holds even when the eigenfunctions $\{u_{\nu}(\mathbf{r}),v_{\nu}(\mathbf{r})\}$ 
and the associated eigenvalues $\{\epsilon_{\nu}\}$ are solutions of the BdG equations (\ref{BdG-equations}) with an arbitrary $\Delta(\mathbf{r})$.
The corresponding form of the ``anomalous'' component $\mathcal{G}_{21}$ reads:
\begin{equation}
\mathcal{G}_{21}(\mathbf{r},\mathbf{r'};\omega_{n}) \, = \, \sum_{n} \, 
\left( \frac{v_{\nu}(\mathbf{r}) \, u_{\nu}(\mathbf{r'})^{*}}{ i \omega_{n} \, - \, \epsilon_{\nu}} \, - \,
\frac{u_{\nu}(\mathbf{r})^{*} \, v_{\nu}(\mathbf{r'})}{ i \omega_{n} \, + \, \epsilon_{\nu}} \right) \,\, .
                                                                                                              \label{eigenfunction-expansion-G-21}
\end{equation}

\noindent
The expressions (\ref{eigenfunction-expansion-G-11}) and (\ref{eigenfunction-expansion-G-21}) allow us to rewrite the grand-canonical average (\ref{thermal-average}) of interest in the alternative form:
\begin{eqnarray}
\langle(H \, & - & \mu N)\rangle_{H_{\mathrm{eff}} \, - \mu N} 
\, = \, 2 \int \! d\mathbf{r} \, \lim_{\mathbf{r'} \rightarrow \mathbf{r}} \, 
\frac{1}{\beta} \sum_{n} \, e^{i \omega_{n} \eta} \, 
\mathcal{H}(\mathbf{r}) \, \mathcal{G}_{11}(\mathbf{r},\mathbf{r'};\omega_{n}) 
                                                                                                      \label{thermal-average-in-terms-of-G} \\
& - & g \int \! d\mathbf{r} \,
\left(\frac{1}{\beta}  \sum_{n} e^{i \omega_{n} \eta}  \, \mathcal{G}_{21}(\mathbf{r},\mathbf{r};\omega_{n}) \right)
\left(\frac{1}{\beta}  \sum_{n'} e^{i \omega_{n'} \eta} \, \mathcal{G}_{21}(\mathbf{r},\mathbf{r};\omega_{n'}) \right)^{*} .                                                                                                \nonumber
\end{eqnarray}

\noindent
To obtain the limiting form of this expression in the BEC regime, we resort to the two integral equations that couple $\mathcal{G}_{11}$ and $\mathcal{G}_{21}$ \cite{PS-2003}, namely,
\begin{eqnarray}
& \mathcal{G}_{11}(\mathbf{r},\mathbf{r'};\omega_{n}) & = \,\,  
\tilde{\mathcal{G}}_{0}(\mathbf{r},\mathbf{r'};\omega_{n}) \, + \, \int \! d\mathbf{r''} \, 
\tilde{\mathcal{G}}_{0}(\mathbf{r},\mathbf{r''};\omega_{n}) \, \Delta(\mathbf{r''}) \,
\mathcal{G}_{21}(\mathbf{r''},\mathbf{r'};\omega_{n})                              \nonumber \\
& \mathcal{G}_{21}(\mathbf{r},\mathbf{r'};\omega_{n}) & = \,\, - \, \int \! d\mathbf{r''} \,  
\tilde{\mathcal{G}}_{0}(\mathbf{r''},\mathbf{r};-\omega_{n}) \, \Delta(\mathbf{r''})^{*} \, 
\mathcal{G}_{11}(\mathbf{r''},\mathbf{r'};\omega_{n})                              \label{two-coupled-integral-equations}
\end{eqnarray}

\noindent
where the non-interacting Green's function $\tilde{\mathcal{G}}_{0}$ satisfies the following equation
\begin{equation}
\left[ i \omega_{n} \, - \, \mathcal{H}(\mathbf{r}) \right] \, 
\tilde{\mathcal{G}}_{0}(\mathbf{r},\mathbf{r'};\omega_{n}) \, = \, \delta(\mathbf{r} - \mathbf{r'}) \,\, .
                                                                                                                \label{equation-for-G0}
\end{equation}

\noindent
We then iterate these integral equations, to expand $\mathcal{G}_{11}$ and $\mathcal{G}_{21}$ to the forth and third order in $\Delta$, respectively.
After a long but straightforward calculation and taking into account the results of Ref.\cite{PS-2003}, we obtain
for the first term on the right-hand side of Eq.(\ref{thermal-average-in-terms-of-G}) the approximate expression:
\begin{eqnarray}
& 2 & \int \! d\mathbf{r} \, \lim_{\mathbf{r'} \rightarrow \mathbf{r}} \, 
\frac{1}{\beta} \sum_{n} \, e^{i \omega_{n} \eta} \, 
\mathcal{H}(\mathbf{r}) \, \mathcal{G}_{11}(\mathbf{r},\mathbf{r'};\omega_{n})              \nonumber \\
& \simeq & \int \! d\mathbf{r} \, \left\{ \left[ \frac{1}{g} \, + \, \frac{m^{2} a_{F}}{8 \pi} 
\left( \mu_{B} \, - \, 2 V(\mathbf{r}) \right) \right] \, |\Delta(\mathbf{r})|^{2} \right.             \nonumber \\
& + & \left. \frac{m^{2} a_{F}}{8 \pi} \, \Delta(\mathbf{r}) \, \frac{\nabla^{2}}{4 m} \, \Delta(\mathbf{r})^{*}
\, - \,  \frac{6 \pi a_{F}}{m} \, \left( \frac{m^{2} a_{F}}{8 \pi} \right)^{2} \, |\Delta(\mathbf{r})|^{4} \right\}
                                                                                             \label{expansion-energy-with-G-11}
\end{eqnarray}

\noindent
where $\mu_{B}$ is the chemical potential for composite bosons defined after Eq.(\ref{Gross-Pitaevskii-equation}).
In a similar fashion we obtain:
\begin{eqnarray}
\frac{1}{\beta} \sum_{n} e^{i \omega_{n} \eta}  \, \mathcal{G}_{21}(\mathbf{r},\mathbf{r};\omega_{n}) 
& \simeq & - \, \left[ \frac{1}{g} \, + \, \frac{m^{2} a_{F}}{8 \pi} \left( \mu_{B} \, - \, 2 V(\mathbf{r}) \right) \right] 
\, \Delta(\mathbf{r})^{*}                                                     \label{expansion-term-with-G-21} \\
& - & \frac{m^{2} a_{F}}{8 \pi} \, \frac{\nabla^{2}}{4 m} \, \Delta(\mathbf{r})^{*} + 
\frac{4 \pi a_{F}}{m} \left( \frac{m^{2} a_{F}}{8 \pi} \right)^{2} |\Delta(\mathbf{r})|^{2} \, 
\Delta(\mathbf{r})^{*} \, ,                                                        \nonumber
\end{eqnarray}

\noindent
which yields for the second term on the right-hand side of Eq.(\ref{thermal-average-in-terms-of-G}) the approximate expression:
\begin{eqnarray}
& - & g \int \! d\mathbf{r} \,
\left(\frac{1}{\beta}  \sum_{n} e^{i \omega_{n} \eta}  \, \mathcal{G}_{21}(\mathbf{r},\mathbf{r};\omega_{n}) \right)
\left(\frac{1}{\beta}  \sum_{n'} e^{i \omega_{n'} \eta} \, \mathcal{G}_{21}(\mathbf{r},\mathbf{r};\omega_{n'}) \right)^{*}                                                                                                \nonumber \\
& \simeq &  \int \! d\mathbf{r} \, \left\{ - \, \left[ \frac{1}{g} \, + \, \frac{m^{2} a_{F}}{4 \pi} 
\left( \mu_{B} \, - \, 2 V(\mathbf{r}) \right) \right] \, |\Delta(\mathbf{r})|^{2} \right.             \nonumber \\
& - & \left. \frac{m^{2} a_{F}}{4 \pi} \, \Delta(\mathbf{r}) \, \frac{\nabla^{2}}{4 m} \, \Delta(\mathbf{r})^{*}
\, + \,  \frac{8 \pi a_{F}}{m} \, \left( \frac{m^{2} a_{F}}{8 \pi} \right)^{2} \, |\Delta(\mathbf{r})|^{4} \right\}
                                                                                             \label{expansion-energy-with-G-21}
\end{eqnarray}

\noindent
where only terms which survive the regularization when $g \rightarrow 0$ have been retained.
Upon combining the two results (\ref{expansion-energy-with-G-11}) and (\ref{expansion-energy-with-G-21}), 
recalling the relation (\ref{Phi-vs-Delta}) between $\Delta(\mathbf{r})$ and the condensate wave function 
$\Phi(\mathbf{r})$, and making explicit the relation
\begin{equation}
\mu \, N \, \simeq \, - \, \frac{\epsilon_{0}}{2} \, N \, + \, \mu_{B} \, \int \! d\mathbf{r} \, |\Phi(\mathbf{r})|^{2}
                                                                                                                             \label{mu-vs-Phi}
\end{equation} 

\noindent
that holds in the BEC limit, we obtain eventually the desired result for the total energy in this limit:
\begin{eqnarray}
\mathcal{E} \, = \, \langle H \rangle_{H_{\mathrm{eff}} \, - \mu N} & \simeq & - \, \epsilon_{0} \, \frac{N}{2} \, + \, 
\int \! d\mathbf{r} \, \left\{ - \, \Phi(\mathbf{r})^{*} \, \frac{\nabla^{2}}{2 m_{B}} \, \Phi(\mathbf{r}) \right.              
                                                                                                                                    \nonumber \\
& + & \left. V_{B}(\mathbf{r}) \, |\Phi(\mathbf{r})|^{2} \, + \, \frac{1}{2} \, \frac{4 \pi a_{B}}{m_{B}} \,
 |\Phi(\mathbf{r})|^{4} \right\}    \,\, .                                                                                     \label{total-energy-GP}
\end{eqnarray}

\noindent
Here, $m_{B} = 2m$ is the mass of the composite bosons, $V_{B}(\mathbf{r}) = 2 \, V(\mathbf{r})$ is the
external potential felt by the composite bosons, and $a_{B} = 2 a_{F}$ is the scattering length associated with their residual mutual 
interaction at the present level of approximation \cite{PS-2003}.
Apart from the term $-\epsilon_{0} N/2$ which accounts for the energy required to form $N/2$ fermion pairs each with binding energy $\epsilon_{0}$, the above expression coincides with the energy of a non-uniform bosonic condensate within the Gross-Pitaevskii theory (cf., e.g., Ref.\cite{PS-book-2003}).
This result proves that the fermionic BdG equations are able to recover a sensible description of the system 
of composite bosons that form in the strong-coupling (BEC) limit, also as far as their total energy is concerned.

As already remarked, the results obtained in terms of the GP theory not only constitute an important benchmark for analogous results obtained from the BdG equations in the strong-coupling (BEC) limit, but may also independently serve to suggest methodologies and approximations to be adopted also for the BdG equations.
To this end, it is convenient to write the analogue of the expression (\ref{quantity-to-be-calculated-fermions}) directly in terms of (composite) bosons.
With reference to the form (\ref{Phi-one-dimensional}) of the bosonic wave function (cf. Appendix C), we obtain:
\begin{eqnarray}
& & \left. \frac{(\mathcal{E} - \mu_{B} N_{B} )}{\mathcal{A}} \right|_{\mathrm{with \, barrier}}  \, - \, 
\left. \frac{(\mathcal{E} - \mu_{B} N_{B} )}{\mathcal{A}} \right|_{\mathrm{no \, barrier}}    \nonumber \\
& = & \int_{- \infty}^{+ \infty} \! dx \, \left\{ \frac{1}{2 \, m_{B}} \, 
\left| \frac{d}{d x} \left( \sqrt{n_{B}(x)} \, e^{i \phi_{B}(x)} \right) \right|^{2} \, + \, 
\frac{q_{B}}{m_{B}} \, n_{B}(x) \, \frac{d \phi_{B}(x)}{dx}  \right.                         \nonumber  \\
& + & \left. V_{B}(x) \, n_{B}(x) \, + \, \frac{g_{B}}{2} \, \left( n_{B}(x) \, - \, n_{B} \right)^{2} \right\}
\, - \, \frac{q_{B}}{m_{B}} \, n_{B} \, \Delta \phi                             \label{quantity-to-be-calculated-bosons}
\end{eqnarray}

\noindent
where $N_{B} = N/2$, $g_{B} = 4 \pi a_{B}/m_{B}$, and $\mu_{B} = g_{B} n_{B} + q_{B}^{2}/(2m_{B})$ at the present 
level of approximation.
Evaluation of the bosonic expression (\ref{quantity-to-be-calculated-bosons}) should thus be fully equivalent 
to evaluation of the fermionic expression (\ref{quantity-to-be-calculated-fermions}) once the strong-coupling (BEC) limit of the latter is considered.

The form (\ref{quantity-to-be-calculated-bosons}) coincides with the expressions (18) and (19) reported in 
Ref.\cite{Hakim-1997}, where the same boundary condition appropriate to a constant flow was adopted.
This boundary condition is enforced by the last term on the right-hand side of 
Eq.(\ref{quantity-to-be-calculated-bosons}), which accounts for the term $j \Delta \phi$ \cite{Anderson-1966}
in the present context and whose origin can be readily understood as follows.
For the bosonic condensate, we write in the absence of the barrier [cf. Eq.(\ref{Phi-one-dimensional})]:
\begin{equation}
\Phi_{\mathrm{nb}}(x) \, = \, \sqrt{n_{B}} \,\, e^{i (q_{B} x + \phi_{\mathrm{nb}}^{B}(x))}     \label{Phi-no-barrier}
\end{equation}

\noindent
where $n_{B}$ is the uniform bosonic density and the phase $\phi_{\mathrm{nb}}^{B}(x) = \lambda_{B} \, x$ (with $\lambda_{B} \ll q_{B}$) is needed to account for the overall phase difference $\Delta \phi$.
In this case, the bosonic part of the total energy (\ref{total-energy-GP}) reduces to:
\begin{equation}
\left. \frac{\mathcal{E}}{\mathcal{A}} \right|_{\mathrm{no \, barrier}} \, = \, \left( \frac{q_{B}^{2}}{2 \, m_{B}} \, n_{B} 
\, + \, \frac{1}{2} \, g_{B} \, n_{B}^{2} \right) D \, + \, \frac{q_{B}}{m_{B}} \, n_{B} \, \lambda_{B} D \,\, .
                                                                                                                \label{energy-bosons-homogeneous} 
\end{equation}

\noindent
The last term on the right-hand side does not depend on the size $D$ of the system since it can be written as the product $j \, \Delta \phi$ \cite{Anderson-1966}, where $j = q_{B} \, n_{B} / m_{B}$ and
\begin{equation}
\Delta \phi \, = \, \int_{x_{A}}^{x_{B}} \! dx \, \frac{d \phi_{\mathrm{nb}}^{B}(x))}{dx} \, = \,
                           \int_{x_{A}}^{x_{B}} \! dx \, \lambda_{B} \, = \, \lambda_{B} D \,\, .     
                                                                                                                        \label{Delta-phi-homogeneous-bosons}
\end{equation}


\vspace{0.3cm}
\begin{center}
\begin{large}
{\bf 5b. Numerical analysis of the energy stability}
\end{large}
\end{center}
\addcontentsline{toc}{subsection}{5b. Numerical analysis of the energy stability}
\vspace{0.1cm} 

A numerical analysis of the bosonic expression (\ref{quantity-to-be-calculated-bosons}) to test the stability of the solutions of the GP equation for point-like bosons was already performed in Ref.\cite{Hakim-1997}, albeit for a delta-like barrier only.
Here, we begin our numerical analysis by extending the results of Ref.\cite{Hakim-1997} to more general barriers of finite width, and then proceed by evaluating the corresponding fermionic expression (\ref{quantity-to-be-calculated-fermions}) so as to consider the whole BCS-BEC crossover.

\begin{center}
\begin{figure}[htc]
\includegraphics[angle=0,width=6.5in]{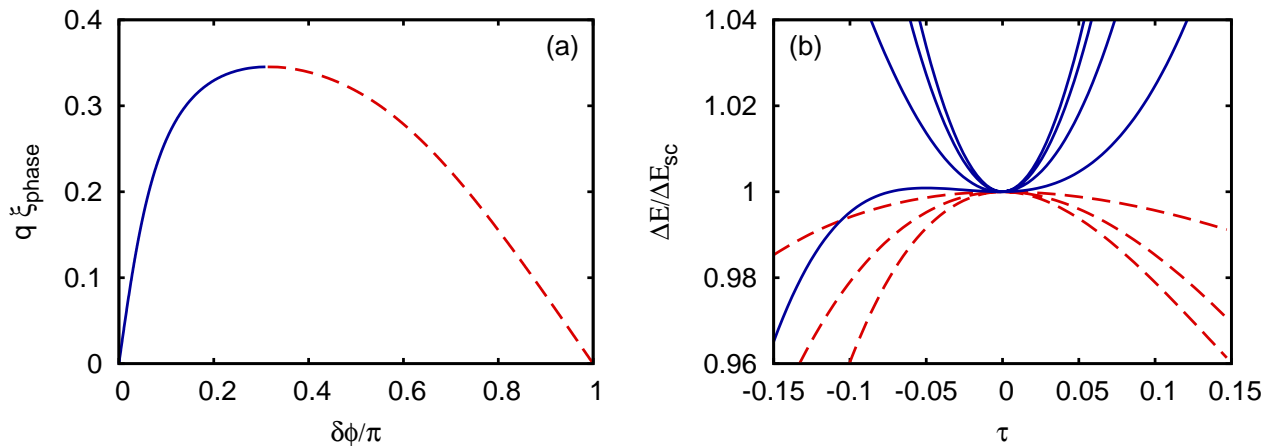}
\caption{(a) Josephson characteristic $q$ vs $\delta \phi$ obtained from the GP equation with a Gaussian barrier of 
              width $\sigma/\xi_{\mathrm{phase}} = 3.0$ and height $V_{0}/E_{F} = 0.7$ (the two branches, respectively 
              to the left and right of the maximum, are evidenced by a different drawing);
              (b) Ratio of the energy differences given by the expression (\ref{quantity-to-be-calculated-bosons}), calculated 
              respectively for the non-self-consistent ($\Delta \mathcal{E}$) and self-consistent 
              ($\Delta \mathcal{E}_{\mathrm{sc}}$) solutions of the GP equation, vs the variable $\tau$ defined in 
              Eq.(\ref{perturbation-from-GP-equation}). The different curves correspond to the values 
              $q \xi_{\mathrm{phase}} = (0.1,0.2,0.3,0.345,0.1,0.2,0.3)$ from top to bottom.}
\label{fig33}
\end{figure}
\end{center}

As an example, we then consider the solution of the GP equation (\ref{Gross-Pitaevskii-equation}) in the presence of a Gaussian barrier of width $\sigma /\xi_{\mathrm{phase}} = 3.0$ and height $V_{0}/E_{F} = 0.7$ (in fermionic terms, this corresponds to the coupling value $(k_{F} a_{F})^{-1} = +3.0$ such that $k_{F} \xi_{\mathrm{phase}} = 3 \sqrt{\pi}/4$).
The resulting Josephson characteristic is shown in Fig.\ref{fig33}(a), whose maximum occurs for 
$\delta \phi_{c}/\pi = 0.3$ and $q_{c} \xi_{\mathrm{phase}} = 0.345$.
Note that in the figure we have represented by a full line the ``left'' branch of the Josephson characteristic when $0 \le \delta \phi \le \delta \phi_{c}$ and by a broken line the ``right'' branch of the Josephson characteristic when $\delta \phi_{c} < \delta \phi \le \pi$, thus anticipating a relevant difference between the two parts of the Josephson characteristic.

Such a difference results evident in Fig.\ref{fig33}(b), where the quantity $\Delta \mathcal{E}$ identified by the right-hand side of Eq.(\ref{quantity-to-be-calculated-bosons}) is reported for several values of $q$ that are consistent with Fig.\ref{fig33}(a), by perturbing the self-consistent solution of the GP equation (\ref{Gross-Pitaevskii-equation}) while keeping the value of the superfluid current uniform in space at the same value $j$ of the self-consistent solution.
Specifically, the perturbed form of the condensate wave function about the self-consistent (sc) solution of the GP equation (\ref{Gross-Pitaevskii-equation}) has been chosen as follows. 
Its amplitude is varied locally according to the expression
\begin{equation}
|\Phi(x)| \, = \, |\Phi_{\mathrm{sc}}(x)| \, + \, \tau \, |\Phi_{\mathrm{sc}}(x=0)| \,\, \left(
\frac{\Phi_{0} - |\Phi_{\mathrm{sc}}(x)|}{\Phi_{0} - |\Phi_{\mathrm{sc}}(x=0)|} \right) \,\, ,  
                                                                                       \label{perturbation-from-GP-equation}
\end{equation}

\noindent
while its phase $\phi_{B}(x)$ adjusts itself locally in order to keep the current uniform.
Here, $\Phi_{0} = \sqrt{n_{0}/2}$ is the bulk value of $|\Phi_{\mathrm{sc}}(x)|$ and $\tau$ is a dimensionless parameter with a typical range $0 \le |\tau| \lapprox 1$.
In this way, the perturbed form is specified only in terms of the value of the varied solution at the center of the barrier
($x=0$).
This procedure, in turn, defines the independent variable 
$\tau = (\Delta |\Phi|/|\Phi_{\mathrm{sc}}|)_{x=0} = (|\Phi(x=0)| - |\Phi_{\mathrm{sc}}(x=0)|)/
|\Phi_{\mathrm{sc}}(x=0)|$ that controls the energy variations reported in Fig.\ref{fig33}(b). 

Other choices of variations about the self-consistent solution might as well have been considered. 
For instance, we could have kept the value of the self-consistent solution at $x=0$ unmodified and varied the spatial extension of the self-consistent solution, still preserving the value of the supercurrent uniform in space.
[We found that the requirement of the current being uniform is essential for a correct identification of the energy stability about the self-consistent solution.]
We again emphasize that the requirement of the current being uniform for all possible variations about the self-consistent
solution necessarily forces the value $\delta \phi$ of the total phase accumulated by the non-self-consistent solution across the barrier to change with respect to the value of the self-consistent solution.
Correspondingly, this change need be considered also for the quantity $\Delta \phi = \delta \phi/2$ appearing on the right-hand side of Eq.(\ref{quantity-to-be-calculated-bosons}) and associated with the solution in the absence of the barrier.

The curves $\Delta \mathcal{E}(\tau)$ reported in Fig.\ref{fig33}(b) correspond to several values of $q$ associated with the self-consistent solution of Fig.\ref{fig33}(a). 
We note, in particular, that the full curves in Fig.\ref{fig33}(b) show a \emph{minimum} for $\tau = 0$ (which is associated with the self-consistent solution) for values of $q$ such that $\delta \phi \le \delta \phi_{c}$ in Fig.\ref{fig33}(a). 
On the other hand, the dashed curves show a \emph{maximum} for $\tau = 0$ for values of $q$ such that $\delta \phi_{c} < \delta \phi$ in Fig.\ref{fig33}(a).
Correspondingly, we conclude that the left part of the Josephson characteristics is \emph{stable} against fluctuations (at least, for the class of fluctuations  here considered), while the right part of the Josephson characteristics is \emph{unstable}.
This conclusion is in agreement with the results of Ref.\cite{Hakim-1997} for a delta-like barrier.

\begin{center}
\begin{figure}[htc]
\includegraphics[angle=0,width=6.5in]{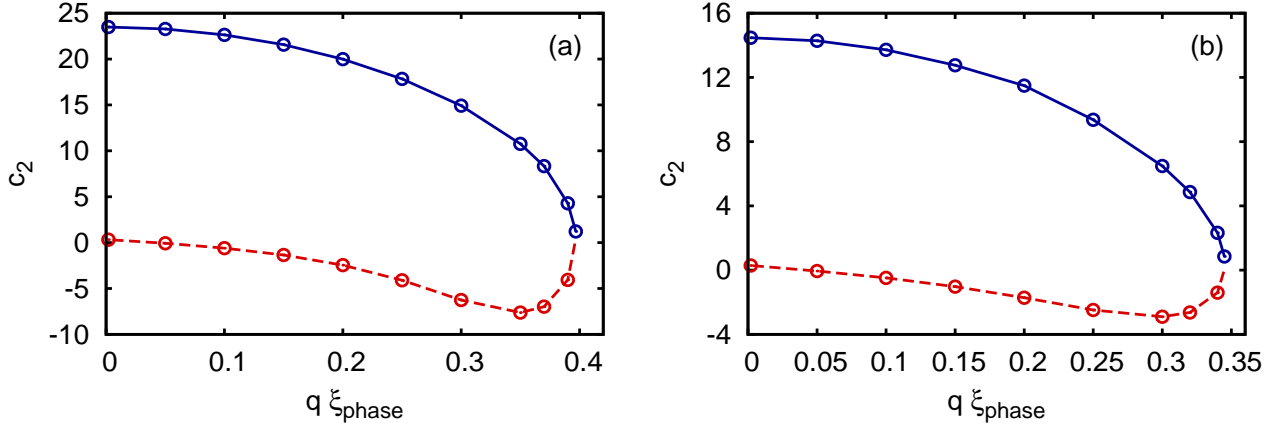}
\caption{Coefficient $c_{2}$ of the polynomial expansion (\ref{polynomial-expansion}) vs $q \xi_{\mathrm{phase}}$ 
              corresponding to the set of curves of Fig.\ref{fig33}(b). Results are reported for two different Gaussian barriers 
              with $V_{0}/E_{F} =0.7$ and:
             (a) $\sigma/\xi_{\mathrm{phase}} = 1.0$;
             (b) $\sigma/\xi_{\mathrm{phase}} = 3.0$.}
\label{fig34}
\end{figure}
\end{center}

In order to fine tune the actual value of $q$ past which the self-consistent solution of the GP equation becomes unstable, 
we report in Fig.\ref{fig34} the behavior vs $q$ of the coefficient $c_{2}$ of the polynomial expansion
\begin{equation}
\frac{\Delta E(\tau)}{\Delta E_{\mathrm{self}}} \, = \, 1 + c_{1} \tau + c_{2} \tau^{2} +
c_{3} \tau^{3} + c_{4} \tau^{4} + c_{5} \tau^{5}                           \label{polynomial-expansion}
\end{equation}

\noindent
which fits best the curves of Fig.\ref{fig33}(b). 
[In general, the need to include terms up to the fifth order in $\tau$ stems from the analysis of the analogous quantity in the
BCS limit, but we leave it here also in the BEC limit for shorting up the discussion.]
Results for two different Gaussian barriers are reported.

\noindent
One sees from these plots that $c_{2}$ remains positive as long as $\delta \phi \le \delta \phi_{c}$, while
it turns negative when $\delta \phi_{c} < \delta \phi$, thus confirming our expectation.
[The feature that $c_{2}$ returns slightly being positive for $\delta \phi \simeq \pi$ is interpreted as due to our numerical error, 
which is estimated from this deviation to be less than $3 \%$.]

Once the stability of the self-consistent solutions of the bosonic GP equation (\ref{Gross-Pitaevskii-equation}) has been established in the presence of a supercurrent for barriers of finite width, we pass to consider the self-consistent solutions of the fermionic BdG equations (\ref{BdG-equations}) for arbitrary couplings across the BCS-BEC crossover and their variations from self-consistency.
To this end, we consider variations of the local gap parameter $\Delta(x)$ still of the form (\ref{perturbation-from-GP-equation}) with $\Delta(x)$ now replacing $\Phi(x)$, which in turn leads to modified values of the wave functions $u_{\nu}(x)$ and $v_{\nu}(x)$ (with the phase $2\phi(x)$ of the local gap parameter being also modified in such a way that the value of the supercurrent is kept uniform).
\begin{center}
\begin{figure}[htc]
\hspace{3cm}\includegraphics[angle=0,width=3.5in]{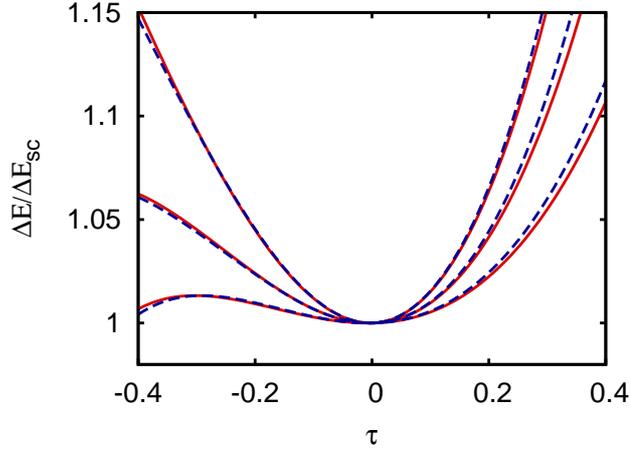}
\caption{The quantity $\Delta \mathcal{E}/\Delta \mathcal{E}_{\mathrm{sc}}$ vs $\tau$ obtained (dashed lines)
              from the solution of the bosonic GP equation as in Fig.\ref{fig33}(b), is compared with the corresponding 
              quantity obtained (full lines) from the solution of the fermionic BdG equations for the coupling 
              $(k_{F} a_{F})^{-1} = +3.0$ and the values $q/k_{F} = (0.01,0.04,0.06)$ from top to bottom.}
\label{fig35}
\end{figure}
\end{center} 
\begin{center}
\begin{figure}[htc]
\includegraphics[angle=0,width=6.5in]{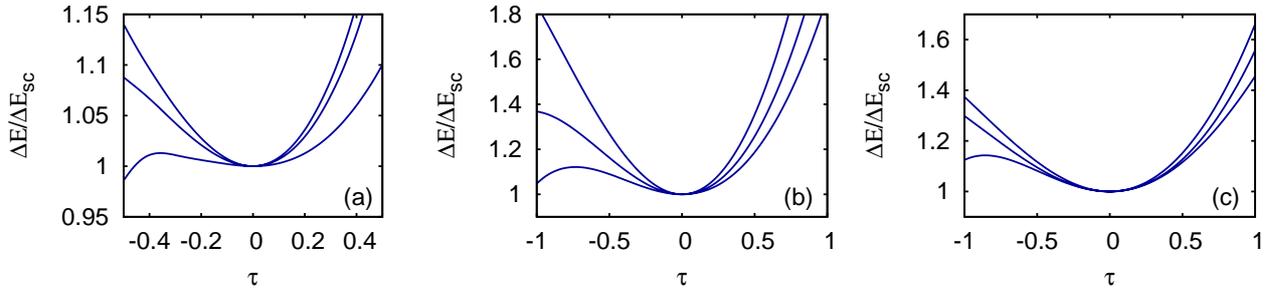}
\caption{The quantity $\Delta \mathcal{E}/\Delta \mathcal{E}_{\mathrm{sc}}$ is plotted vs $\tau$ for a Gaussian barrier 
              of width $\sigma k_{F} = 4.0$ and height $V_{0}/E_{F} = 0.1$ and three values of the coupling $(k_{F} a_{F})^{-1}$ 
              across the BCS-BEC crossover: (a) +1.0; (b) 0.0; (c) -1.0. 
              For each coupling, the results corresponding to three values of $q/k_{F}$ are reported (from top to bottom):
              (a) $0.01,0.04,0.06$ (with $q_{c}/k_{F}=0.061$);
              (b) $0.01,0.15,0.20$ (with $q_{c}/k_{F}=0.21$);
              (c) $0.01,0.05,0.07$ (with $q_{c}/k_{F}=0.076$).}
\label{fig36}
\end{figure}
\end{center} 
\vspace{-1.0cm}

Following a procedure that we have adopted several times in this paper, we begin by exploring the BCS-BEC crossover for the quantity $\Delta \mathcal{E}(\tau)$ by considering again a coupling value ($(k_{F} a_{F})^{-1} = + 3.0$) deep in the BEC region, in such a way that a direct comparison can be established with the results obtained previously via the GP equation.

We thus report in Fig.\ref{fig35} the results for $\Delta \mathcal{E}(\tau)$ using a Gaussian barrier of width 
$\sigma k_{F} = 5.0$ and height $V_{0}/E_{F} = 0.03$, obtained for different values of the supercurrent and using the GP equation (dashed lines) as well as the BdG equations (full lines).
The good agreement resulting from these two sets of curves gives us confidence in the accuracy of the solution of the BdG equations even as far as the quantity $\Delta \mathcal{E}(\tau)$ is concerned.
Here and in the following, only values of $q$ such that $\delta \phi < \delta \phi_{c}$ corresponding to the stable solutions
will be considered.

Accordingly, we report in Fig.\ref{fig36} the curves $\Delta E(\tau)$ for the three characteristic couplings
$(k_{F} a_{F})^{-1} = (+1.0, 0.0, -1.0)$ across the BCS-BEC crossover, 
using a Gaussian barrier of width $\sigma k_{F} = 4.0$ and height $V_{0}/E_{F} = 0.1$.
Curves corresponding to three different values of $q$ are reported for each coupling. 
In all cases, we have verified that \emph{the energy stability of the solutions is lost as soon as $q$ approaches its critical value $q_{c}$ corresponding to the top of the Josephson characteristic} for the given barrier.
More precisely, to extract the accurate values of $q$ at which the instability in $\Delta \mathcal{E}(\tau)$ begins to manifest, we have also studied the $q$-evolution of the coefficient $c_{2}$ in a polynomial fit of the type (\ref{polynomial-expansion}).

\begin{center}
\begin{figure}[htc]
\hspace{3cm}\includegraphics[angle=0,width=3.5in]{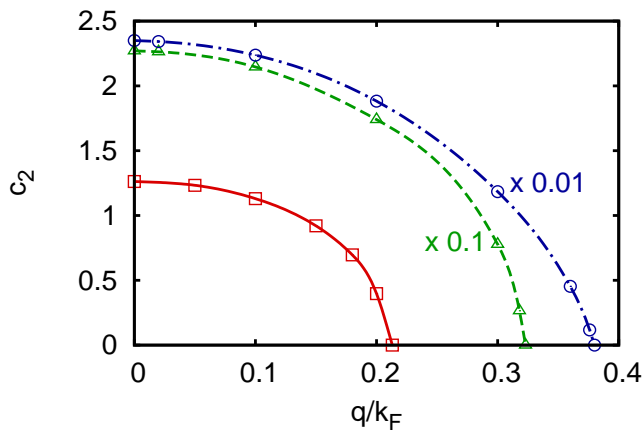}
\caption{Coefficient $c_{2}$ of the polynomial expansion (\ref{polynomial-expansion}) vs the wave vector $q$ of 
              the supercurrent at unitarity for a Gaussian barrier of width $\sigma k_{F} = 4.0$ and decreasing values 
              of the height $V_{0}/E_{F}$: $0.1$ (red squares); $0.01$ (green triangles); $0.001$ (blue circles).
              [The data have been multiplied by the indicated factors in order to make a common drawing of the three curves.]
              The critical value $q_ {c}/k_{F} = 0.380$ at which $c_{2}$ changes sign in the limit $V_{0}/E_{F} \rightarrow 0^{+}$     
              compares favorably with the corresponding value $q_ {c}^{\mathrm{Landau}}/k_{F} = 0.397$ resulting from the    
              Landau criterion at unitarity.}
\label{fig37}
\end{figure}
\end{center}

The above analysis can be repeated for progressively smaller barrier heights, up to the point that the conditions occur for the Landau criterion discussed in sub-section 4f to emerge.
To this end, we plot in Fig.\ref{fig37} the behavior of the coefficient $c_{2}$ of the polynomial expression (\ref{polynomial-expansion}) vs $q$ at unitarity for progressively decreasing values of $V_{0}/E_{F}$, while keeping the width $\sigma k_{F}$ of the Gaussian barrier fixed.
We see from this figure that, when the critical value $q_{c}$ (at which the coefficient $c_{2}$ changes sign) approaches the Landau value $q_{c}^{\mathrm{Landau}}$ as soon as $V_{0}/E_{F} \rightarrow 0^{+}$ for the given coupling, the curves $\Delta \mathcal{E}(\tau)$ become indeed progressively more unstable (this process becomes actually quicker as soon as 
$q_{c}^{\mathrm{Landau}}$ is approached). 

\noindent
These results conclude our analysis about the energy stability of the self-consistent solutions of the BdG equations in the presence of a barrier across
which a supercurrent flows.
This analysis, in turn, reinforces the validity of these mean-field solutions, to the extent that they are thus expected to somewhat survive the inclusion of fluctuations beyond mean field.
In addition, the same analysis, when extrapolated to vanishingly small barriers, has also been proved to be consistent with the insurgence of the Landau instability in our numerical solutions of the BdG equations.
This is \emph{a priori} a nontrivial result, which then reinforces the validity of both analyses we have made about the energy and Landau instabilities.
 
\vspace{0.5cm}  
\section{Concluding remarks}

In this paper, we have presented the results of a systematic numerical solution of the Bogoliubov-de Gennes equations at zero temperature for superfluid fermions with an arbitrary mutual interaction, adopting a spatial geometry (the presence of a barrier of finite width) and a boundary condition (the presence of a steady superfluid flow) which are characteristic of the Josephson and related effects. Reaching full self-consistency in the calculations has proved essential to obtain meaningful results, especially when approaching the BEC side of the crossover.

This work represents the first systematic study of the Josephson and related effects throughout the BCS-BEC crossover, a topic which is actively studied at present both experimentally and theoretically. 
Extending the study of the Josephson effect with \emph{continuity} from the BCS (weak-coupling) limit, where it has been traditionally confined, across the unitary (intermediate-coupling) region and up to the BEC (strong-coupling) limit, where coherence effects based on (composite) bosons eventually emerge, has allowed us to cast the physics of the Josephson effect under a more general and sometimes  different perspective. 
Specific questions can be focused on in this context, which would not even emerge otherwise. 
We refer, in particular, to our findings about the role played by the Landau criterion for the lost of superfluidity when the relevant elementary excitations are generated in the limit of a vanishing barrier, and about the role played by the Andreev bound states in the presence of a finite barrier. 

Our emphasis on the achievement of self-consistency in the solutions of the BdG equations has made the \emph{nonlinear\/} effects typical of a collective flow to emerge through the self-consistent requirement itself, while at each step of self-consistency the BdG equations were solved by scattering methods which are standard in the solution of the linear Schr\"{o}dinger equation.

Implementation of the computational scheme has required us to develop rather sophisticated numerical procedures. Also for this reason, it was important to rely on independent ``benchmarks'' with which the outcomes of the numerical calculations could be confronted.
In the BCS (weak-coupling) limit of the crossover, when the size of the Cooper pairs by far exceeds the barrier width, the requirement that all barriers should be assimilated to a Dirac-delta barrier has represented a stringent test on the calculation.
In the BEC (strong-coupling) limit of the crossover, on the other hand, the results of the numerical solution of the fermionic Bogoliubov-de Gennes equations have been favorably confronted with those obtained by the independent solution of the Gross-Pitaevskii equation for the composite bosons that form in that limit. This joining of the physics of the fermionic BdG equations to that of the bosonic GP equation represents \emph{per se} an important step forward toward a unified understanding of coherence effects related to the broken gauge symmetry, especially as far as their practical implementation is concerned.

The results obtained in this paper could be extended along several directions, besides a straightforward application of the present approach to more complicated geometries. In this context, an extension of the present approach to \emph{finite temperature}, while still spanning the fermionic coupling from the weak (BCS) to the strong (BEC) limits, would require one to consider (at least) pairing fluctuations beyond mean field. In this way one should be able to recover the correct physics about the Bose-Einstein transition temperature in the BEC limit, as well as the physics of the pseudo-gap above the critical temperature in the unitary region. To achieve this goal, a formulation of the theory in terms of spatially dependent single-particle Green's functions should be preferred, in an analogous fashion to what we did in sub-section 5a relatively to the expression for the total energy in the BEC limit.

Consideration of the \emph{time dependence} for the Josephson effect would also considerably broaden the impact and application range of the theoretical results. This is partly due to the experimental setup utilized with ultra-cold Fermi atoms, whereby the presence of localized traps makes it preferable the use of initial and/or boundary conditions leading to the onset of time-dependent effects. In addition, a general study of non-equilibrium time-dependent phenomena, which are intrinsically related to the building up of coherence and the onset of decoherence in macroscopic quantum systems, would constitute a significant task that could focus on the importance that macroscopic quantum phenomena are expected to acquire in the planning of future technology.
In this context, one may anticipate that the time-dependent versions of the BdG equations for fermions and the GP equation for bosons will serve as starting points for the study of time-dependent macroscopic quantum phenomena throughout the BCS-BEC crossover, along similar lines to those followed in the present paper for the study of time-independent phenomena. In this respect, the stationary results obtained in the present paper could be used as initial and/or final conditions for time-dependent effects.

\section{Acknowledgments}

Discussions with G. Deutscher, R. Hulet, W. Ketterle, F. Pistolesi, S. Stringari, and A. Tagliacozzo are gratefully acknowledged.
This work was partially supported by the Italian MIUR under research contracts PRIN-2005 ``Ultracold Fermi Gases and Optical Lattices'' and PRIN-2007 ``Ultracold Atoms and Novel Quantum Phases''.

\section{Appendix A.  A conditionally convergent integral}

In sub-section 2b we have discussed in detail the procedure for constructing a complete set of eigensolutions of the BdG equations (\ref{BdG-equations}) in the presence of a supercurrent, to be used in the gap equation (or self-consistency condition) (\ref{self-consistency}) as well as in the number (\ref{density}) and current (\ref{current}) equations.
Apart form the bound states which may exist below the continuum threshold, in each of the remaining five energy ranges specified in Figs.\ref{fig4}(a) and \ref{fig4}(b) we have used the appropriate wave vectors 
$k^{(n)}$ in the outermost spatial intervals to label the continuum eigensolutions subject to outgoing boundary conditions.

Under these circumstances, we have found it appropriate to normalize the continuum eigensolutions in terms of the wave-vector components $k=k_{x} \equiv k_{\perp}$ and $(k_{y},k_{z}) \equiv \mathbf{k}_{\parallel}$, where $\perp$ and $\parallel$ identify the directions ``parallel'' and ``orthogonal'' to the surface that fixes the geometry of the three-dimensional barrier
(recall that $k_{\perp}$ coincides, by definition, with $k_{\ell}$ of Eq.(\ref{eigeinvalue-E}) in the outermost intervals where $V_{\ell}=0$).
In the numerical solutions of the BdG equations the wave vector $\mathbf{k}_{\parallel}$ appears as a parameter through the definition of the reduced chemical potential (\ref{mu-tilde}).
The scattering problem is thus implemented by fixing the value of $\mathbf{k}_{\parallel}$ first and then solving for the values of $k_{\perp}$ which extend in principle from $- \infty$ to $+ \infty$.

With these premises, it would appear natural to perform the integral over the three-dimensional wave vector
$\mathbf{k}=(k_{\perp},\mathbf{k}_{\parallel})$, by integrating first over $k_{\perp}$ and then over $\mathbf{k}_{\parallel}$.
Nonetheless, the opposite order of the integrations could, at least in principle, be chosen as well.
We shall show in the following, however, that the order of the integrations matters considerably, with the result that the intergral turns out to be convergent only when the integration over 
$k_{\perp}$ is performed first.
On physical grounds, this difference stems by the different role played by the presence of the barrier when one is considering the two alternative limits, namely, of keeping $\mathbf{k}_{\parallel}$ fixed while letting $|k_{\perp}|$ to increase without bound, or doing the opposite of keeping $|k_{\perp}|$ fixed while letting $\mathbf{k}_{\parallel}$ to increase without bound.
In the first situation the presence of the barrier should become progressively irrelevant, while in the second one the barrier keeps influencing the scattering wave functions no matter how large $\mathbf{k}_{\parallel}$ is.

Let us refer specifically to the gap equation at zero temperature in its regularized version (\ref{final-regularized-gap-equation}).
Apart from the discrete sum over the bound states (if any), the sum over the index $\nu$ therein reduces to an integral over 
$\mathbf{k}=(k_{\perp},\mathbf{k}_{\parallel})$, where in the integral over $k_{\perp}$ (which extends from $- \infty$ to $+ \infty$) the integrand takes alternative forms depending on the energy range it is spanning.
For instance, in range 4 of Fig.\ref{fig4}(a) when $\tilde{\mu} > 0$, $k_{\perp}$ coincides with 
$k^{(2)}$ in the left branch and with $k^{(1)}$ in the right branch of that figure, and the wave functions take 
the form (\ref{Psi-electron-right-range-3}) or (\ref{Psi-electron-left-range-3}), in the order.
A similar situation occurs in range 7 of Fig.\ref{fig4}(b) when $\tilde{\mu} < 0$.

\begin{center}
\begin{figure}[htc]
\hspace{3cm}\includegraphics[angle=0,width=3.5in]{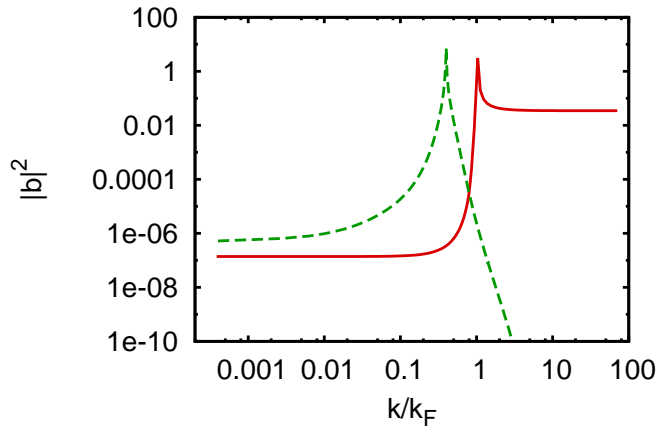}
\caption{Magnitude square of the coefficient $b$ of the wave function (\ref{Psi-vs-Upsilon}) in $x=0$ corresponding 
              to setting $a_{1}=1$ in the outermost left interval. The two curves correspond to keeping fixed either 
              $k_{\perp}=k_{F}$ (full line) or $|\mathbf{k}_{\parallel}|=k_{F}$ (dashed line) and taking the independent variable be 
              $|\mathbf{k}_{\parallel}|$ or $k_{\perp}$, respectively. 
              For definiteness, we have considered the coupling value $(k_{F} a_{F})^{-1} = -1.0$ for which $\mu>0$, while 
              $\tilde{\mu}$ can acquire both signs.
              The calculation has been done for a Gaussian barrier with $\sigma k_{F} = 3.5$ and $V_{0}/E_{F}= 0.2$, and for
              a current with $q/k_{F}=0.03$.}
\label{fig38}
\end{figure}
\end{center}

To evidence the effects of the barrier in the two alternative situations discussed above (that is, of either keeping $\mathbf{k}_{\parallel}$ fixed while letting $|k_{\perp}|$ increasing without bound or doing the opposite), we plot as an example in Fig.\ref{fig38} the (square magnitude of the) coefficient $b$ of the wave function (\ref{Psi-vs-Upsilon}) in the central spatial interval (namely, at $x=0$ where the potential barrier is maximum), which corresponds to an electron-like excitation impinging from the left (that is, with $a_{1}=1$ in the outermost left interval).
We see from this figure that, when $\mathbf{k}_{\parallel}$ is kept fixed and $k_{\perp}$ is allowed to grow without bound
(dashed line), $|b|^{2}$ becomes exponentially small, thus showing that the barrier has no influence on the wave functions in this limit.
In the reverse situation, when $k_{\perp}$ is kept fixed and $|\mathbf{k}_{\parallel}|$ is allowed to grow without bound (full line), $|b|^{2}$ approaches a constant value that depends on $k_{\perp}$ and is in general not negligible.

We consider first the convergence of the integral over $k_{\perp}$ in 
Eq.(\ref{final-regularized-gap-equation}) in the barrier region, for large values of $|k_{\perp}|$ and given value of $\mathbf{k}_{\parallel}$.
In this limit, we have just concluded that the barrier does not influence the wave function appreciably.
For instance, in the expression (\ref{Psi-vs-Upsilon}) the coefficient $a \rightarrow1$ while all other coefficients become negligible.
It is then sufficient to consider the expressions (\ref{u2-v2-discretized} ) for the coefficients 
$u_{\ell}$ and $v_{\ell}$ entering the wave function $\Upsilon_{\ell}^{(1)}(x=0;q,E)$, in which we rewrite:
\begin{equation}
E + \xi_{k_{\ell}-q} = 
\frac{(k_{\ell}^{2} + \mathbf{k}_{\parallel}^{2})}{2m} + V_{\ell} - \mu_{0} + 
\sqrt{\left(\frac{(k_{\ell}^{2} + \mathbf{k}_{\parallel}^{2})}{2m} + V_{\ell} -
\mu_{0} \right)^{2} + \Delta_{\ell}^{2}}   \,\, .                 \label{E-plus-csi}
\end{equation}

\noindent
[Note that far from the barrier where $V_{\ell}=0$, $\Delta_{\ell}=\Delta_{0}$, and 
$k_{\ell}=k_{\perp}$, this expression reduces to that for the homogeneous case with $q=0$.]
For large values of $|k_{\perp}|$ (and therefore of $k_{\ell}$) we then obtain:
\begin{equation}
u_{\ell} \, v_{\ell} \, \simeq \, \frac{\Delta_{\ell}}{2 \, \epsilon_{\ell}} \, - \, 
\frac{\Delta_{\ell}^{3}}{4 \, \epsilon_{\ell}^{3}} \, + \, \cdots     \label{approximate-uv-inside-barrier}
\end{equation}

\noindent
where we have introduced the short-hand notation 
$\epsilon_{\ell} = (k_{\ell}^{2}+\mathbf{k}_{\parallel}^{2})/(2m) + V_{\ell} - \mu_{0}$.
Setting further $k_{\ell}^{2} = k_{\perp}^{2} + (k_{\ell}^{2} - k_{\perp}^{2}) \equiv k_{\perp}^{2}
+ \delta_{\ell}^{2}$, we expand:
\begin{eqnarray}
\frac{1}{\epsilon_{\ell}} = \left( \frac{\mathbf{k}^{2}}{2m} - \mu_{0} +
\frac{\delta_{\ell}^{2}}{2m} + V_{\ell} \right)^{-1}  & \simeq &
\left( \frac{\mathbf{k}^{2}}{2m} - \mu_{0} \right)^{-1}            \nonumber \\
& - & \left(\frac{\delta_{\ell}^{2}}{2m} + V_{\ell} \right) 
\left( \frac{\mathbf{k}^{2}}{2m} - \mu_{0} \right)^{-2}            \label{approximate-1-epsilon}
\end{eqnarray}

\noindent
with $\mathbf{k}^{2} = k_{\perp}^{2} + \mathbf{k}_{\parallel}^{2}$.
Equation (\ref{approximate-uv-inside-barrier}) thus becomes:
\begin{equation}
u_{\ell} \, v_{\ell} \simeq \frac{\Delta_{\ell}}{2 \left( \frac{\mathbf{k}^{2}}{2m} - \mu_{0} \right)} 
- \frac{\Delta_{\ell} \left(\frac{\delta_{\ell}^{2}}{2m} + V_{\ell} \right)}
{2 \left( \frac{\mathbf{k}^{2}}{2m} - \mu_{0} \right)^{2}} 
- \frac{\Delta_{\ell}^{3}}{4 \, \left( \frac{\mathbf{k}^{2}}{2m} - \mu_{0} \right)^{3}} \, + \, \cdots \,\, .
                                                                        \label{more-approximate-uv-inside-barrier}
\end{equation}

\noindent

Here, the first term on the right-hand side when inserted into 
Eq.(\ref{final-regularized-gap-equation}) cancels the leading term of the second term on the right-hand side of that equation.
In the numerator of the second term on the right-hand side of 
Eq.(\ref{more-approximate-uv-inside-barrier}), on the other hand, when $|k_{\perp}|$ is large enough we may neglect $\delta_{\ell}^{2}/(2m)$ in comparison with $V_{\ell} \ne 0$.
In this limit, the leading term on the right-hand side of Eq.(\ref{final-regularized-gap-equation}) reduces to:
\begin{eqnarray}
& - & \, \frac{1}{(2 \pi)^{2}} \, \int_{k_{c}}^{\infty} \! dk_{\parallel} \, k_{\parallel} \, 
\int_{k_{c}}^{\infty} \! dk_{\perp} \,  \frac{\Delta_{\ell} \, V_{\ell}}
{2 \left( \frac{\mathbf{k}^{2}}{2m} - \mu_{0} \right)^{2}}           \nonumber \\
& \approx & 
- \, \frac{m^{2} \, \Delta_{\ell} \, V_{\ell}}{8 \pi^{2}} \, \int_{k_{c}}^{\infty} \! dk_{\parallel} \, k_{\parallel} \, \frac{1}{\left( k_{\parallel}^{2} \, - \, 2m \, \mu_{0} \right)^{3/2}}
                                                                                                \label{integral-k-perp-first-ultraviolet}
\end{eqnarray}

\noindent
where for convenience we have introduced in both integrals a common lower cutoff $k_{c}$ (such that 
$k_{c}^{2} > 2m \mu_{0}$ when $\mu_{0} >0$) since we are here interested only in the ultraviolet convergence of the integrals.
This result proves analytically that the double integral in Eq.(\ref{final-regularized-gap-equation}) 
is definitely convergent when the integration over $k_{\perp}$ is performed first.
We have also verified from the fully numerical calculation that the same $k_{\parallel}^{-2}$ power-law behavior of 
Eq.(\ref{integral-k-perp-first-ultraviolet}) results in the ultraviolet for the integrand of the last integral over
$k_{\parallel}$ in Eq.(\ref{final-regularized-gap-equation}) inside the barrier (outside the barrier, on the other hand, an even faster decay results since the subtraction of the homogeneous contribution in Eq.(\ref{final-regularized-gap-equation}) is now more effective
there).

Let's, finally, consider the opposite case when $k_{\perp}$ is kept fixed and 
$|\mathbf{k}_{\parallel}|$ is allowed to grow without bound.
In this limit, we have argued that the effect of the barrier cannot altogether be neglected, 
since the scattering coefficients are in general non negligible for large values of $k_{\perp}$ where they approach \emph{constant} values that depend on $k_{\perp}$ (cf. Fig.\ref{fig38} for $|b|^{2}$ inside the barrier as an example).
In this case, it is sufficient to demonstrate that the double integral over $k_{\perp}$ and
$\mathbf{k}_{\parallel}$ in Eq.(\ref{final-regularized-gap-equation}) does not converge when the integration over $\mathbf{k}_{\parallel}$ is performed first, by considering that equation far from the barrier (for example, for $x \approx - \infty$) as this was actually the most favorable situation for the convergence of the integral when the integration over $\mathbf{k}_{\perp}$ was performed first.
We may also consider, for instance, the simple case of a rectangular barrier of width $L$ and height $V_{0}$ for which the scattering coefficients can be calculated analytically when 
$\mathbf{k}_{\parallel}^{2}/(2m)$ is the largest energy scale in the problem.
In the place of Eq.(\ref{integral-k-perp-first-ultraviolet}), we obtain the following expression for the leading term on the right-hand side of Eq.(\ref{final-regularized-gap-equation}) when 
$x \approx - \infty$:
\begin{equation}
\frac{1}{(2 \pi)^{2}} \, \int_{k_{c}}^{\infty} \! dk_{\perp} \, 
Re \left\{ b_{1}(k_{\perp}) e^{-i2k_{\perp}x} \right\}
\int_{k_{c}}^{\infty} \! dk_{\parallel} \, k_{\parallel} \,  \frac{\Delta_{0}}
{\left( \frac{\mathbf{k}^{2}}{2m} - \mu_{0} \right)}                    \label{integral-k-parallel-first-ultraviolet}
\end{equation}

\noindent
where 
\begin{equation}
|b_{1}(k_{\perp})|^{2} \, = \, \frac{(m V_{0})^{2} \, \left( 1 \, - \, \cos 2 L \sqrt{k_{\perp}^{2} - 2m V_{0}} \right)}
{2k_{\perp}^{4} - 8 k_{\perp}^{2} (m V_{0}) + (m V_{0})^{2} 
\left( 1 \, - \, \cos 2 L \sqrt{k_{\perp}^{2} - 2m V_{0}} \right)}             \label{b1-square}
\end{equation}

\noindent
and we may assume $k_{c} > \sqrt{2m V_{0}}$.
It is clear from the expression (\ref{integral-k-parallel-first-ultraviolet}) that the integral over 
$k_{\parallel}$ therein diverges logarithmically in the ultraviolet.

These results prove our initial statement that the order of the integrations over $k_{\perp}$ and 
$k_{\parallel}$ in the gap equation (\ref{final-regularized-gap-equation}) matters,
owing to the different role played by the barrier when either $k_{\perp}$ or $k_{\parallel}$ are allowed to grow first without bound.
Consistently with these results, we have implemented the numerical solution of the BdG equations (\ref{BdG-equations}) by integrating first over $k_{\perp}$, as it would anyway appear more natural on physical grounds.

\vspace{0.5cm} 
\section{Appendix B.  An analytic integral for the total current with a Dirac-delta barrier}

We have seen in sub-section 2c while discussing the \emph{non-self-consistent\/} solution of the BdG equations for a 
delta-like barrier, that the total current can be cast in the form [cf. Eq.(\ref{partial-current-delta-barrier})]:
\begin{equation}
J(\delta \phi) \, = \, \frac{1}{2 \pi} \, \int_{0}^{k_{F}} \! d k_{\parallel} \, k_{\parallel} \, J(\delta \phi;k_{\tilde{\mu}})
           \, = \, \frac{\Delta_{0} \, \sin \delta \phi}{4 \pi} \, \int_{0}^{k_{F}} \! d k_{\parallel} \, k_{\parallel}
           \frac{ T_{k_{\tilde{\mu}}}}{ \sqrt{ 1 - T_{k_{\tilde{\mu}}} \sin^{2} \, \delta \phi/2} }                               \label{total-current-delta-barrier}
\end{equation}

\noindent
where
\begin{equation}
T_{k_{\tilde{\mu}}} \, = \, \frac{1}{1 \, + \, \tilde{\mathcal{Z}}_{\tilde{\mu}}^{2}} \, = \,
\frac{1} {1 \, + \, \frac{ m^{2} \mathcal{Z}^{2} }{ k_{\tilde{\mu}}^{2} }} \, = \, 
\frac{1 \, - \, \mathbf{k}^{2}}{\frac{1}{T_{k_{F}}} \, - \, \mathbf{k}^{2}}  \,\, .                                     \label{T-k-mu-manipulated}
\end{equation}

\noindent
Here, we have set $\mathbf{k} = \mathbf{k}_{\parallel}/k_{F}$ and
\begin{equation}
T_{k_{F}} \, = \, \frac{1} {1 \, + \, \frac{ m^{2} \mathcal{Z}^{2} }{ k_{F}^{2} }} 
\, = \, \frac{1}{1 \, + \, \tilde{\mathcal{Z}}_{F}^{2}} \,\le \, 1 \,\, .                                                           \label{definition-T-F}
\end{equation}

\noindent
Simple manipulations then bring Eq.(\ref{total-current-delta-barrier}) into the form:
\begin{eqnarray}
J(\delta \phi) & = & \frac{k_{F}^{2} \, \Delta_{0} \, \sin \delta \phi}{4 \pi} \, 
\frac{T_{k_{F}}}{\sqrt{1 - T_{k_{F}} \sin^{2} \delta \phi/2}}                                  \nonumber \\
& \times & \int_{0}^{1} \! d k \, k \,\, \frac{( 1 \, - \, k^{2})}{\sqrt{1 \, - \, T_{k_{F}} \, k^{2}} \, 
\sqrt{1 \, - \, \frac{T_{k_{F}} \cos^{2} \delta \phi/2}{1 \, - \, T_{k_{F}} \sin^{2} \delta \phi/2} \, k^{2}}}  \,\, .
                                                                           \label{total-current-delta-barrier-manipulated}
\end{eqnarray}

\noindent
Identifying the two new quantities 
\begin{equation}
X = T_{k_{F}}  \,\,\,\,\,\,\,\,\,\,\,\,\,\,  \mathrm{and}  \,\,\,\,\,\,\,\,\,\,\,\,\,\, 
Y = \frac{T_{k_{F}} \cos^{2} \delta \phi/2}{ 1 - T_{k_{F}} \sin^{2} \delta \phi/2}                  \label{variables-X-Y}
\end{equation}

\noindent 
and changing the integration variable from $k$ to $t = k^{2}$, Eq.(\ref{total-current-delta-barrier-manipulated}) 
reduces eventually to the compact form:
\begin{equation}
J(\delta \phi) \, = \, \frac{k_{F}^{2} \, \Delta_{0} \, \sin \delta \phi}{8 \pi} \, 
\frac{T_{k_{F}}}{\sqrt{1 - T_{k_{F}} \sin^{2} \delta \phi/2}} \,\, 
\int_{0}^{1} \! d t \,\, \frac{( 1 \, - \, t)}{\sqrt{1 \, - \, X \, t} \, \sqrt{1 \, - \, Y \, t}}  \,\, .
                                                                    \label{total-current-delta-barrier-simplified}
\end{equation}
\begin{center}
\begin{figure}[t]
\hspace{3cm}\includegraphics[angle=0,width=3.5in]{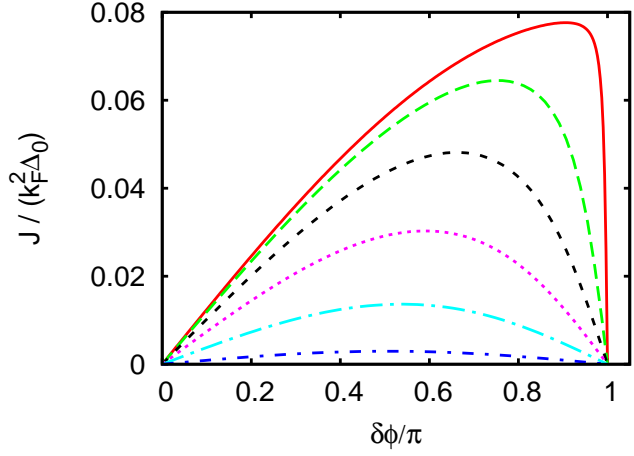}
\caption{Josephson characteristics obtained from the expression (\ref{total-current-delta-barrier-compact}) for 
              intermediate values of $T_{k_{F}}$ defined by Eq.(\ref{definition-T-F}). 
              Here, $\tilde{\mathcal{Z}}_{F}=(0.01,0.10,0.25,0.50,1.00,2.50)$ from the top to the bottom curve.}
\label{fig39}
\end{figure}
\end{center} 
In this expression one recognizes an integral representation of the Appell hypergeometric function of two variables \cite{WW}
\begin{equation}
F_{1}(\alpha,\beta,\beta',\gamma;X,Y) \, = \, 
\frac{\Gamma(\gamma)}{\Gamma(\alpha) \Gamma(\gamma - \alpha)}
\int_{0}^{1} \! d t \,\, t^{\alpha - 1} \, 
\frac{(1 - t)^{\gamma - \alpha -1}}{(1 - t \, X)^{\beta} \,\, (1 - t \, Y)^{\beta'}}
                                                \label{Appell-hypergeometric-function-integral}
\end{equation}

\noindent
when $\alpha = 1$, $\beta = \beta' = 1/2$, and $\gamma = 3$.
In Eq.(\ref{Appell-hypergeometric-function-integral}), $\Gamma(z)$ is the Euler's Gamma Function 
(such that $\Gamma(n+1)=n!$ for integer values of its argument) and by definition 
$(a)_{n}=a(a+1)(a+2) \cdots (a+n-1)$ with $a_{0}=1$.
Alternatively, the Appell hypergeometric function admits the double serie representation:
\begin{equation}
F_{1}(\alpha,\beta,\beta',\gamma;X,Y) \, = \, \sum_{m=0}^{\infty} \, \sum_{n=0}^{\infty} \,
\frac{(\alpha)_{n+m} \, (\beta)_{m} \, (\beta')_{n} }{ m! \, n! \, (\gamma)_{n+m}} \,\, X^{m} \, Y^{n}
                                             \label{Appell-hypergeometric-function-series}
\end{equation}

\noindent
which is absolutely convergent for $|X|<1$ and $|Y|<1$.
In this way, the total current (\ref{total-current-delta-barrier-simplified}) takes its final form:
\begin{equation}
J(\delta \phi) \, = \, \frac{k_{F}^{2} \, \Delta_{0} \, \sin \delta \phi}{16 \pi} \, 
\frac{T_{k_{F}}}{\sqrt{1 - T_{k_{F}} \sin^{2} \delta \phi/2}}  \,\, F_{1}\left(1,\frac{1}{2},\frac{1}{2},3;X,Y\right)  
                                                                    \label{total-current-delta-barrier-compact}
\end{equation}

\noindent
with $X$ and $Y$ defined in Eq.(\ref{variables-X-Y}).

Recall that the expression (\ref{total-current-delta-barrier-compact}) holds for \emph{any} value of $T_{k_{F}}$ in the interval $0 \le T_{k_{F}} < 1$.
In particular, in the limit of a weak barrier when $T_{k_{F}} \rightarrow 1$ Eqs.(\ref{variables-X-Y})
give $X \rightarrow 1$ and $Y \rightarrow 1$, so that $F_{1}(1,1/2,1/2,3;X \rightarrow 1,Y \rightarrow 1) = 2$ and the expression 
(\ref{total-current-delta-barrier-compact}) reduces to the result (\ref{total-current-delta-barrier-weak}) obtained in the text by a different method.
Analogously, in the limit of a strong barrier when $T_{k_{F}} \approx \tilde{\mathcal{Z}}_{F}^{-2} \ll 1$ we may take $X = 0$ and $Y = 0$ in the arguments of the hypergeometric function, so that $F_{1}(1,1/2,1/2,3;X = 0,Y = 0) = 1$ and the expression (\ref{total-current-delta-barrier-compact}) reduces to the result (\ref{total-current-delta-barrier-strong}) of the text.

More generally, for intermediate values of $T_{k_{F}}$ the Josephson characteristics 
(\ref{total-current-delta-barrier-compact}) can be obtained by computing numerically the hypergeometric function therein in terms
of the series representation (\ref{Appell-hypergeometric-function-series}), as shown in Fig.\ref{fig39}. 
This plot shows the progessive evolution from $J(\delta \phi) \propto \sin \delta \phi$ for a strong barrier (that is, small $T_{k_{F}}$) to
$J(\delta \phi) \propto \sin \delta \phi/2$ for a weak barrier (that is, large $T_{k_{F}}$).
Only the results obtained from Eq.(\ref{total-current-delta-barrier-compact}) for a strong barrier survive, however, the 
inclusion of self-consistency in the calculation, as discussed extensively in sub-section 2c of the text.

\vspace{0.5cm} 
\section{Appendix C.  Absence of the reflected wave in the Josephson tunneling}

In this Appendix, we analyze the flow of a Bose-Einstein condensate against a potential barrier at zero temperature 
as described by the Gross-Pitaevskii equation, by evidencing the analogies and differences with the scattering problem 
of a single particle against the same barrier as described by the ordinary Schr\"{o}dinger equation. 

To this end, we begin by rewriting the Gross-Pitaevskii equation (\ref{Gross-Pitaevskii-equation}) 
entirely in terms of bosonic quantities as follows:
\begin{equation}
- \frac{\nabla^{2}}{2 m_{B}} \, \Phi(\mathbf{r}) + V_{B}(\mathbf{r}) \, \Phi(\mathbf{r}) +
g_{B} \, |\Phi(\mathbf{r})|^{2} \, \Phi(\mathbf{r}) 
\, = \, \mu_{B} \, \Phi(\mathbf{r})                                     \label{G-P-equation-point-bosons}                  
\end{equation}

\noindent
where we have set $g_{B} = 4 \pi a_{B} / m_{B}$.
Were not for the nonlinear term proportional to $g_{B}$ in Eq.(\ref{G-P-equation-point-bosons}), the structure of this equation would be identical to that of the (time-independent) Schr\"{o}dinger equation for a particle of mass $m_{B}$ subject to the potential $V_{B}(\mathbf{r})$ and with eigenvalue $\mu_{B}$.
The nonlinear term, however, is responsible for the occurrence of phenomena which would not be possible according to the ordinary Schr\"{o}dinger equation.

In particular, in what follows we shall focus on the occurrence of a reflected wave when a current flow impinges against a potential barrier.
While an ordinary scattering process against a fixed potential as described by the Schr\"{o}dinger equation  unavoidably results in a reflected wave (barring exceptional cases for special values of the energy), we shall argue that the flexibility introduced by the nonlinear term in the Gross-Pitaevskii equation may lead to \emph{a complete absence of the reflected wave\/}.
This is because the wave function $\Phi(\mathbf{r})$ is able to adapt itself locally in the vicinity of the potential barrier 
$V_{B}(\mathbf{r})$, with the result of shaping the induced potential proportional to $g_{B}$ in such a way to \emph{counteract the effect of the barrier} as far as the the presence of the reflected wave is concerned.
In this context, the absence of the reflected wave in the description of the condensate corresponds to the occurrence of the stationary Josephson effect of interest in this paper.
It turns out that this effect can occur provided the current does not exceed a critical value, a condition which is also familiar for the Josephson effect.

For a potential barrier with a slab geometry of the type considered in this paper, $V_{B}(\mathbf{r})$ depends only on the coordinate 
$x$ orthogonal to the barrier and the solution of Eq.(\ref{G-P-equation-point-bosons}) can be cast in the form:
\begin{equation}
\Phi(x) \, = \, \sqrt{n_{B}(x)} \,\, e^{i \phi_{B}(x)} \,\, e^{i q_{B} x}         \label{Phi-one-dimensional}
\end{equation}

\noindent
where $n_{B}(x) = |\Phi(x)|^{2}$ is the local bosonic density.
[Note that, for a barrier with a slab geometry, the coordinates parallel to the barrier disappear from the solution of the GP equation when the current flow is taken orthogonal to the barrier. This marks a difference from the solutions of the fermionic 
BdG equations, for which the coordinates parallel to the barrier play a crucial role especially in the BEC limit where they contribute in an essentially way to the formation of the composite bosons.] 
To the solution (\ref{Phi-one-dimensional}) there corresponds the current
\begin{equation}
j(x) \, = \, \frac{n_{B}(x)}{m_{B}} \, \left( q_{B} \, + \, \frac{d \phi_{B}(x)}{d x} \right) \,\, .           \label{current-GP}
\end{equation}

Far from the barrier, we expect $n_{B}(x)$ to recover its bulk value $n_{B}^{0} $ and $\phi_{B}(x)$ to reach constant values 
(say, $\phi_{B}(x \rightarrow - \infty) = 0$ and $\phi_{B}(x \rightarrow + \infty) = \phi_{B}$), such that the current (\ref{current-GP})
is given by $n_{B}^{0} \, q_{B} / m_{B}$ (corresponding to a superfluid of density $n_{B}^{0} $ flowing with velocity
$V = q_{B} / m_{B}$).
To mantain this value everywhere, the phase $\phi_{B}(x)$ in Eq.(\ref{current-GP}) has to adjust itself to the local variation of the density $n_{B}(x)$.
Under these conditions, the bosonic chemical potential in Eq.(\ref{G-P-equation-point-bosons})
has the value $\mu_{B} = g_{B} n_{B}^{0} + q_{B}^{2}/(2 m_{B})$.
Note also here that in the absence of the barrier nothing would, in principle, prevent $q_{B}$ (and thus the current) to grow without bound owing to Galilean invariance.
It is just the presence of the barrier, by breaking the translational symmetry of the system, to introduce an upper limit on the allowed 
values of $q_{B}$.
 
We thus consider the one-dimensional version of Eq.(\ref{G-P-equation-point-bosons}) 
\begin{equation}
- \frac{1}{2 m_{B}} \, \frac{d^{2} \Phi(x)}{d x^{2}} + V_{B}(x) \, \Phi(x) + g_{B} \, |\Phi(x)|^{2} \, \Phi(x) 
\, = \, \mu_{B} \, \Phi(x)                                                \label{G-P-equation-one-dimensional}                  
\end{equation}

\noindent
subject to the boundary condition \cite{Hakim-1997}:
\begin{equation}
\Phi(x) \, = \, \sqrt{n_{B}} \, e^{i q_{B} x} \, e^{i \phi_{B}(x \rightarrow \pm \infty)} 
\,\,\,\,\,\,\,\,\,\,\,\,  \mathrm{when}  \,\,\,\,\,\,\,\,\,\,\,\,  x \rightarrow \pm \infty  \,\, .    
                                                                                                      \label{boundary-conditions}
\end{equation}

\begin{center}
\begin{figure}[htc]
\includegraphics[angle=0,width=6.5in]{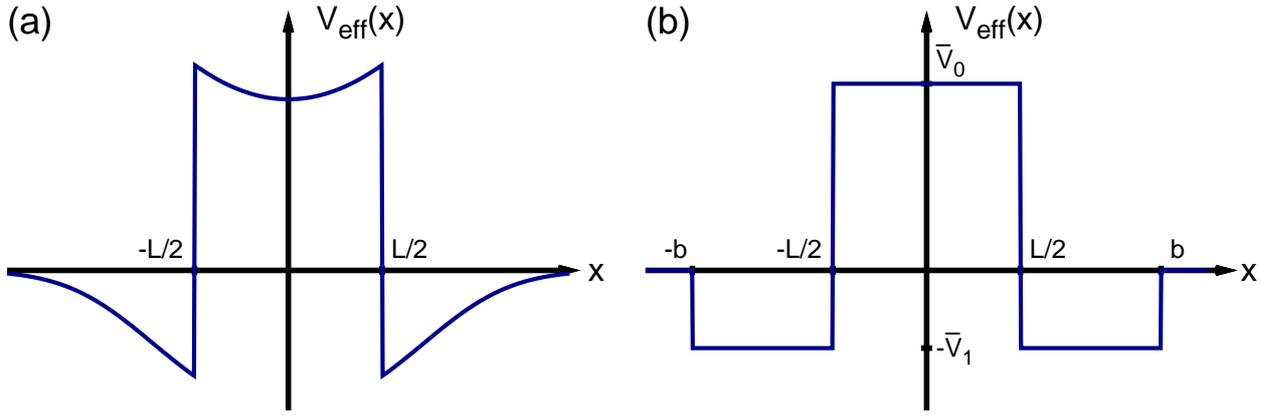}
\caption{(a) Generic shape of the effective potential $V_{\mathrm{eff}}(x)$ of Eq.(\ref{effective-potential}) when the 
              external potential $V_{B}(x)$ has the form of a rectangular barrier;
              (b) Simplified model potential (\ref{simplified-model-potential}), showing the essential features of being repulsive 
              in the central region and attractive outside.}
\label{fig40}
\end{figure}
\end{center}

\noindent
Suppose that we have found such a solution, which we label $\Phi_{q_{B}}(x)$ for brevity.
Recalling that $\mu_{B} = g_{B} n_{B}^{0} + q_{B}^{2}/(2 m_{B})$, we can manipulate
Eq.(\ref{G-P-equation-one-dimensional}) and cast it in the form of a Schr\"{o}dinger equation 
with eigenvalue $q_{B}^{2} / (2 m_{B})$:

\begin{equation}
- \frac{1}{2 m_{B}} \, \frac{d^{2} \Phi_{q_{B}}(x)}{d x^{2}} 
+ V_{\mathrm{eff}}(x) \, \Phi_{q_{B}}(x)  
\, = \, \frac{q_{B}^{2}}{2 m_{B}} \, \Phi_{q_{B}}(x)           \label{G-P-one-dimensional-manipulated}                  
\end{equation}

\noindent
with the \emph{effective potential}
\begin{equation}
V_{\mathrm{eff}}(x) \, = \, V_{B}(x) \, - \, g_{B} \, n_{B}^{0} \,
 \left( 1 \, - \, \frac{n_{B}(x)}{n_{B}^{0}} \right)                     \label{effective-potential}      
\end{equation}

\noindent
where $n_{B}(x) = |\Phi_{q_{B}}(x)|^{2}$.
Typically, $|\Phi_{q_{B}}(x)|$ (and consequently $n_{B}(x)$) is lowered from its bulk value over a length scale 
$\xi_{\mathrm{phase}}$ about the barrier, thereby producing an effective potential (\ref{effective-potential}) with the generic shape reported in Fig.\ref{fig40}(a).
The characteristic feature of this potential, which is relevant to our argument, is that it is repulsive in the central region and attractive outside.

To make the solution of the Schr\"{o}dinger equation (\ref{G-P-one-dimensional-manipulated}) with a potential of this shape
tractable, we further assimilate the potential to the simplified model form depicted in Fig.\ref{fig40}(b), namely,
 
\begin{equation} 
V_{\mathrm{eff}}(x) \, = \, \left\{ \begin{array}{ll}
\bar{V}_{0} & \mbox{if \,\, $|x| < L/2$} \\
- \bar{V}_{1} & \mbox{if \,\, $L/2 < |x| < b$} \\
0 & \mbox{otherwise}  
\end{array} \right.\label{simplified-model-potential}
\end{equation}
 
\noindent
with $\bar{V}_{0} >0$ and $\bar{V}_{1} >0$.
For this potential, five spatial regions can be identified where the solutions of Eq.(\ref{G-P-one-dimensional-manipulated}) have the local form
\begin{equation}
\Phi_{\ell}(x) \, = \, A_{\ell} \, e^{i k_{\ell} x} \, + \, B_{\ell} \, e^{- i k_{\ell} x} 
\,\,\,\,\,\,\,\,\,\,\,\,\,\,\,\,\,\,  (\ell=1,2,\cdots,5)                                    \label{local-solutions}
\end{equation}

\noindent
with $k_{1} = k_{5} = q_{B} \equiv \sqrt{2 m_{B} E}$, 
$k_{2} = k_{4} = \sqrt{2 m_{B} (E + \bar{V}_{1})}$, and
$k_{3} = \sqrt{2 m_{B} (E - \bar{V}_{0})}$.

Requiring the continuity of the functions (\ref{local-solutions}) and of their derivatives at each interface, we can express the rightmost coefficients $(A_{5},B_{5})$ in terms of the leftmost ones $(A_{1},B_{1})$ via the transfer matrix $\mathbf{Y}$ and write:
\begin{equation}
\left( \begin{array}{c} A_{5} \\ B_{5} \end{array} \right) \, = \, 
\left( \begin{array}{cc} Y_{11} & Y_{12} \\ Y_{21} & Y_{22}  \end{array} \right)
\left( \begin{array}{c} A_{1} \\ B_{1} \end{array} \right)               \label{transfer-matrix}
\end{equation}

\vspace{0.3cm}
\noindent
with $\det \mathbf{Y} = 1$.
For a wave impinging from the left we can set $A_{1}=1$ and $B_{5}=0$, yielding
$B_{1} = - Y_{21} / Y_{22}$ and $A_{5} = 1 / Y_{22}$
(the correct normalization $A_{1} \rightarrow \sqrt{n_{B}^{0}}$ will be imposed only at the end).
In addition, current conservation across the different regions requires  
$|B_{1}|^{2} + |A_{5}|^{2} = 1$, yielding $|Y_{22}|^{2} - |Y_{21}|^{2} = 1$.
In this specific example, the condition $B_{1} = 0$ for the \emph{absence of the reflected wave} (which we argued above to correspond to the occurrence of the Josephson tunneling) is then achieved when the matrix element $Y_{21}$ vanishes.

Further progress can be obtained only by numerical calculations.
To reduce the number of parameters in the calculation, and keep contact at the same time with the original problem, we take for 
$V_{B}(x)$ a rectangular barrier of height $V_{0}$ and width $L$ and relate the values of $\bar{V}_{0}$ and $\bar{V}_{1}$ in Eq.(\ref{simplified-model-potential}) to the original expression (\ref{effective-potential}) by approximating therein
$ \bar{V}_{0} \approx  V_{0} - g_{B} n_{B}^{0} ( 1 - n_{B}(0) / n_{B}^{0})$ and
$ \bar{V}_{1} \approx - g_{B} n_{B}^{0} ( 1 - n_{B}(\bar{x}) / n_{B}^{0})$, where $\bar{x}$ is an average value of $x$ in the attractive region of the effective potential.
To simplify the treatment even further, we may also take $n_{B}(\bar{x}) \approx n_{B}(0)$ so that
$ \bar{V}_{0} \approx  V_{0} - \alpha \, g_{B} n_{B}^{0}$ and 
$ \bar{V}_{1} \approx - \alpha \, g_{B} n_{B}^{0}$ with $0 < \alpha < 1$.
For fixed values of $V_{0}$, $L$, and $g_{B} n_{B}^{0}$, we can then solve for the scattering problem with a chosen set of values of $\alpha$ and $q_{B}$, thus determining the matrix elements of the transfer matrix (\ref{transfer-matrix}) vs the parameter $b$ of Fig.\ref{fig40}(b).
Quite generally, one finds that the equation $|B_{1}|(b) = 0$ admits (infinitely many) solutions only when $q_{B}$ is smaller than an upper value $q_{B}^{c}$, and that out of these solutions only the first two (that we label $I$ and $II$) with $0 < b^{(I)} < b^{(II)}$ are acceptable, as they correspond to spatial profiles of the wavefunctions which are consistent with the assumed shape of the effective potential in Fig.\ref{fig40}(a).

As a specific example, we report the numerical results obtained for a barrier with $V_{0} = 5 \, g_{B} n_{B}^{0}$ and 
$L = 1.5 \, \xi_{\mathrm{phase}}$ where $\xi_{\mathrm{phase}} = 1/\sqrt{4 m_{B} g_{B} n_{B}^{0}}$.
For such a strong barrier we expect $n_{B}(0) \approx 0$, so that we can take $\alpha = 1$ in the parametrization of the effective potential, yielding $\bar{V}_{0} =  4 \, g_{B} n_{B}^{0}$ and $ \bar{V}_{1} = - g_{B} n_{B}^{0}$.
For a given value of $q_{B}$, we first determine the two solutions $b^{(I)}(q_{B})$ and $b^{(II)}(q_{B})$ of the equation 
$|B_{1}|(b) = 0$ discussed above, in correspondence to which we obtain two distinct values for the coefficient $A_{5} = 1 / Y_{22}$.
Writing further $A_{5} = |A_{5}| \, \exp \{ i \phi_{B} \}$, we eventually determine the two corresponding values $\phi_{B}^{(I)}$ and $\phi_{B}^{(II)}$ of the phase.

\begin{center}
\begin{figure}[htc]
\includegraphics[angle=0,width=6.5in]{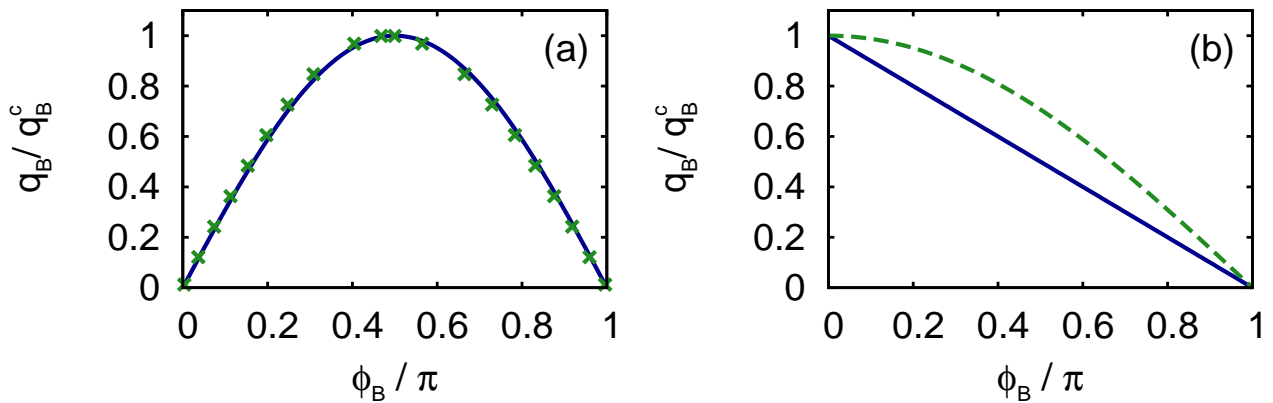}
\caption{Josephson characteristic $q_{B}(\phi_{B})$ obtained from the solution of the Schr\"{o}dinger-type equation 
              (\ref{G-P-one-dimensional-manipulated}) with an effective potential of the shape of Fig.\ref{fig40}(b) and  
              corresponding to a: 
              (a) Strong barrier (whose parameters are specified in the text); (b) Vanishingly small barrier.
              In the latter case, the result of the approximate calculation (full line) is compared with the corresponding solution 
              of the Gross-Pitaevskii equation (dashed line).}
\label{fig41}
\end{figure}
\end{center}

By this procedure, we end up with the Josephson characteristic $q_{B}(\phi_{B})$ plotted in Fig.\ref{fig41}(a), which 
can be nicely fitted by the expected relation $q_{B}(\phi_{B}) = q_{B}^{c} \, \sin \phi_{B}$ with 
$q_{B}^{c} = 8.26 \times 10^{-3} \, \xi_{\mathrm{phase}}^{-1}$.
Note that for the same rectangular barrier, the numerical solution of the original Gross-Pitaevskii equation (\ref{G-P-equation-one-dimensional}) yields the same 
functional relation of $q_{B}$ vs $\phi_{B}$ with the value $q_{B}^{c} = 14.0 \times 10^{-3} \, \xi_{\mathrm{phase}}^{-1}$.

It is interesting to consider also the opposite case of a vanishingly small barrier, so that 
$\bar{V}_{0} = \bar{V}_{1} = - \alpha \, g_{B} n_{B}^{0}$ and the effective potential of Fig.\ref{fig40}(b) reduces to a square well of depth $|\bar{V}_{1}|$ and width $2 b$.
In this case, only three spatial regions need be considered for the scattering problem and one 
can readily solve analytically for the transfer matrix of Eq.(\ref{transfer-matrix}). One obtains:
\begin{equation}
Y_{21} = \frac{i}{2} \left( \frac{q_{B}}{\sqrt{q_{B}^{2} + 2 m_{B} |\bar{V}_{1}|}} -
\frac{\sqrt{q_{B}^{2} + 2 m_{B} |\bar{V}_{1}|}}{q_{B}} \right)
\sin (2 b \sqrt{q_{B}^{2} + 2 m_{B} |\bar{V}_{1}|})                           \label{Y-21}         
\end{equation}

\noindent
which vanishes for $b = b^{(I)} = 0$ and 
$b = b^{(II)} = \pi / (2 \sqrt{q_{B}^{2} + 2 m_{B} |\bar{V}_{1}|})$.
Only the second solution is of interest here.
In the present case, the value of the parameter $\alpha$ entering $|\bar{V}_{1}|$ cannot be determined \emph{a priori} 
but need be determined self-consistently. 
To this end, it is sufficient to know the expression of the wave function in the central region, namely,
\begin{equation}
|\Phi_{2}(x)|^{2} \, = \, 1 \, - \, \frac{2 m_{B} |\bar{V}_{1}|}{q_{B}^{2} + 2 m_{B} |\bar{V}_{1}|}
\, \, \cos^{2} (x \sqrt{q_{B}^{2} + 2 m_{B} |\bar{V}_{1}|}) \,\, ,   \label{square-wave-function-central}
\end{equation}

\noindent
whose average value in the interval $- b^{(II)} < x < b^{(II)}$ equals:
\begin{equation}
\langle |\Phi_{2}|^{2} \rangle \, = \, 1 \, - \, 
\frac{m_{B} |\bar{V}_{1}|}{q_{B}^{2} + 2 m_{B} |\bar{V}_{1}|}  \,\, .  
                                                                                \label{average-wave-function-central}
\end{equation}

\noindent
Recalling that by our definition $\alpha = 1 - \langle |\Phi_{2}|^{2} \rangle$,  
Eq.(\ref{average-wave-function-central}) yields eventually the result:
\begin{equation}
\alpha \, = \, \frac{1}{2} \, - \, \frac{q_{B}^{2}}{2 m_{B}} \, \frac{1}{g_{B} n_{B}^{0}} 
                                                                                \label{self-consistent-alpha}
\end{equation}

\noindent
where the assumption $0 < \alpha$ requires that $q_{B} < q_{B}^{c} = \sqrt{m_{B} g_{B} n_{B}^{0}}$.
Recalling that $g_{B} n_{B}^{0} = m_{B} s^{2}$, we thus obtain that the superfluid velocity 
$V = q_{B} / m_{B}$ cannot exceed the sound velocity $s$ of the original bosonic system, 
as required by the Landau criterion.

Finally, from the matrix element $Y_{11}$ we can also determine the phase $\phi_{B}$ of the 
non-vanishing coefficient in the rightmost region, and obtain for the Josephson characteristic the result:
\begin{equation}
\phi_{B} \, = \, \pi \left( 1 \, - \, \frac{q_{B}}{\sqrt{q_{B}^{2} + 2 m_{B} |\bar{V}_{1}|}} \right)
\, = \, \pi \left( 1 \, - \, \frac{q_{B}}{q_{B}^{c} } \right) \,\, .                    \label{q-linear}
\end{equation}

\noindent
This expression is plotted in Fig.\ref{fig41}(b) together with the corresponding result of the Gross-Pitaevskii equation, 
which gives $q_{B} = q_{B}^{c} \, \cos (\phi_{B} / 2)$ for the same ``second branch'' of solutions.
In spite of the crudeness of the model that we have adopted for the effective potential 
$V_{\mathrm{eff}}(x)$, it is remarkable that our simplified solution captures the essential 
features of the exact solution, especially as far as the ``second branch'' is concerned (which 
always proves the most difficult one to be obtained).

\vspace{0.5cm} 
\section{Appendix D. Some useful manipulations for the numerical solution of the 
             Gross-Pitaevskii equation}

It was stressed in sub-section 2d that one keystone of the present approach to the Josephson effect throughout the BCS-BEC crossover is the possibility of confronting the numerical calculations based on 
the fermionic BdG equations (\ref{BdG-equations}) in the strong-coupling (BEC) limit with the independent solution of the Gross-Pitaevskii equation (\ref{Gross-Pitaevskii-equation}), provided some reasonable physical conditions are satisfied \cite{PS-2003}.
The conditions for the agreement between the two alternative (fermionic and bosonic) calculations were also discussed in sub-section 2d (cf., in particular, Fig.\ref{fig10} therein), and were repeatedly explored in Section 4 while discussing the results for the Josephson and related effects with an SsS barrier.
The relationship with the Gross-Pitaevskii equation was explored, in addition, in Appendix C for providing an alternative interpretation of the Josephson effect, whereby the role of self-consistency was emphasized as leading to a destructive interference for the wave reflected by the barrier.

In this context, it is worth mentioning that the solution of the one-dimensional version 
(\ref{G-P-equation-one-dimensional}) of the Gross-Pitaevskii equation which describes a steady flow 
past an obstacle was discussed with some length in the literature, even providing analytic results in some limits 
(cf. Ref.\cite{Hakim-1997} and references quoted therein).
As a matter of fact, these analytic results have been utilized in our calculations whenever possible.
Yet, in most circumstances we had to solve numerically the equation 
(\ref{G-P-equation-one-dimensional}) in the presence of a barrier with a slab geometry and a finite current.
To this end, some preliminary formal manipulations of this equation have proven useful to speed its numerical solution. 
We report them here for completeness.

Recalling the boundary condition (\ref{boundary-conditions}) and the modified value of the chemical potential $\mu_{B} = g_{B} n_{B}^{0} + q_{B}^{2}/(2 m_{B})$ in the presence of a current, we begin by dividing both members of Eq.(\ref{G-P-equation-one-dimensional}) by $g_{B} \, (n_{B}^{0})^{3/2}$ and introduce the notation $\xi_{\mathrm{phase}} = (4 m_{B} g_{B} \, n_{B}^{0})^{-1/2}$, $b(x) = V_{B}(x)/g_{B} \, n_{B}^{0}$, and $\tilde{q} = q_{B} \, \xi_{\mathrm{phase}}$, as well as the rescaled variable $y = x / \xi_{\mathrm{phase}}$.
Setting further $\Phi(y)/(n_{B}^{0})^{1/2}  \equiv f(y) = f_{0}(y) \, \exp\{i \phi_{B}(y)\} \, \exp\{i \tilde{q} y\}$
(where $f_{0}(y)$ is a real function - cf. Eq.(\ref{Phi-one-dimensional})), the one-dimensional Gross-Pitaevskii equation (\ref{G-P-equation-one-dimensional}) then reduces to the form:
\begin{equation}
2 \, f''(y) \, + \, \left( 1 \, + \, 2 \, \tilde{q}^{2} \, - \, b(y) \right) \, f(y) \, - \, f_{0}(y)^{2} \, f(y) \, = \, 0 \,\, .
                                                                                     \label{simplified-Phi-one-dimensional}
\end{equation}

\noindent
Here, the second derivative of $f(y)$ can be expressed in terms of $f_{0}(y)$,  $\phi_{B}(y)$, and
$\tilde{q}$, such that the real and imaginary parts of Eq.(\ref{simplified-Phi-one-dimensional}) can 
be separated, in the order, as follows:
 \begin{equation}
\left\{ 
\begin{array}{l}
2 \, f_{0}''(y) \, - \, 2 \, f_{0}(y) \, \left( \tilde{q} \, + \, \phi_{B}'(y) \right)^{2} \, + \,
f_{0}(y) \, \left( 1\, + \, 2 \, \tilde{q}^{2} \, - \, b(y) \right) \, - \, f_{0}(y)^{3} \, = \, 0   \\
2 \, f_{0}'(y) \, \left( \tilde{q} \, + \,  \phi_{B}'(y) \right) \, + \, f_{0}(y) \, \phi_{B}''(y) \, = \, 0  \,\, .
\end{array}
\right.                                                       \label{real-imaginary-parts-GP-one-dimensional}
\end{equation}

\noindent
We note that the second of these equations corresponds to the condition of a uniform current.
From the expression (\ref{current-GP}) of the current, one writes, in fact, for this condition: 
\begin{equation}
f_{0}(y)^{2} \, \left( \tilde{q} \, + \, \phi_{B}'(y) \right) \, = \, \tilde{q}  \,\, ,        
                                                                                             \label{simplified-continuity-equation} 
\end{equation}

\noindent
whose derivative yields the second of Eqs.(\ref{real-imaginary-parts-GP-one-dimensional}).
We can then conveniently replace the second of Eqs.(\ref{real-imaginary-parts-GP-one-dimensional}) 
by Eq.(\ref{simplified-continuity-equation}).

Further manipulation can be done in order to remove the function $\phi_{B}(y)$ from the first of 
Eqs.(\ref{real-imaginary-parts-GP-one-dimensional}), by making it to depend only on the function $f_{0}(y)$.
Upon multiplying both sides of that equation by $f_{0}(y)^{3}$, the second term on its right-hand side
becomes $-2 f_{0}(y)^{4} ( \tilde{q} + \phi_{B}'(y))^{2} = -2 \tilde{q}^{2}$ according to 
Eq.(\ref{simplified-continuity-equation}).
In the place of Eqs.(\ref{real-imaginary-parts-GP-one-dimensional}), we thus solve explicitly the two
following equations:
\begin{equation}
\left\{ 
\begin{array}{l}
f_{0}(y)^{3} \, \left[ 2 \, f_{0}''(y) \, + \, f_{0}(y) \, \left( 1\, + \, 2 \, \tilde{q}^{2} \, - \, b(y) \right) 
                              \, - \, f_{0}(y)^{3} \right] \, = \, 2 \, \tilde{q}^{2}   \\
f_{0}(y)^{2} \, \left( \tilde{q} \, + \, \phi_{B}'(y) \right) \, = \, \tilde{q}  \,\, .
\end{array}
\right.                                                                \label{final-equations-from-GP}
\end{equation}

\noindent
Once the first of these equations is solved for $f_{0}(y)$, the second one yields:

\begin{equation}
\phi_{B}(y) \, = \,  \tilde{q} \, \int_{- \infty}^{y} \! dy' \,\, \frac{\left( 1 \, - \, f_{0}(y')^{2} \right)}{f_{0}(y')^{2}}
\end{equation}
 
\noindent
where we have assumed $\phi_{B}(y = - \infty) = 0$ as usual.

Finally, the order of the first of Eqs.(\ref{final-equations-from-GP}) can be lowered in the spatial regions 
where $b(y)$ is constant.
To this end, we multiply both sides of that equation by $f_{0}'(y)$, such that it can be rewritten in the form:
\begin{equation}
\frac{d}{d y}  f_{0}'(y)^{2} \, + \, \frac{(1\, + \, 2 \, \tilde{q}^{2} \, - \, b(y))}{2} \, 
\frac{d}{d y} f_{0}(y)^{2} \, - \, \frac{1}{4} \frac{d}{d y} f_{0}(y)^{4} \, + \, 
\tilde{q}^{2} \frac{d}{d y} \frac{1}{f_{0}(y)^{2}} \, = \, 0 \,\, .            \label{real-GP-manipulated-once-more}
\end{equation}

\noindent
In particular, outside a rectangular barrier $b(y)=0$, so that  Eq.(\ref{real-GP-manipulated-once-more}) reduces to
the form:
\begin{equation}
f_{0}'(y)^{2} \, + \, \frac{(1\, + \, 2 \, \tilde{q}^{2})}{2} \, f_{0}(y)^{2} \, - \, \frac{1}{4} f_{0}(y)^{4} \, + \, 
\tilde{q}^{2} \frac{1}{f_{0}(y)^{2}} \, = \, \frac{1}{4} \, + \, 2 \, \tilde{q}^{2}     \,\, ,      
                                                                                                               \label{real-GP-manipulated-constant}
\end{equation}

\noindent
where the value of the constant on the right-hand side has been obtained by considering either one of the limits $y \rightarrow \pm \infty$ where $f_{0}(y \rightarrow \pm \infty) = 1$ and 
$f_{0}'(y \rightarrow \pm \infty)=0$.

Once the numerical solution of the one-dimensional Gross-Pitaevskii equation in the presence of a current 
is obtained along these lines, comparison with the corresponding solution of the fermionic BdG equations in the strong-coupling (BEC) limit is eventually made by recalling the relations $m_{B}=2m$, 
$n_{B}^{0}=n_{0}/2$, $q_{B}=2q$, $V_{B}(x)=2V(x)$, and $a_{B}=2a_{F}$, as well as the rescaling 
(\ref{Phi-vs-Delta}) between $\Phi(x)$ and $\Delta(x)$.

\vspace{0.5cm} 
\section{Appendix E. Failure of the transfer-matrix method for solving the BdG equations}

When solving for an ordinary one-dimensional scattering problem described by the Schr\"{o}dinger equation with a finite-range potential, it is often convenient to resort to the so-called 
\emph{transfer matrix} whereby the amplitudes of the wave function at the farthest right of the potential are expressed in terms of those at its farthest left.
In particular, when the profile of the potential is approximated by a step-like curve over a number of intervals 
as we did in sub-section 2b, we can solve the Schr\"{o}dinger equation by elementary methods in each of these intervals and connect the solutions in contiguous intervals via the continuity conditions.
In this way, the amplitudes of the wave function in the $\ell$-th interval can be expressed in terms of those of the $(\ell-1)$-th interval, and by recursion in terms of those of the interval with $\ell=1$ at the farthest left.  
Reaching eventually the interval with $\ell=M$ at the farthest right, one ends up with an operative definition of the transfer matrix.
The advantage of this method is that one always deals with matrices of a small size ($2 \times 2$
for the ordinary Schr\"{o}dinger equation); the disadvantage is that these matrices needs to be multiplied for a large ($M-1$) number of times to obtain the required transfer matrix.

When dealing with a one-dimensional scattering problem described by the BdG equations as in the present paper, it is then natural to ask whether it might be convenient to organize its solution in terms of the transfer matrix, which again considers the sequence of continuity conditions taken one at a time instead of grouping them altogether in a single equation as we did in sub-section 2b 
[cf. Eq.(\ref{equation-AW=B})].
To obtain the transfer matrix, the continuity conditions (\ref{continuity-conditions}) at the point $x_{\ell}$ between the $\ell$-th and $(\ell+1)$-th intervals have to be rewritten in the form
\begin{equation}
W_{\ell + 1} \, = \, \mathbf{M}_{\ell + 1}(x=x_{\ell})^{-1} \, \mathbf{M}_{\ell}(x=x_{\ell}) \, W_{\ell}  
                                                                                               \label{continuity-conditions-rewritten}
\end{equation}

\noindent
where $W_{\ell}$ is the column vector (\ref{column-vector-Wl}), such that eventually:
\begin{equation}
W_{\ell = M} \, = \, \left(  \prod_{\ell' = 1}^{M-1} \, 
 \mathbf{M}_{\ell' + 1}(x=x_{\ell'})^{-1} \, \mathbf{M}_{\ell'}(x=x_{\ell'}) \, \right) W_{\ell = 1} 
                                                                                               \label{transfer-matrix-BdG}
\end{equation}

\noindent
which defines the $4 \times 4$ transfer matrix for the BdG equations.
The unknowns to be determined by solving for Eq.(\ref{transfer-matrix-BdG}) can again be
identified by using ``outgoing boundary conditions'' as we did in sub-section 2b.
For instance, when one selects an electron-like excitation impinging from the left with
$(a_{1}=1,d_{1}=0,b_{M}=0,c_{M}=0)$, an electron-like excitation impinging from the right with
$(a_{1}=0,d_{1}=0,b_{M}=1,c_{M}=0)$, a hole-like excitation impinging from the left with 
$(a_{1}=0,d_{1}=1,b_{M}=0,c_{M}=0)$, or a hole-like excitation impinging from the right with 
$(a_{1}=0,d_{1}=0,b_{M}=0,c_{M}=1)$, in all these cases the unknowns of the inhomogeneous system (\ref{transfer-matrix-BdG}) are $(b_{1},c_{1},a_{M},d_{M})$.
When looking instead for bound-state solutions with $(a_{1}=0,d_{1}=0,b_{M}=0,c_{M}=0)$, the
system (\ref{transfer-matrix-BdG}) becomes homogeneous in the same unknowns
$(b_{1},c_{1},a_{M},d_{M})$.
All other coefficients in the intermediate regions with $\ell = (2,\cdots,M-1)$ can be determined, 
for instance, from $(b_{1},c_{1})$ by applying the condition (\ref{continuity-conditions-rewritten}) recursively from each interval to the next.
Finally, the expedient introduced in Eqs.(\ref{Psi-vs-Upsilon-rescaled}) and (\ref{continuity-conditions-rescaled}) can 
also be used here for the same reasons.
The expression (\ref{continuity-conditions-rewritten}) is thus rewritten as follows:
\begin{equation}
\tilde{W}_{\ell + 1} \, = \, \mathbf{M}_{\ell + 1}(x=0)^{-1} \, \mathbf{M}_{\ell}(x=x_{\ell} - x_{\ell - 1}) \, 
\tilde{W}_{\ell}      \,\, ,                                            \label{continuity-conditions-rewritten-rescaled}
\end{equation}

\noindent
such that the values of the variable $x$ entering the phase factors $\exp\{i k_{\ell}^{(n)}x\}$ in
the expression (\ref{column-vector-Mn}) do not exceed the size of the largest interval.

By this procedure, one has traded the large [$4(M-1) \times 4(M-1)$] size of the single matrix $\mathbf{A}$ in
Eq.(\ref{equation-AW=B}) with the [$(M-1)$ times] repeated multiplication in Eq.(\ref{transfer-matrix-BdG})
of a small [$4 \times 4$] matrix.
Although this trading may, in principle, seem favorable, in practice it turns out not to be the case
because there occurs a severe limitation on the number $n$ of iterations of the expression
(\ref{continuity-conditions-rewritten}) that one can actually perform (and therefore on the maximun number $M$ of zones 
one can rely on).
Following a similar line of reasoning to that introduced at the end of Section 3, we can take $\mathrm{Im}\{k\}/k_{F}$ at most of the order $500$, such that $(500)^{2n} \lapprox e^{500}$ as required by standard numerical precision.
This estimate yields $n \approx 40$, which is indeed too small a number for a satisfactory determination of the gap profiles that has to be sustained in the cycles of self-consistency.
The numerical attempts we have done along these lines have failed accordingly.
For the above reasons, we have stuck with the method discussed at length in Section 2, which solves for all continuity conditions at once as evidenced by Eq.(\ref{equation-AW=B}).

\vspace{0.5cm} 
\section{Appendix F. Failure of the tunneling Hamiltonian for describing the Josephson effect in the BCS-BEC crossover}
              
The theory of electron tunneling in superconductors has historically been based on the use of the tunneling 
Hamiltonian \cite{Mahan}, following the original treatment by Josephson on coherent pair tunneling 
\cite{Josephson}.
Only more recently the BdG equations have, in fact, been utilized to calculate physical quantities related to 
the Josephson effect \cite{BTK}, especially as far as its connection with the Andreev-Saint-James bound states 
is concerned. 
In this Appendix, we wish to comment on the reasons why attempts to use the tunneling Hamiltonian are bound 
to fail when trying to describe the Josephson effect throughout the BCS-BEC crossover.

Within the tunneling Hamiltonian formalism, the fermions at the left (L) and right (R) of a barrier are assumed to be independent from each other, in such a way that they are described by destruction $c_{\mathbf{k},\sigma}$ and creation  $c^{\dagger}_{\mathbf{k},\sigma}$ operators of wave vector $\mathbf{k}$ and spin component $\sigma$ (with $\mathbf{k}=\mathbf{k}_{L}$ at the left and $\mathbf{k}=\mathbf{k}_{R}$ at the right of the barrier)
which anticommute with each other.
The barrier itself provides a coupling between these two otherwise independent systems, via the so-called 
\emph{tunneling Hamiltonian}:
\begin{equation}
H_{\mathrm{t}} \, = \, \sum_{\mathbf{k}_{L},\mathbf{k}_{R},\sigma} \, \left( 
t_{\mathbf{k}_{L}\mathbf{k}_{R}} \, c^{\dagger}_{\mathbf{k}_{L},\sigma} \, c_{\mathbf{k}_{R},\sigma} \, + \, 
t_{\mathbf{k}_{L}\mathbf{k}_{R}}^{*} \, c^{\dagger}_{\mathbf{k}_{R},\sigma} \, c_{\mathbf{k}_{L},\sigma} \right)
                                                                                                                    \label{tunneling-Hamiltonian}
\end{equation}

\noindent
where $t_{\mathbf{k}_{L}\mathbf{k}_{R}}$ is the tunneling matrix element.
A full account of the tunneling processes would then require one to specify the dependence of this matrix elements on its arguments in some detail.

This problem is readily overcome in conventional superconductors, since the tunneling takes place over a 
narrow energy range of the order of the Debye energy about the Fermi surface.
The fermions involved in tunneling have thus wave vectors near the Fermi wave vectors $k_{F}^{L}$ and $k_{F}^{R}$ 
on the two sides of the barrier, so that it is appropriate to consider the tunneling matrix elements in
Eq.(\ref{tunneling-Hamiltonian}) as being approximately independent of their arguments \cite{Mahan}.
In this case, the Debye energy provides a natural cutoff for the integrals which enter the expression 
of the maximum Josephson current $J_{0} $ (in zero voltage), namely,
\begin{equation}
J_{0} \, = \, 2 \, \Delta_{L} \, \Delta_{R} \, \sum_{\mathbf{k}_{L},\mathbf{k}_{R}} \, 
\frac{|t_{\mathbf{k}_{L}\mathbf{k}_{R}}|^{2}}{E_{\mathbf{k}_{L}} \, E_{\mathbf{k}_{R}} \,
(E_{\mathbf{k}_{L}} \, + \, E_{\mathbf{k}_{R}})}              \label{maximum-current-tunneling-Hamiltonian}
\end{equation}

\noindent
and the integrals are evidently finite even in three dimensions when one approximates 
$|t_{\mathbf{k}_{L}\mathbf{k}_{R}}|^{2} \approx |t_{k_{F}^{L}k_{F}^{R}}|^{2}$.
[Equivalently, within the above approximation one may transform the wave-vector integrals into energy integrals, whereby the density of state is taken at the Fermi level consistently with the BCS (weak-coupling) limit,
in such a way that the ensuing integrals converge like in two dimensions.]

It is, however, clear that this approximation does not apply when considering the BCS-BEC crossover, for which no upper 
momentum cutoff can be introduced in order to account for the formation of the composite bosons.
In this case, the integral on the right-hand side of Eq.(\ref{maximum-current-tunneling-Hamiltonian}) would be
ultraviolet divergent, unless the factor $|t_{\mathbf{k}_{L}\mathbf{k}_{R}}|^{2}$ therein vanishes rapidly enough 
for large $|\mathbf{k}_{L}|$ and $|\mathbf{k}_{R}|$.
Making that integral to converge would thus require a resonable knowledge of the tunneling properties for fermions impinging on the barrier with large energies, knowledge which could only come from a separate study of the scattering on the barrier of superfluid fermions with all possible energies.
This study is naturally provided by approaching the problem directly in terms of the BdG equations, as we have consistently done in this paper, thus making it useless any further recourse to the tunneling Hamiltonian approach.


\end{document}